%% file: Thesis.tex
\documentclass[12pt,twoside,letterpaper]{mitthesis}
\usepackage{graphics}

\begin{document}

\pagestyle{empty}
\include{Cover}

\cleardoublepage

\pagestyle{plain}

\tableofcontents
\newpage


\cleardoublepage
\include{Introduction}

\cleardoublepage
\include{DarkMatter}

\cleardoublepage
\include{SUSY}

\cleardoublepage
\include{CosmicRays}

\cleardoublepage
\include{Detector}

\cleardoublepage
\include{DataAnalysis}

\cleardoublepage
\include{SusySignals}
\cleardoublepage
\include{Backgrounds}

\cleardoublepage
\include{Limits}

\cleardoublepage
\include{Results}

\cleardoublepage
\include{Future}
\cleardoublepage
\appendix
\include{Fitting}
\cleardoublepage
\include{Geomagnetic}

\cleardoublepage
\include{Propagation}
\cleardoublepage
\include{ExtraPlots}

\cleardoublepage
\bibliography{bib}

\end{document}

%% file: Cover.tex
\title{A Search for Z=$-1$ Dark Matter Annihilation Products in Cosmic Rays with AMS-01}
\author{Gray Rybka}
\department{Department of Physics}
\degree{Doctor of Philosophy}
\degreemonth{June}
\degreeyear{2007}
\thesisdate{May 30, 2007}

\supervisor{Peter Fisher}{Professor}
\chairman{Thomas Greytak}{Associate Department Head for Education}
\maketitle
\cleardoublepage

\begin{abstractpage}

The majority of mass in the universe has not been observed optically and is termed dark matter.  The supersymmetric neutralino provides an interesting dark matter candidate, which may self-annihilate in our galaxy, producing particles visible in the cosmic ray spectrum.  During a ten day space shuttle flight, the AMS-01 detector recorded over 100 million cosmic ray events.  This analysis searches for the products of neutralino annihilation in the AMS-01 Z=-1 spectrum, and uses the results to place limits on which supersymmetric and dark matter halo distribution models are compatible.

\end{abstractpage}

\cleardoublepage

\section*{Acknowledgments}

	I would like to thank a number of people who have made this thesis possible and my graduate experience more enjoyable:  The AMS Collaboration for completing the AMS-01 experiment long before I joined.  My adviser, Prof. Peter Fisher, who is also my role model as a scientist.  Prof. Ulrich Becker for imparting some of his immeasurable detector wisdom to me.  Sa Xiao, who gently pointed out the most egregious errors in my analysis so that I might correct them.  Kathryn Oseen-Senda for her moral support and editing advice.  Igor Moskalenko, who revealed to me the secrets of GALPROP.  Christine Titus and Laurence Barrin for keeping the labs running smoothly.  Finally I'd like to thank to my fellow graduate students, LNS personnel, friends and family for their support during my graduate career.

%% file: Introduction.tex
\chapter{Introduction}
\section{Two Questions in Physics}

	Through a number of different astrophysical observations it has become clear that a large quantity of gravitating matter in the universe is electromagnetically invisible \cite{ref:pdg}.  The total quantity of dark matter in our galaxy can be estimated through studies of the motions of stars and galaxies, and the amount of dark matter in the universe can be measured with the cosmic microwave background \cite{ref:wmap}.  The composition of dark matter, however, is still very much an open question.  Current theories suggest that it is particulate, massive, and non-baryonic, but beyond that, little is known.

	The results of nearly all present-day particle experiments are correctly predicted by the Standard Model \cite{ref:pdg}.  At energies slightly higher than are currently accessible at particle accelerators, however, the equations that govern Standard Model interactions break down.  There are numerous theories that extend the Standard Model to solve this problem, but at present the correct theory is unknown.

	Certain types of supersymmetric (SUSY) theories solve the high energy problems with the Standard Model and also predict the existence of the neutralino, a heavy, stable, neutral particle that should still remain from the Big Bang.  When the models are properly tuned, the total mass of these remnants add up to roughly the same mass that is attributed to dark matter, potentially answering both an astrophysical question and a particle physics question at the same time \cite{ref:ssdm}.  The neutralino is its own antiparticle, which means that there should be a steady annihilation of these dark matter neutralinos producing a detectable signal in our galaxy.  Solving two deep problems at once and having a testable prediction makes the neutralino a tempting experimental quarry.

\section{Previous and Ongoing Dark Matter Searches}
	
	A heavy, stable, neutral particle is not unique to supersymmetry.  In general, the supersymmetric neutralino is categorized as a Weakly-Interacting Massive Particle (WIMP).  There are two primary ways to search for WIMPs: directly and indirectly.  Direct searches hope to observe the rare event of a nucleus recoiling after being struck by a WIMP.  Indirect searches require the assumption that WIMPs are slowly annihilating in the galaxy and look for remnants of these annihilations.
	
	Recent direct searches, such as CDMS \cite{ref:cdmsnospin}, have used cooled semiconductor detectors as the source of nuclei to collide with WIMPs.  Future detectors will use larger and even more sensitive semiconductors \cite{ref:supercdms}, liquid noble gases \cite{ref:xenon10}, supercritical fluids \cite{ref:PICASSO}, and direction sensitive gas detectors \cite{ref:DRIFT}.  The goal of all of these methods is to discriminate the rare low momentum-transfer collisions between WIMPs and nuclei from charged particle and neutron backgrounds.  The rate in these experiments is dependent on WIMP-nucleon cross section and mass, as well as the dark matter density at Earth.

	Indirect searches look for a variety of WIMP annihilation products from a variety of locations: charged particles from annihilations nearby in the galaxy \cite{ref:HEAT1}, gamma rays from the center of the galaxy \cite{ref:EGRET1}, and neutrinos from annihilation of WIMPs trapped inside the sun \cite{ref:ICECUBE}.  The signal in these experiments depends on WIMP self-annihilation cross section, mass, branching ratio to the particle being observed, and the dark matter distribution throughout the galaxy.  The fact that indirect and direct searches depend on different interaction cross sections as well as the dark matter density over different length scales makes them complementary, though it makes comparison between direct and indirect results difficult and model dependent.  This analysis will be an indirect dark matter search.

\section{Goals}

	This thesis details an attempt to find or put limits on supersymmetric WIMP annihilation using the Alpha Magnetic Spectrometer 1 (AMS-01) experiment.  The analysis starts with a set of representative models in supersymmetric parameter space and, along with certain assumptions as to the distribution of dark matter in our galaxy and the propagation of cosmic rays, predicts the electron and antiproton signals at Earth from neutralino annihilation from each of these models.  It then describes a search for these signals against the cosmic ray background in the Z=$-1$ spectrum taken by the AMS-01 experiment.  From these results, limits on the possible distribution of dark matter in our galaxy are made, the limits are compared with results from other dark matter experiments, and directions for future experiments are suggested.

	The argument is laid out in the following chapters:
\begin{description}
\item[Chapter 2: Dark Matter] briefly describes evidence for the existence of dark matter, as well as several theories as to how it is distributed in galaxies.
\item[Chapter 3: Supersymmetry] explains the primary reasons supersymmetry is an attractive theory and how representative models were chosen out of such a large variation of possible input parameters.
\item[Chapter 4: Cosmic Rays] summarizes relevant knowledge about where cosmic rays, the primary background for this experiment, come from and how it is thought that they behave while traveling through the galaxy.
\item[Chapter 5: The AMS-01 Detector] describes the experiment from which all data for this analysis is drawn: its mission, its design, and details of its construction relevant to the analysis.
\item[Chapter 6: Data Analysis] covers the steps taken to turn raw detector events into a spectrum of Z=-1 particles and the simulation done to determine how a particle flux at Earth looks in the detector.
\item[Chapter 7: SUSY Signals] begins the analysis in earnest, describing the methods used to turn a supersymmetric model into an expected particle flux above Earth.
\item[Chapter 8: Backgrounds] discusses the way cosmic ray backgrounds were fitted to the data, and how the charge misidentified event spectrum was estimated and subtracted from the Z=-1 spectrum.
\item[Chapter 9: Limits] describes the simultaneous fitting of both neutralino annihilation signal and background to the data and how upper bounds on signal strength were found.
\item[Chapter 10: Results] presents the result of the aforementioned fitting in the form of an upper bound on the boost factor from dark matter distribution, and compares these results with those from other recent dark matter experiments.  It then explains the implications of these limits and under what conditions they are expected to hold true.
\item[Chapter 11: Recommendations for Future Work] summarizes the lessons learned in this analysis and makes suggestions for improvements in future experiments and theory.
\end{description}

\begin{figure}[htp]
\centering
\includegraphics[width=16cm]{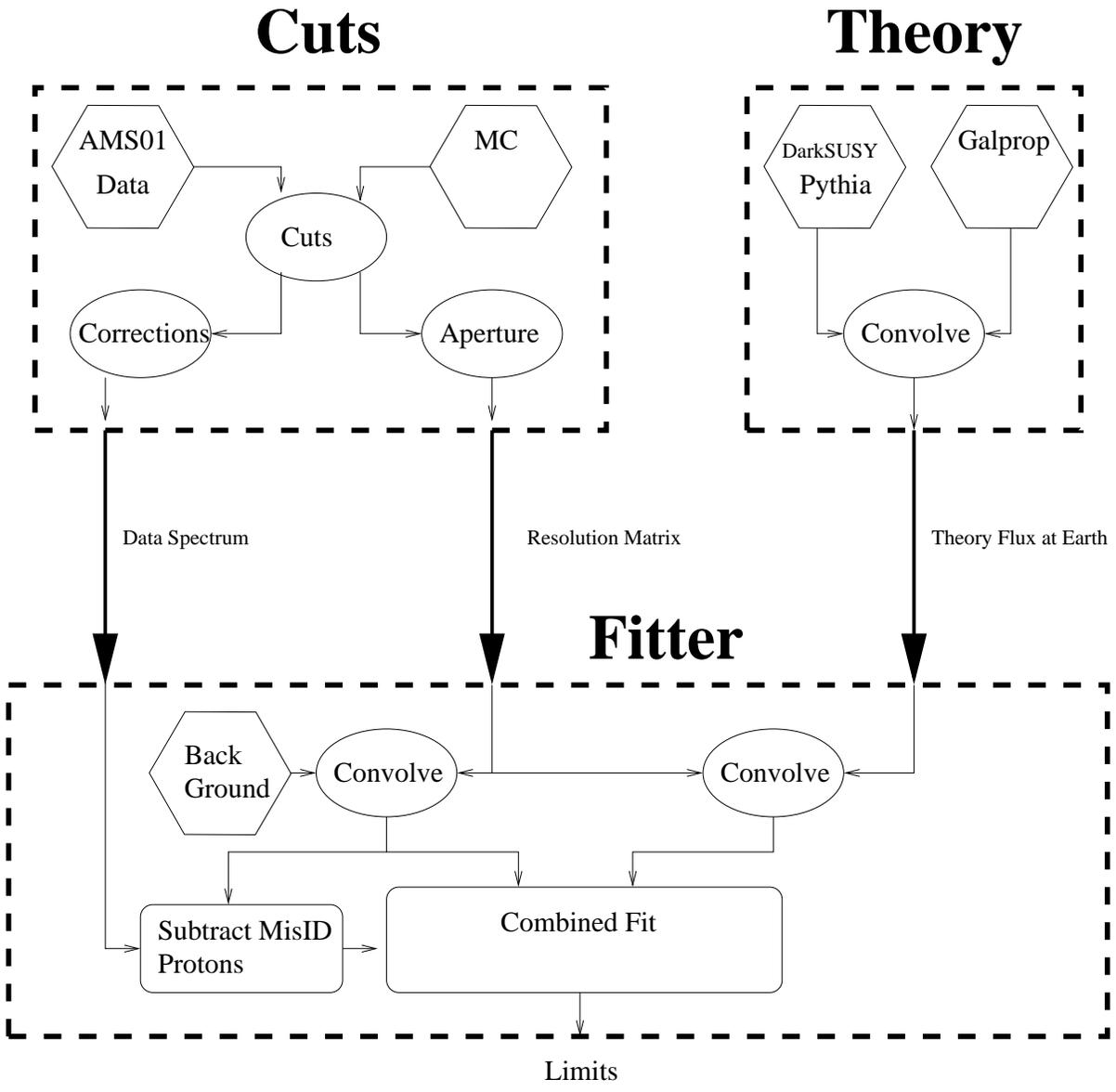}
\caption[Analysis Flowchart]{The Analysis Flowchart.  This shows the important steps in this analysis and the flow of information between steps.}
\label{fig:intro:flowchart}
\end{figure}

The flow of information in the analysis is shown diagrammatically in Fig.~\ref{fig:intro:flowchart}.

%% file: DarkMatter.tex
\chapter{Dark Matter}

	Wherever the mass-to-light ratio is measured in the universe, it is found to be greater than expected from gas and stars alone \cite{ref:galaxybook}.  This excess mass is called dark matter.  This chapter will briefly go over a few pieces of evidence supporting the existence of dark matter and some suggested candidates for dark matter.  Finally, models for the amount and distribution of dark matter in our galaxy will be discussed.
	
\section{Evidence for Dark Matter}

	The earliest hints of dark matter came from F.~Zwicky's \cite{ref:zwicky} observations of the Coma cluster.  The velocity dispersion of galaxies in the Coma cluster indicated a far higher mass than was expected from the total light output of the cluster.  More recent studies which fit the amount of X-ray emitting gas to the gravitational potential of clusters like Coma indicate that around 85\% of the mass of these clusters is in the form of dark matter \cite{ref:clusterreview}.

	Current theories of structure formation in the universe after the Big Bang suggest that the structure of temperature deviations in the angular distribution of the cosmic microwave background is highly sensitive to the dark matter content of the universe.  With too little dark matter the balance between radiation pressure and gravitation on baryonic matter would cause less structure to form than is observed in the cosmic microwave background.  With too much dark matter, its gravitation would overcome the radiation pressure and cause more structure than is observed.
The recent measurement of the cosmic microwave background by the WMAP experiment are well explained by the $\Lambda$CDM model, which describes a flat universe with both dark energy and cold dark matter.  The WMAP measurement indicates that there is roughly 6 times as much dark matter as baryonic matter in the universe \cite{ref:wmap}.

		If stars were the primary source of mass in galaxies then the enormous density of stars at the core would make their average rotational velocity fall off with radius as $1/\sqrt{R}$ toward the edges of the galaxy.  Rotational velocities in almost all galaxies, however, are independent of or rise with the radius out to the furthest measurable points \cite{ref:GalaxyRotCurves}.  This indicates the mass distribution of the galaxy is not the same as the light distribution, that is to say, stars are not the primary source of mass.  The simplest mass distribution that produces a flat rotation curve is an isothermal halo: the distribution an isothermal, collisionless gas takes when gravitationally bound by its own mass \cite{ref:galaxybook}.  This distribution is likely a result of how the halo was formed, rather than an indication that the dark matter is thermalized \cite{ref:nfw}.
	
\section{Constraints on Dark Matter Candidates}

	Astrophysical measurements put constraints on the properties of dark matter.  The degree to which structure is found in the universe indicates that the constituents of dark matter must be non-relativistic or ``cold" at the time of galaxy formation.  Relativistic dark matter would have flowed out from baryonic clumps and caused less structure in the universe than is presently seen \cite{ref:ssdm}.  Additionally, because dark matter is not visible, it must have no electromagnetic charge.  Strongly interacting dark matter would bind to quarks, creating composite charged particles that are not seen, so dark matter may not have color charge either.  Dark matter may still interact with other particles through the weak force, however.
	
	The most prosaic solution to the problem of dark matter would be large, dark clumps of normal baryonic matter, called Massive Compact Halo Objects (MACHOs).  Searches for MACHOs, such as the MACHO experiment, look for gravitational lensing from these massive objects, but have not found nearly enough to account for dark matter, making MACHOs an unlikely scenario \cite{ref:pdg}.

	Neutrinos have very weak interactions with other matter, and were initially thought a good candidate for dark matter.  Unfortunately, the mass of standard model neutrinos is so low that they were relativistic at the beginning of galaxy formation, conflicting with the structure observations.  Also, the expected total neutrino mass from big bang nucleosynthesis is less than 1.4\% of the total mass in the universe, insufficient to account for dark matter \cite{ref:pdg}. 

	A more exotic candidate for dark matter would be a Weakly-Interacting Massive Particle (WIMP).  There are a number of WIMP candidates motivated by post-standard model particle physics.  One of these is the lightest supersymmetric particle (LSP), which will be the focus of this analysis.  A search for WIMPs can be conducted either directly, by observing a collision between a WIMP and a nucleus, or indirectly, by relying on the assumption that WIMPs are their own antiparticles and searching for the products of WIMP on WIMP annihilation \cite{ref:ssdm}.

	This analysis will focus on WIMP dark matter, but there are also non-WIMP dark matter candidates, such as the axion.  Axions provide a way to explain why the strong force is CP-invariant \cite{ref:axionorig}.  Dark matter axions are expected to have masses of a few $\mathrm{\mu eV}$.  The current limit on the density of (DFSZ) axions near Earth with masses from 1.9 to 3.3 $\mathrm{\mu eV}$  is 3 $\mathrm{GeV/cm^3}$ from the ADMX experiment, which searches for the process of an axion decaying to two photons \cite{ref:axions}.  Better limits, possibly excluding axions as the sole component of dark matter, are expected within a few years.
	
\section{Distribution of Dark Matter in our Galaxy}

	The vertical motions of stars with respect to the galactic disk are affected by the density of matter in the disk.  By measuring these motions, the mass to light ratio nearby in our galaxy is found to be 50\% higher than one would expect from only stars, gas, and dust, indicating the presence of dark matter near our solar system \cite{ref:galaxybook}.  An isothermal sphere that accounts for the velocity dispersion of nearby stars indicates that the local density of dark matter is in the range $\rho_0 = 0.2 - 0.4\ \mathrm{GeV\ cm^{-3}}$.  A flattened halo would only increase this density \cite{ref:dmdensity}.

\begin{figure}
\centering
\begin{tabular}{|c|c|c|}
\hline
\multicolumn{3}{|c|}{Scenarios for the distribution} \\
\multicolumn{3}{|c|}{of dark matter near Earth} \\
\hline
Smooth & Clumpy & Caustics \\
\hline
\includegraphics[width=4cm]{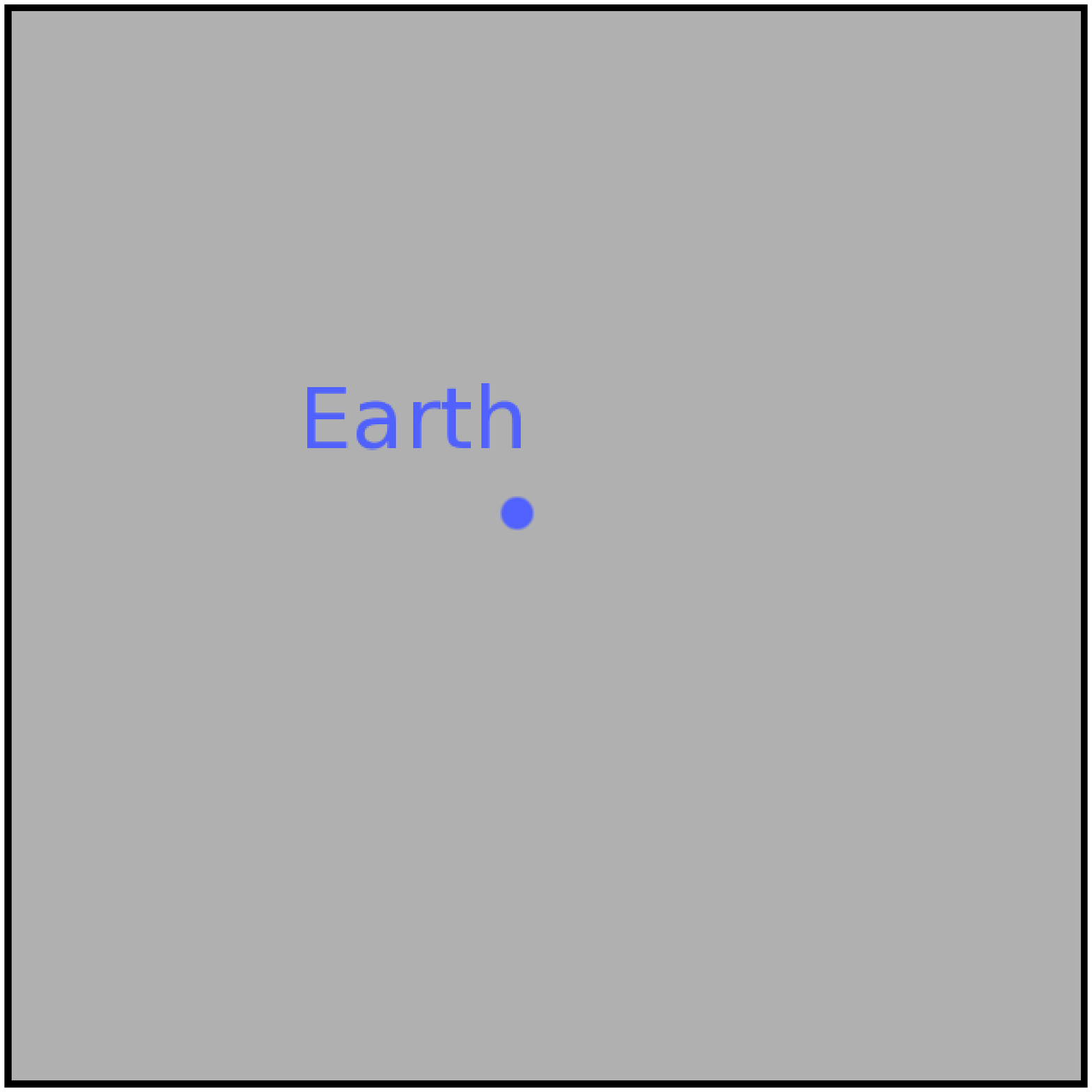} &
\includegraphics[width=4cm]{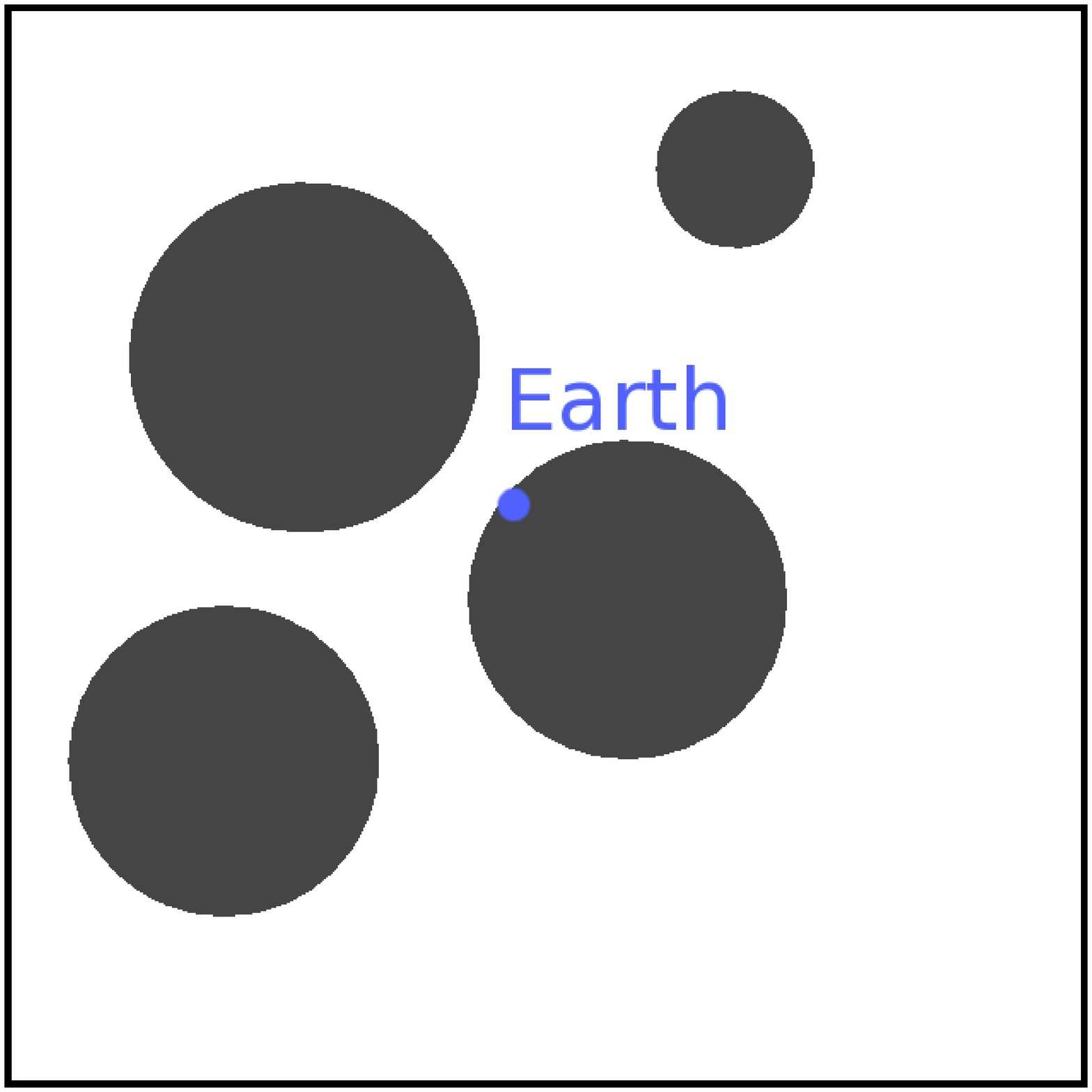} &
\includegraphics[width=4cm]{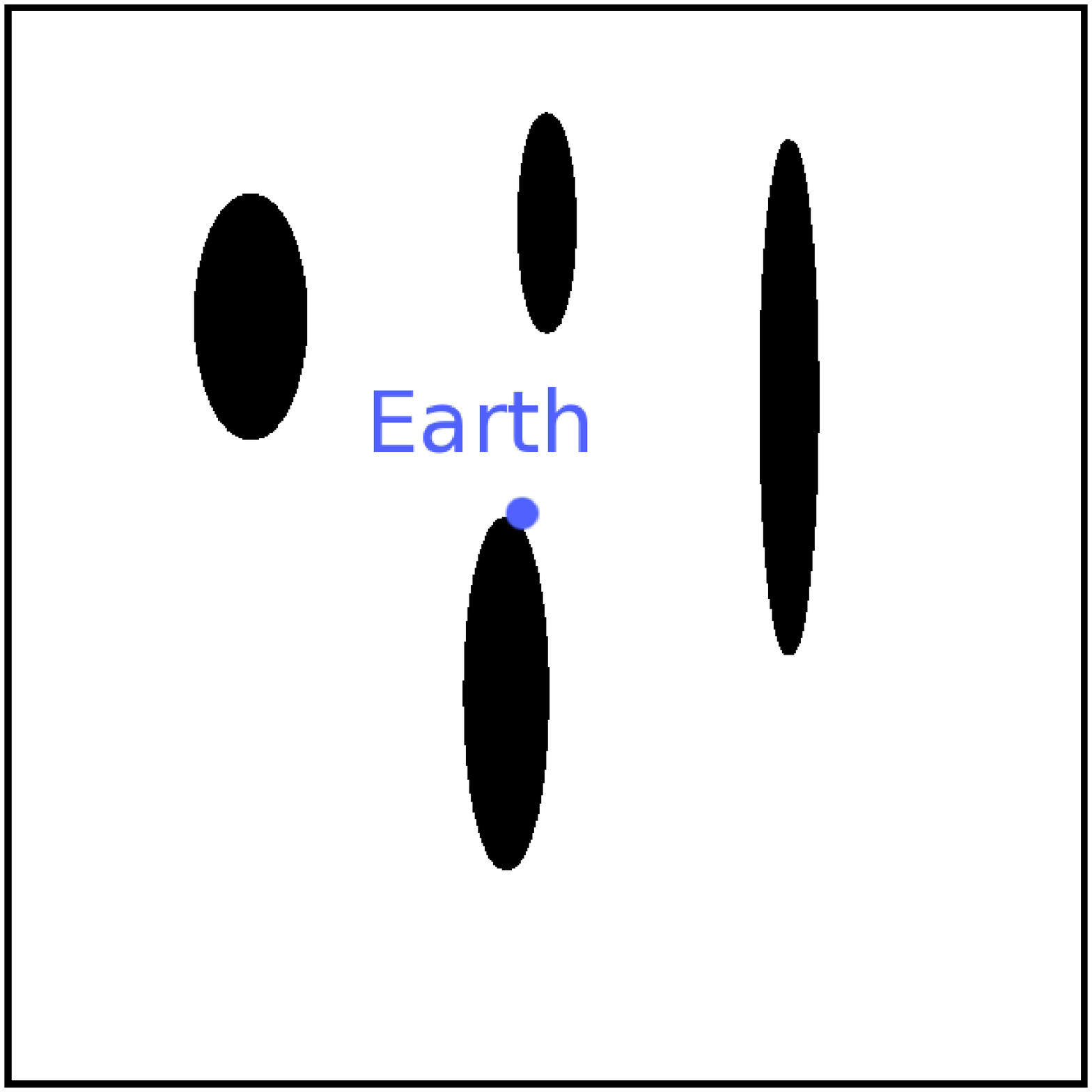} \\
\hline
Structure destroyed by tides. & Structure survives tides. & Structure deformed by tides. \\
Distribution has no clumps. & Distribution consists of & Distribution consists of \\
 & self-gravitating clumps. & stretched structures with \\
 &  & extremely dense cores. \\
\hline
Boost $=$ 1 & Boost $>$ 1 & Boost $\gg$ 1 \\
\hline
\end{tabular}

\caption[Dark Matter Distribution Scenarios]{ The dark matter near Earth my be smooth or clumpy.  These figures are meant to represent the possible distributions of dark matter within a few hundred parsecs of Earth.  If dark matter is clumpy, whether the Earth is in a clump or not is very important to direct detection.}
\label{fig:dm:clumtypes}
\end{figure}

		Though the galactic scale density distribution of dark matter is mostly fixed by the motion of stars in the galaxy, whether dark matter has structure at smaller scales is unknown.  Initial density fluctuations after the Big Bang are expected to produce clumps of dark matter at all scales \cite{ref:bertclumps}.  These clumps may still remain in the galaxy, or may have been wiped out by gravitational tides.  It is also possible that the clumps were compressed by tides \cite{ref:caustics}.  These possibilities are shown in Fig.~\ref{fig:dm:clumtypes}.  As the WIMP annihilation rate is proportional to dark matter density squared, the quality of this substructure that is important to indirect detection experiments is the spatially averaged density squared.  The lowest value this quantity can take is achieved from a smooth isothermal distribution, so it is useful to define the Boost Factor of a dark matter distribution model as $B=\langle\rho^2\rangle_{\mathrm{model}}/\langle\rho^2\rangle_{\mathrm{smooth}}$.  Thus if a particular type of WIMP would have an average annihilation rate of $R$ for a smooth isothermal sphere, it has a rate of $B\times R$ for a clumpy dark matter model with boost factor $B$.  The boost factors from clumps surviving galaxy formation could be in the hundreds\cite{ref:earlyclumps}, and even larger if these clumps have been compressed by gravitational tides.

		While an isothermal sphere has infinite density at its center, there are various ways to modify it to give a realistic, finite density.  This analysis will use the cored isothermal halo distribution \cite{ref:dmdensity}.  It is shown in \cite{ref:galpropgreens}, that the behavior of the dark matter distribution at the core does not significantly affect the signals of charged particles from dark matter annihilation seen at Earth.
		
		This analysis will take the density of dark matter in the region of the solar system to be $0.3\ \mathrm{GeV\ cm^{-3}}$, and treat any increase (or decrease) in this local density as part of the boost factor.  Thus if, for example,  the dark matter density is measured more accurately later and found to be $0.2\ \mathrm{GeV\ cm^{-3}}$, all quoted boost factor upper bounds should be increased by $(9/4)$.

%
%

%% file: SUSY.tex
\chapter{Supersymmetry}

	The supersymmetric neutralino is one of the most promising candidates for WIMP dark matter.  This chapter presents the supersymmetric theories used in this analysis, and the requirements dark matter places on supersymmetry.

\section{Supersymmetry and Particle Physics}

	The Standard Model of particle physics has two serious flaws: the Higgs mass is divergent because of loop diagrams and there is no explanation for the large variations in the strengths of the fundamental forces.  Supersymmetric theories may fix this by adding a new symmetry to the Poincar\'{e} group that forces each known particle to be paired to a superpartner particle with 1/2 less spin \cite{ref:firstsusy}.  Superpartners cancel out the loop divergences in the Higgs mass caused by their standard model partner, and cause force strengths to unify at high energies \cite{ref:susybook}.

	While supersymmetry is a component of many complicated theories, it is useful to consider the minimal supersymmetric extension to the standard model, the MSSM.  In the MSSM, the number of standard model particles are doubled: squarks are the superpartners of quarks, sleptons are the superpartners of leptons, photinos are the partners of photons and so on.  Unbroken supersymmetry yields superparticles with the same masses as their standard model partners.  Since no superpartners have been observed in searches extending to masses over 100GeV, the supersymmetry must be broken at a high energy, causing superpartners to be considerably more massive than their standard model partners.  All of the additional superpartner masses and couplings must be specified, along with parameters that describe how the supersymmetry is broken, to completely describe the MSSM (124 total parameters is a popular choice) \cite{ref:pdg}. By assuming little mixing in the supersymmetric sector, a more practical version of the MSSM can be formed with only 6 free parameters \cite{ref:ssdm}.

	Another supersymmetric model considered here is minimal supergravity (mSUGRA).  mSUGRA is the result of requiring the standard model to have a local supersymmetry that is compatible with general relativity.  It is effectively a further constrained version of the MSSM, and can be described by four parameters and the choice of sign on another parameter \cite{ref:pdg}.  Even as few as this still allows for a large variation in the predictions of supersymmetry.

	Supersymmetry manifests itself in particle experiments in a variety of ways.  It is hoped that sleptons, squarks, and other superparticles may be directly observed at the LHC \cite{ref:lhcsusy}.  Indirectly, the small effects of SUSY particles in loop diagrams may be observed in precision measurements of processes like $b\rightarrow s\gamma$ and the anomalous magnetic moment of the muon \cite{ref:pleasingmsugra}.  A combination of several different measurements is needed to determine all supersymmetric parameters, particularly as current measurements indicate no significant deviation from the standard model, and exclude very small regions of supersymmetric parameter space. 

\section{Supersymmetry and Dark Matter}

\begin{figure}[htp]
\centering
\includegraphics[width=15cm]{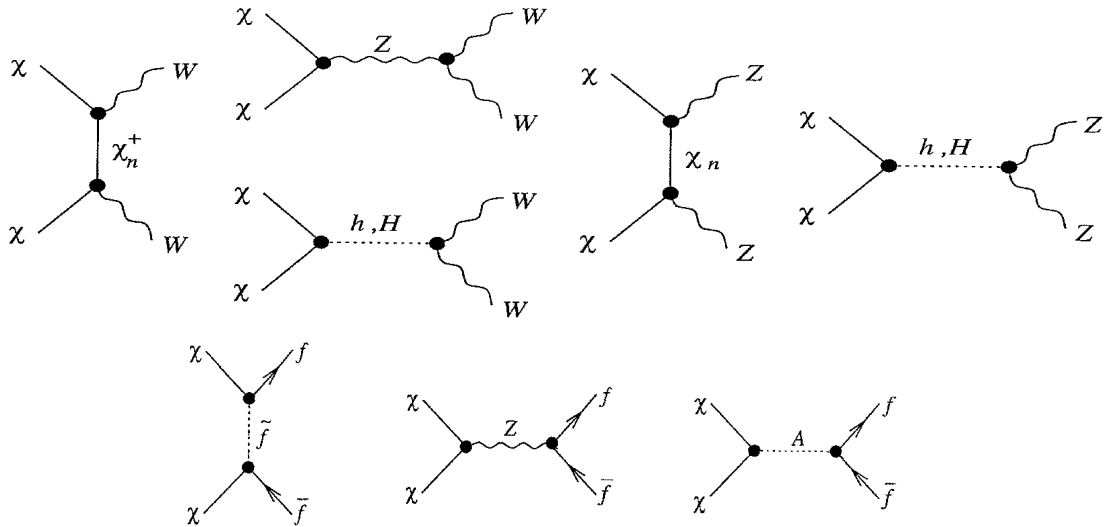}
\caption[Neutralino Annihilation]{Tree-level Feynman diagrams for SUSY neutralino annihilation into Ws, Zs, and fermions, from \cite{ref:ssdm}}
\label{fig:susy:feyndiags}
\end{figure}

   In some theories of SUSY, a proton can decay into leptons by way of superparticles.  The present proton lifetime lower bound is $2\times10^{29}$ years even for neutrino decay modes, so this process is clearly either very rare or nonexistent \cite{ref:snopdecay}.  Thus it has been posited that superparticles can only be produced or destroyed in pairs, preventing this mechanism of proton decay.  This is known as R-parity and has the interesting side effect that the lightest supersymmetric particle (LSP) cannot decay. Thus any LSPs produced in the Big Bang would still be present in the universe.  A neutral, massive LSP would then it would make an excellent candidate for WIMP dark matter.  The neutralino, a superposition of gauginos (the superpartners of bosons), fits this description.  The neutralino also has the fortunate quality of being its own antiparticle, making it capable of self-annihilation \cite{ref:ssdm}.  This analysis will focus on annihilation into standard model particles, as illustrated in Fig.~\ref{fig:susy:feyndiags}.

\begin{figure}[htp]
\centering
\includegraphics[width=16cm]{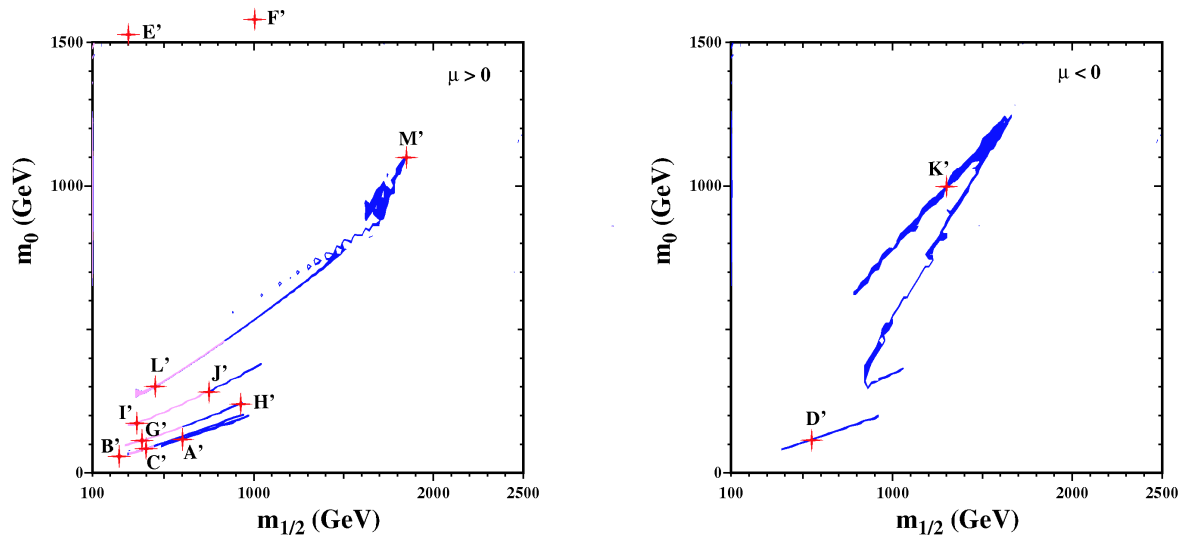}
\caption[mSUGRA Parameter Space]{mSUGRA parameter space. Parameters consistent with the WMAP 3-year data are shaded, and letters mark representative benchmarks from \cite{ref:sspointsupdated}.}
\label{fig:susy:benchmarks}
\end{figure}

   The expected total mass of SUSY neutralinos remaining from the Big Bang can be calculated for a given set of supersymmetric parameters.  Assuming dark matter is composed solely of neutralinos, measurements of the total dark matter mass of the universe can be used to put constraints on valid supersymmetric parameters.  The constraints on the parameters of the mSUGRA model from the WMAP data, from \cite{ref:sspointsupdated}, are shown in Fig.~\ref{fig:susy:benchmarks}.  Despite the reduction in parameter space, the neutralino mass and interactions still have considerable room to vary.  Other measurements are required to fix the properties of the neutralino.
   
	The neutralino mass, annihilation cross section, and favored annihilation channels are all strongly dependent on the values chosen for supersymmetric parameters.  This analysis uses the mSUGRA parameter sets defined in \cite{ref:sspoints} and updated in \cite{ref:sspointsupdated}, labeled Model A--Model M (The updated points were referred to as A'--M', but in this work the prime will be dropped for convenience, see Fig. \ref{fig:susy:benchmarks}).  These models are benchmarks, chosen to be representative of possible neutralino composition and interactions that are consistent with both the WMAP 3-year data and current accelerator limits, though they may not to cover the entire range of possible supersymmetric parameters.  A number of other mSUGRA and MSSM parameter sets suggested by other experiments will also be examined in the results chapter.

	An amount of calculation is involved to derive the neutralino properties from a parameter set.  These calculations are performed by the software package DarkSUSY \cite{ref:darksusy} and the results for neutralino masses, annihilation cross sections, and branching fractions shown in Table \ref{table:ss:dstable}.

\begin{table}[htp]
\centering
\input{dstable.tex}
\caption[Results from DarkSUSY]{Results from DarkSUSY.  Branchings are given as a fraction of total cross section.}
\label{table:ss:dstable}
\end{table}

%% file: dstable.tex
\begin{tabular}{|c|c|c|c|c|c|c|c|}
\hline
Model & $M_{\chi}$ (GeV) & $<\sigma_{\chi\chi}v> (\mathrm{cm^3s^{-1}})$ & $\rightarrow W^{+}W^{-}$ & $\rightarrow ZZ$ & $\rightarrow b\bar{b}$ & $\rightarrow t\bar{t}$ & $\rightarrow \tau^{+}\tau^{-}$ \\
\hline
A & 243 & $4.7 \times 10^{-29}$ & 0.004 & 0.001 & 0.105 & 0.876 & 0.014 \\
B & 95 & $6.8 \times 10^{-28}$ & 0.034 & 0     & 0.767 & 0     & 0.199 \\
C & 158 & $1.3 \times 10^{-28}$ & 0.015 & 0.003 & 0.695 & 0     & 0.287 \\
D & 212 & $6.9 \times 10^{-29}$ & 0.001 & 0.001 & 0.254 & 0.026 & 0.718 \\
E & 112 & $3.1 \times 10^{-27}$ & 0.865 & 0.106 & 0.027 & 0     & 0.002 \\
F & 419 & $1.3 \times 10^{-27}$ & 0.034 & 0.025 & 0.003 & 0.938 & 0     \\
G & 148 & $7.2 \times 10^{-28}$ & 0.003 & 0.001 & 0.720 & 0     & 0.276 \\
H & 388 & $2.8 \times 10^{-29}$ & 0     & 0     & 0.938 & 0.002 & 0.059 \\
I & 138 & $3.6 \times 10^{-27}$ & 0.001 & 0     & 0.945 & 0     & 0.054 \\
J & 309 & $3.3 \times 10^{-28}$ & 0     & 0     & 0.981 & 0     & 0.018 \\
K & 554 & $1.3 \times 10^{-26}$ & 0     & 0     & 0.889 & 0.001 & 0.110 \\
L & 181 & $8.4 \times 10^{-27}$ & 0     & 0     & 0.992 & 0     & 0.008 \\
M & 794 & $2.7 \times 10^{-27}$ & 0     & 0     & 0.890 & 0     & 0.110 \\
\hline
\end{tabular}

%% file: CosmicRays.tex
\chapter{Cosmic Rays}

	Experiments looking for charged particle signals from space must be able to pick out their signals against the cosmic ray background.  Charged cosmic rays consist primarily of protons (86\%) and ionized helium (11\%), though they contain almost all stable charged nuclei, as well as electrons, positrons, and antiprotons \cite{ref:pdg}.  Of interest to this analysis are the proton, electron, and antiproton spectra in the region from 1GeV to 1 TeV.  

\begin{figure}[htp]
\centering
\includegraphics[width=12cm]{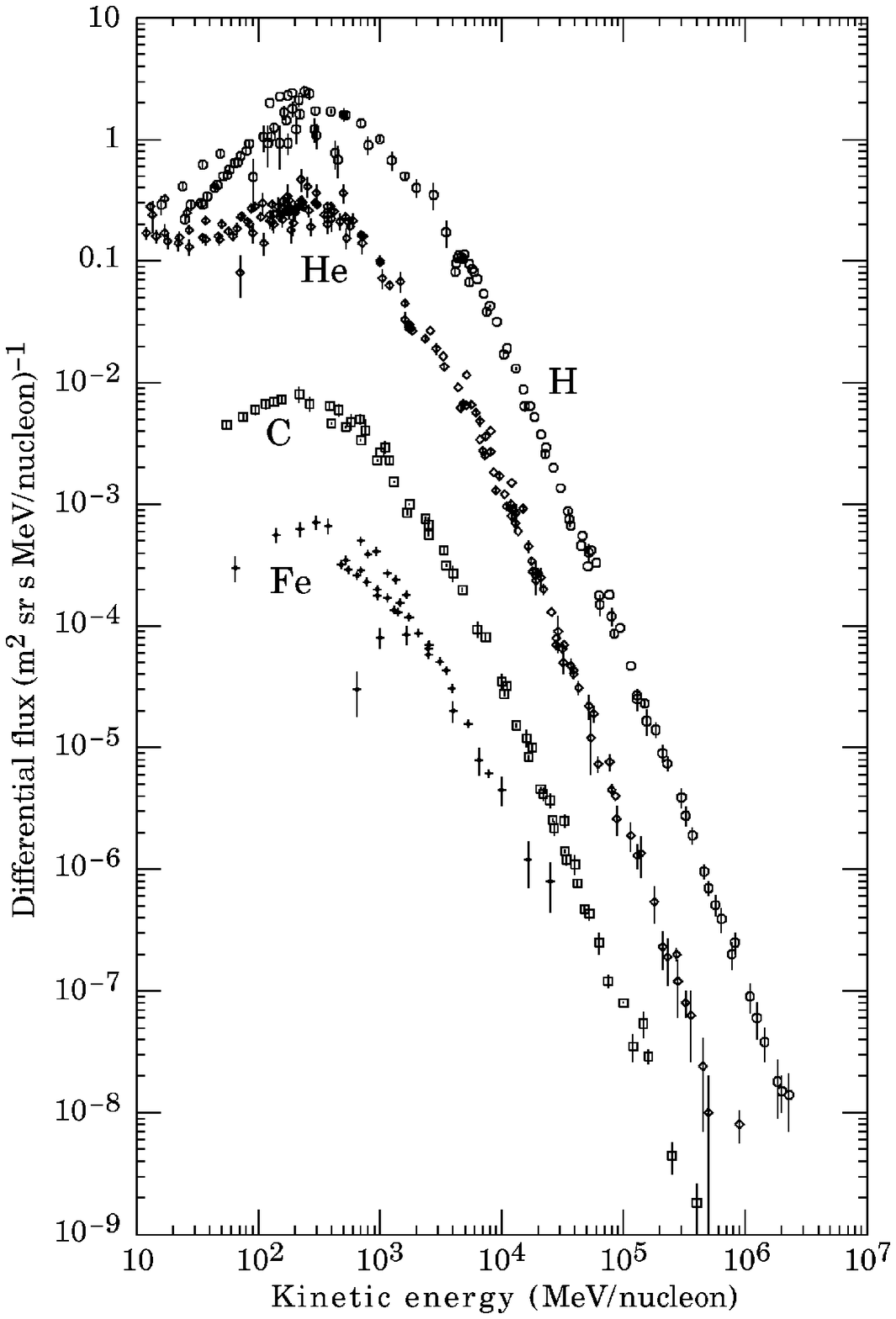}
\caption[Cosmic Ray Nuclei Spectra.]{Cosmic Ray Nuclei Spectra, from \cite{ref:pdg}.}
\label{fig:cr:craynuclei}
\end{figure}

\begin{figure}[htp]
\centering
\includegraphics[width=12cm]{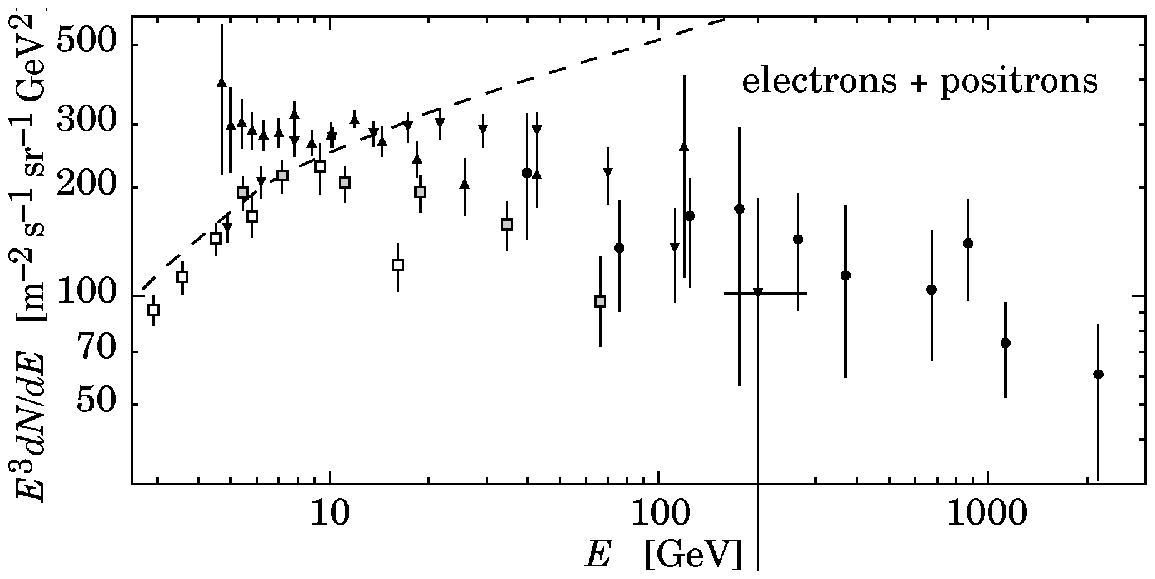}
\caption[Cosmic Ray Electron Spectra.]{Cosmic Ray Electron Spectra as measured from in variety of experiments, compiled in \cite{ref:pdg}.}
\label{fig:cr:crayelectrons}
\end{figure}

	The so-called primary cosmic rays are thought to have been ejected during supernova explosions.  They gain energy through Fermi acceleration \cite{ref:fermi} passing through the shocks \cite{ref:shocks} of the supernova.  This gives their energy spectrum a power law shape as they spread throughout the galaxy.	The proton spectrum above 1 GeV is roughly a power law with spectral index of about $-2.7$ (Fig. \ref{fig:cr:craynuclei}).  The electron spectrum is one percent as intense as the proton spectrum at 10 GeV and is a power law with a steeper index, close to $-3.1$ (Fig. \ref{fig:cr:crayelectrons}).  

	There are also secondary cosmic rays.  These cosmic rays are formed by collisions of primary cosmic rays with the interstellar medium.  Most cosmic antiprotons are thought to be secondary in nature \cite{ref:galpropparams}.  The antiproton spectrum is $2\times 10^{-4}$ as intense as the proton spectrum at 10 GeV \cite{ref:pdg}.

	Once away from their source, cosmic rays drift through the galaxy under the influence of galactic magnetic fields, losing energy due to collisions with other particles as well as through synchrotron radiation.  This process is important for those searching for dark matter annihilation signals as it affects both the signal and the background, changing their spectral shape.

	While traveling through the galaxy, charged cosmic rays follow magnetic field lines, which are twisted by motions of astrophysical plasmas.  This makes the trajectory of cosmic rays like a random walk, thus the process can be described as diffusion.  During their trip through the galaxy, the cosmic rays lose energy from ionization and synchrotron radiation and may gain energy energy passing through magnetic turbulence or shocks \cite{ref:thegalprop}.  This has the effect of steepening the power law index for the cosmic ray backgrounds.  For a monoenergetic cosmic ray source, this would cause the detected cosmic ray spectrum to be blurred and extend from somewhat above the produced energy all the way down to the lowest detectable energy.

	Once cosmic ray particles enter our solar system, they interact with particles and magnetic fields generated by the sun.  This process, known as solar modulation, has a tendency to decrease the intensity of the cosmic ray flux at energies below 1 GeV/nucleon as particles are ``blown away" by the solar wind.

	The canonical approximation for this effect is known as the Force Field Approximation \cite{ref:thesmpaper}.  This approximates solar wind as magnetic inhomogeneities streaming from the sun and scattering cosmic rays.  This yields the equation relating the cosmic ray flux at a point inside the solar system to the flux outside:
\begin{equation}
	J(r,E)=\frac{E^2-E_0^2}{(E+|q|\Phi)^2-E_0^2}J(r_{\infty},E+|q|\Phi)
\end{equation}
where $J$ is the cosmic ray differential flux, $E$ is the cosmic ray energy, $r$ is the distance from the sun, $E_0$ being the mass of the cosmic ray particle in question, $\Phi$ is the effective potential at Earth from the solar wind pushing outwards, and $r_{\infty}$ indicates outside the solar system.  This approximation is convenient and fairly accurate for energies above 1 GeV and will be used for this analysis.
	
	The Force Field Approximation has a number of shortcomings that must be addressed.  Primarily, the parameter $\Phi$ is not well known and ranges from 500~MV to 1000~MV during the eleven year solar cycle.  Additionally, the Force Field Approximation assumes solar effects are spherically symmetric and independent of charge sign.  More complicated solar models exist to address these issues, but are not yet completely developed enough to be useful to this analysis \cite{ref:galpropparams}.  The choice of $\Phi$ for this analysis is described in Chapter 7.

	Cosmic rays are also affected by the Earth's magnetic field.  Low energy particles will be turned away except at high latitudes \cite{ref:stormer}.  The cosmic ray spectrum accessible to a particle detector will be dependent on its latitude and height \cite{ref:logairv1}.  Approximating the Earth's field as a displaced dipole, the lowest momentum accessible to a detector in low Earth orbit is given by:
	
\begin{equation}
\frac{p}{|Z|}>(\mathrm{59.6[GeV/c]} )\frac{\cos^4\lambda}{(1+\sqrt{1-Q \cos^3\lambda\cos\phi_{EW}})^2}
\label{eqn:cutoff}
\end{equation}

where $p/|Z|$ is the cutoff rigidity, $Q$ is the sign of the particle's charge, $\lambda$ is the geomagnetic latitude of the detector, and $\phi_{EW}$ is the east-west component of the zenith angle of the incident trajectory ($\phi_{EW}>0$ particles come from the east, $\phi_{EW}<0$ particles come from the west, and $\phi_{EW}=0$ particles come from directly overhead).  Particles detected with a momentum below the cutoff given by this equation are not likely to be of cosmic origin.  The geomagnetic cutoff equation is derived in Appendix B. 

	One irregularity in the Earth's magnetic field warrants special attention.  The region above the South Atlantic has a particularly weak magnetic field.  This region is called the South Atlantic Anomaly, and the weak field there leads to an especially low geomagnetic cutoff. Thus, the flux of low energy particles in the South Atlantic Anomaly is unusually high.  This high flux can overwhelm the trigger of the detector, so data from this region was removed \cite{ref:gpthesis}.

%% file: Detector.tex
\chapter{The AMS-01 Detector}

	The AMS experiment is designed primarily to search for antimatter in cosmic rays \cite{ref:AMS01Mag}.  The AMS-01 experiment was flown on space shuttle flight STS-91 for 10 days in June of 1998 and recorded over 100 million cosmic rays.  AMS-01 on STS-91 was flown as a precursor mission for the AMS-02 experiment, which is intended to be mounted on the International Space Station.  The AMS-01 experiment is fully detailed in \cite{ref:AMS01}, but this chapter will briefly cover the components that are important for this analysis.

\begin{figure}[htp]
\centering
\includegraphics[width=14cm]{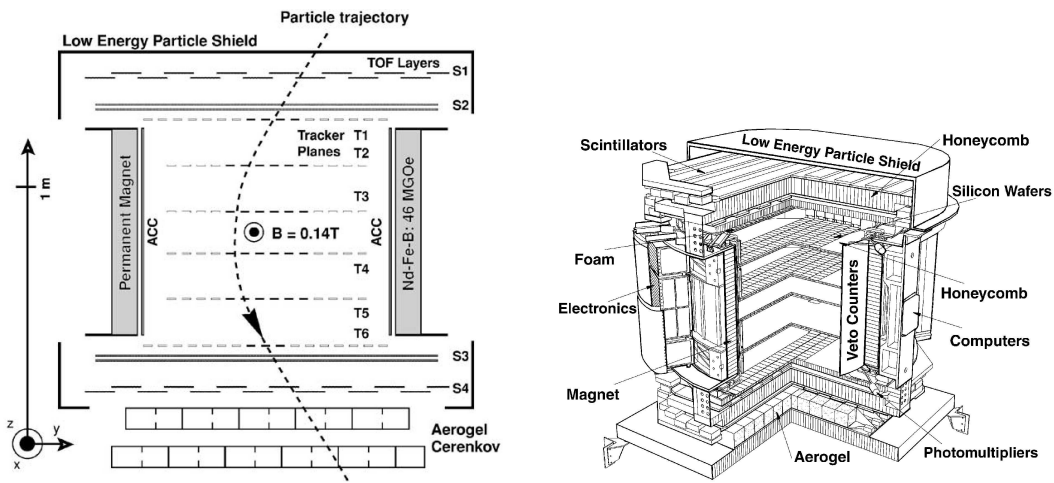}
\caption[The AMS-01 Detector]{The AMS-01 detector sketch and schematic, from \cite{ref:AMS01}.}
\label{fig:det:amsschem}
\end{figure}

	AMS-01 consisted of a time of flight (TOF) system, a permanent magnet containing a silicon tracker system, anticoincidence counters, a threshold \v{C}erenkov counter, and a DAQ system.  Their arrangement is shown in Fig.~\ref{fig:det:amsschem}. 
	
\section{Time of Flight}

	The Time of Flight system consisted of 4 layers, 2 above the tracker and 2 below.  Each layer consisted of 14 plastic scintillator bars that were 1 cm thick and 11 cm wide, and ranged from 72 to 136 cm long.  The bars were optically coupled to 3 photomultipliers on each end.  The hodoscope arrangement was alternated so a 3 dimensional hit position could be constructed if two consecutive layers were hit.  A less accurate hit position could also be constructed from a single layer hit by comparing time differences of the phototube signals at either end of the bar \cite{ref:AMS01TOF2}.

	The amplitude and timing of the signals from the phototubes were recorded and also used for the trigger.  The speed of the particle and whether it was going upwards or downwards was calculated from the difference in signal times between planes, while the absolute charge was determined from the signal magnitudes.  The time resolution for each signal was better than 120ps, which gave a velocity uncertainty ($\delta\beta/\beta$) of 6\% for particles with rigidity greater than 16GV.  The time of flight for a particle was about 5 ns, which resulted in an expected upward/downward going particle separation of 1 in $10^{11}$ \cite{ref:AMS01TOF}.  This is three orders of magnitude larger than the number of particles recorded during the flight, so it will be assumed that there were no upward/downward errors made.  More information on the TOF can be found in \cite{ref:AMS01TOF} and \cite{ref:AMS01TOF2}.

\section{Tracking System}
	The tracking system was composed of a layered silicon detector inside of a permanent magnet.	The magnet bent the trajectory of incident particles so that the tracker could measure their rigidity.  The AMS-01 magnet was a cylinder with a 1.5 kG internal dipole field, but a very small outside dipole moment \cite{ref:AMS01Mag}.  It was composed of 64 sectors of Nd–Fe–B permanent magnets to make a structure 800 mm long with an inner diameter of 1115 mm, giving it a maximum bending power of $0.15\ \mathrm{T\ m^2}$.

	The tracker consisted of six microstrip silicon detector layers which spanned one meter from the first to the last.  Each layer consisted of ladders made of seven to fifteen silicon sensors that ran parallel to the field of the magnet. Sensors were read out from both sides, and the strips on one side ran perpendicular to the other.  The p-doped side (S-side) measured position in the bending plane
with a readout pitch of 110\ $\mathrm{\mu m}$ and the n-doped side (K side) measured the position in the nonbending plane with a pitch of 208\ $\mathrm{\mu m}$.  Silicon layers were 300\ $\mathrm{\mu m}$ thick, with carbon fiber skins that were 220\ $\mathrm{\mu m}$ thick on layers inside the magnet and 700\ $\mathrm{\mu m}$ thick on layers outside the magnet.  On average, a normally incident particle went through 0.65\% of a radiation length worth of material per tracker plane.  The 58,368 redout channels of the tracker were processed by VA\_hdr\footnote{Integrated Detector \& Electronics AS (IDE AS), Veritasveien 9, N-1322 H\o vick, Norway, commercialized as VA\_hdr/AMS.64} chips, high-dynamic range amplifier/multiplexer ASICs with a charge-sensitive amplifier and shaper before being digitized and stored by tracker data reduction cards.

	During operation, charge collected on nearby strips was grouped into a cluster and the weighted mean of the charge distribution was used to determine the location where the particle had passed through the tracker plane.  This yielded a position resolution of 10\ $\mathrm{\mu m}$\ in the bending plane and 30\ $\mathrm{\mu m}$ in the non-bending plan.  The position resolution translated to a momentum resolution of 9\% for protons in 1--10 GeV range \cite{ref:AMS01Tracker}.   This resolution was worse for particles with lower momenta due to multiple scattering.  At higher momenta, the resolution was proportional to the momentum giving a resolution of nearly 100\% for 100GeV particles.
	
	A measurement of total energy loss in the tracker was used to determine the absolute charge of the particle passing through the tracker.  A combined charge measurement of the TOF and tracker was estimated to be accurate to 1 in $10^{7}$ in distinguishing Z=1 nuclei from Z=2 nuclei \cite{ref:AMS01}.  During the flight, the tracker alignment calibration was constantly monitored to an accuracy of 5 $\mathrm{\mu m}$ with a laser alignment system.  More detailed descriptions of the design and performance of the tracker can be found in \cite{ref:AMS01Tracker}, \cite{ref:AMS01}, and \cite{ref:CristHardName}.

\section{Additional Components}
	
	Just inside the magnet were the 16 anticoincidence counters (ACC), 10 mm thick plastic scintillator detectors that were placed horizontally to detect particles that did not pass entirely through the tracker or that were created in interactions between particles and the magnet material.  Events which triggered the ACC were discarded with the hardware veto as they were likely to have been measured incompletely. 

	Below the lower TOF layer was an Aerogel Threshold \v{C}erenkov Counter that was used to separate protons from positrons and antiprotons from electrons at energies below 3.5 GeV.  As this analysis deals only with higher energy particles, information from the ATC was not used \cite{ref:atc}.

	The electronics trigger of AMS-01 consisted of three levels: Fast, Level-1, and Level-3.  The Fast trigger required the signals in the top and bottom TOF planes to be coincident within 200\ $\mathrm{\mu s}$ \cite{ref:gpthesis}.  The Level-1 trigger required that the TOF was hit in a place that would indicate a particle path that passed through the tracker, that these hits were unique on both the top and bottom planes, that there was at least one additional TOF hit on plane 2 or 3, and that there was no signal in the ACC.  The Level-3 trigger required that at least 3 tracker clusters lay along the expected particle trajectory \cite{ref:trigger}.  During the flight, events which passed all three triggers were recorded to hard disks located on the space shuttle.

\section{Flight}
\begin{figure}[htp]
\centering
\includegraphics[width=6cm]{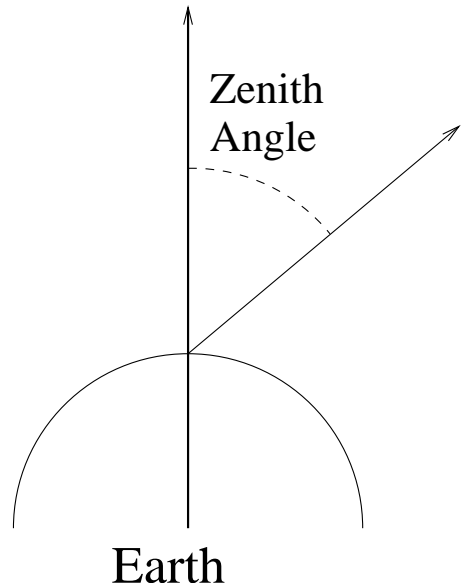}
\caption[Definition of Zenith]{Definition of Zenith Angle.}
\label{fig:det:zenithdef}
\end{figure}

\begin{figure}[htp]
\centering
\includegraphics[width=12cm]{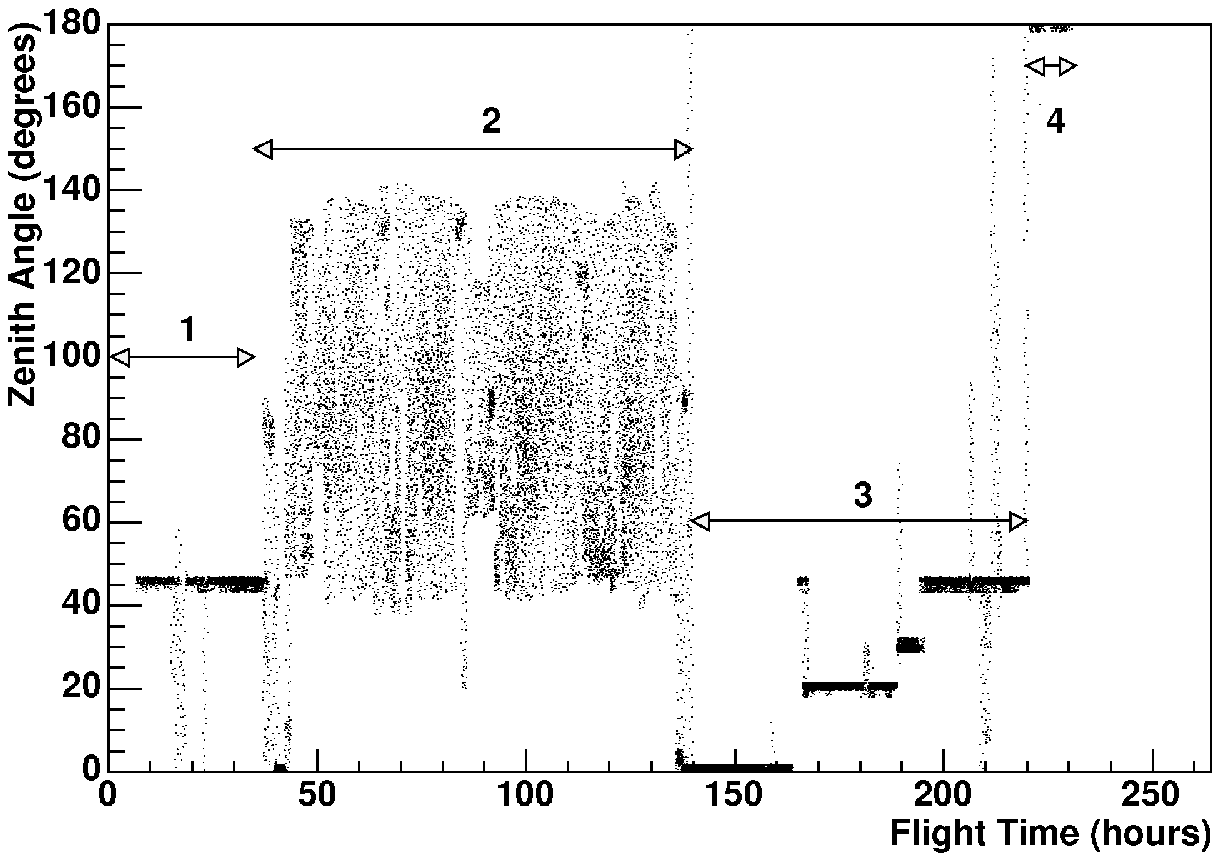}
\caption[Zenith angle of AMS-01]{Zenith angle of AMS-01 as a function of time, from \cite{ref:gpthesis}.}
\label{fig:det:zenith}
\end{figure}

	The flight aboard Space Shuttle Discovery lasted from June 2 to June 12, 1998.  The data from the 10 day flight can be divided into 4 periods \cite{ref:gpthesis}:
\begin{enumerate}
\item From the beginning of data taking a few hours after launch on June 3, for 25 hours until the shuttle docked with the space station Mir.
\item The four days during which Discovery was docked with Mir.
\item The nearly five days of upright orientation after separating from Mir.
\item The last 9 hours of flight during which the shuttle was inverted and AMS-01 pointed towards Earth.
\end{enumerate}
	The zenith angle (Fig.~\ref{fig:det:zenithdef}) fluctuated quickly during period two when the shuttle was docked with Mir and many particles may have interacted with the station before passing through AMS-01 \cite{ref:reycothesis}.  These particles likely lost energy and created secondary particles, making data from that period difficult to interpret.  With the shuttle inverted, cosmic rays passed through the bottom of the shuttle's cargo bay before reaching AMS-01, causing data from this period to suffer from the same issues.  Therefore, this analysis only uses data from the first and third of these periods, during which AMS-01 covered a range of zenith angles from 0 to 45 degrees (Fig.~\ref{fig:det:zenith}).  This led to roughly 94 hours of active data taking time used in this analysis. 

%% file: DataAnalysis.tex
\input{numbers.tex}

\chapter{Data Analysis}
	
	This chapter describes the procedure used to turn the detector data from periods one and three into a momentum spectrum of positive unit charge and negative unit charge particles and how Monte Carlo was used to predict how the cosmic ray flux would appear in the detector.  Over 100 million events were recorded during the 10 day flight of AMS-01.  After cuts, there were \protonspassedcuts particles identified as Z=+1 and \electronspassedcuts particles identified as Z=-1.

\section{Monte Carlo}

\begin{figure}[htp]
\centering
\includegraphics[width=14cm]{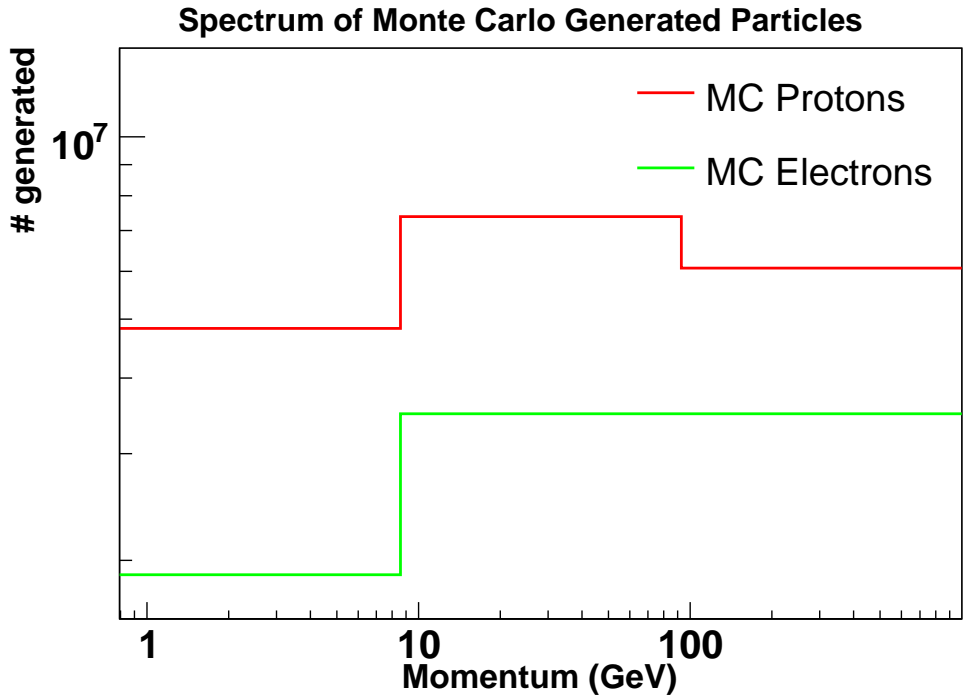}
\caption[Monte Carlo Generated Particles]{The number of Monte Carlo particles generated for each momentum bin.  More protons were generated than electrons to study the effect of protons misidentified as electrons.}
\label{fig:da:mcgenspectrum}
\end{figure}

	The AMS-01 detector was simulated using the GEANT 3 Monte Carlo package \cite{ref:Simulation}.  There were \mcprotoncount protons and \mcelectroncount electrons with momenta from 1 GeV/c to 1 TeV/c were generated with this simulation.  The spectra of generated particles can be seen in Fig.~\ref{fig:da:mcgenspectrum}.  They are piecewise logarithmically flat, with more low momentum protons produced to minimize statistical errors on misidentified charge.  The output format of the simulation was exactly the same as that from real data and both were analyzed with exactly the same software.

\section{Event Reconstruction}

	When triggered, each subdetector recorded the properties of the particle that passed through it.  These properties were grouped into events and were stored in HBOOK ntuples \cite{ref:DataFormats} that were later converted into ROOT trees for analysis.  For this analysis, the two most important properties were the particle's charge and rigidity.  Charge magnitude was determined by choosing the most likely integer charge to have produced the energy deposition observed in the TOF and tracker, while rigidity and charge sign were determined from a fit to the tracker hits.

	Two methods of fitting were used to determine the particle rigidity from the four to six tracker hits.  The first was a simple fit of a circular track to the hits giving a rough estimate of the rigidity.  However, the magnetic field inside the tracker had inhomogeneities of up to 20\%, and 100 GeV electrons have an expected scattering angle of $1\times 10^{-4}$ radians each time they pass through a tracker plane \cite{ref:pdg}.  Therefore, the second method used to determine rigidity was a numerical integration method that accounted for magnetic field inhomogeneities and multiple scattering (referred to as ``FastFit") \cite{ref:AMS01Tracker}.  The FastFit method was similar to that described in \cite{ref:FastFit}.  The tracker fit was combined with the TOF hit locations to determine the entry angle of the particle in the detector.

The following parameters from event reconstruction were used to classify each recorded event:

\begin{itemize}
\item
The time the event was recorded.
\item
The shuttle's position and orientation above Earth at the time of the event.
\item
The detector electronics live time at the time of the event.
\item
Whether anticoincidence counters were triggered during the event.
\item
The entry angle of the particle in the detector, as determined by a combination of the tracker and TOF track fits. 
\item
The number of particles and tracks found by the track reconstruction algorithm.
\item
The number of tracker planes used in the reconstruction of the track.
\item
The charge determined from energy deposition in the tracker planes.
\item
The rigidity of the particle in the tracker as determined by the FastFit algorithm.
\item
The rigidity of the particle calculated using the first and last 3 trackers hits, as determined by the FastFit algorithm.  These were called the half-rigidities.
\item
The rigidity of the particle in the tracker as determined by a simple circular fit to the track.
\item
The $\chi^2$ of the FastFit algorithm.
\item
The number of TOF planes used to determine the $\beta$ of the particle.
\item
The most likely charge determined by the energy deposition in the TOF planes.
\item
The $\chi^2$ of the determined $\beta$ of the particle from the TOF with respect to spacial hits.
\item
The $\chi^2$ of the determined $\beta$ of the particle from the TOF with respect to the time the hit was recorded.
\item
The properties of the hits and clusters recorded in the tracker at the time of the event.
\end{itemize}
\section{Event Selection}
\subsection{Preselection Cuts}
	During preselection, the events were scoured for indications that they might have been measured incorrectly, poorly understood, or were simply detector noise.

	First, events were removed if they occurred while the shuttle was docked with MIR, as the particle spectrum could be polluted by spallation of cosmic rays off of MIR, showering pions and muons into the detector.  Events were also removed if they occurred when the shuttle was passing through the South Atlantic Anomaly, or if the detector live time was less than 35\%, to avoid errors from pileup in the TOF or tracker electronics.

	Events with no reconstructed particle or track were removed, as were events with more than one reconstructed particle to prevent confusion between tracks.  Events were required to have used at least 4 tracker planes and 3 TOF planes in their reconstruction to obtain accurate momentum and velocity measurements and uncertainties.

	If the particle triggered the anticoincidence counters or was measured to have come from beneath the detector, the event was cut.  This insured that all particles passed through the entire detector from top to bottom, like the example event shown in Fig.~\ref{fig:det:amsschem}.  Events reconstructed as having entered the detector at an angle of more than $40^\circ$ may have escaped the detector without hitting all of the tracker planes, and were therefore cut.  Events were only accepted while the detector was pointing away from Earth, within $50^\circ$ of zenith to avoid particles that may have been produced in the Earth's atmosphere.

	If the charge magnitude measured from the TOF energy deposition was not equal to the charge magnitude from the tracker energy deposition, or if the upper half rigidity and lower half rigidity in the tracker had different signs, the event was cut.  Either of these indicated an inconsistency in determining the particle charge.

	Finally, events were only kept if they were above the expected geomagnetic cutoff rigidity at the shuttle's latitude.  The cutoff for protons and electrons was calculated with Formula \ref{eqn:cutoff} for particles coming from direction with the highest cutoff.  Events were only accepted if they had a momentum at least $30\%$ higher than this value, to avoid errors from inaccurate shuttle postion, inexact magnetic field approximation, or overestimated momentum.

\subsection{Selection Cuts}

	Selection cuts attempted to keep as many electrons as possible while removing events that passed preselection but may have been protons misidentified as electrons.  Monte Carlo events were used to choose cuts that would segregate the two.  The weights of Monte Carlo events were adjusted so that they represented a sample from a solar modulated power law spectrum that mimicked the expected cosmic ray background flux.  Cuts on the parameters were then chosen so as to maximize the number of detected electrons times the ratio of electrons to misidentified protons (signal squared over noise) over the fit range (4 GeV--160 GeV).

\begin{figure}[htp]
\centering
\includegraphics[width=14cm]{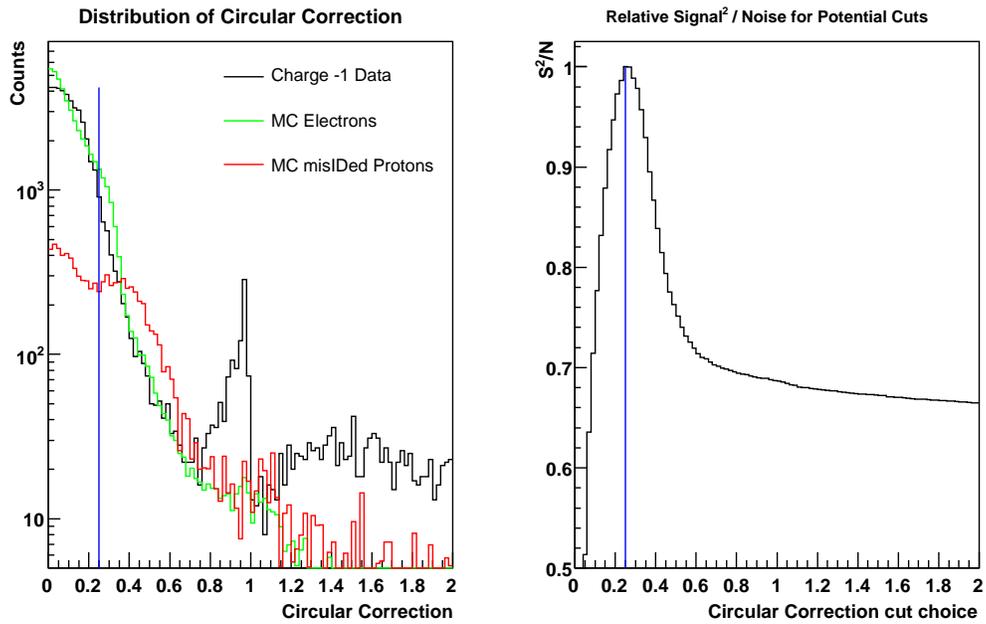}
\caption[Circular Correction Cut]{Determining the cut on circular correction.  \emph{Left:} The distribution of circular correction for data and Monte Carlo events. \emph{Right:}  The expected signal squared to noise ratio versus the circular correction cut choice.  To maximize the expected signal squared to noise, events with a circular correction of 0.26 or greater were cut.}
\label{fig:da:circledifference_var}
\end{figure}

	
	For some events, the circular rigidity was very different from the FastFit rigidity.  This likely indicated that there was something wrong with the track or the fitting algorithm for these events.  To remove events of this type, the ``circular correction'' was defined as the ratio:
\begin{equation}
\mathrm{circular\ correction}\equiv| \frac{\mathrm{rigidity}_{\mathrm{FastFit}} - \mathrm{rigidity}_{\mathrm{circular}}}{\mathrm{rigidity}_{\mathrm{FastFit}}} |
\end{equation}
	A large circular correction indicates that the two track fitting methods disagree on the measured particle's rigidity.  Circular correction for Monte Carlo and data is plotted in Fig.~\ref{fig:da:circledifference_var}, along with the Monte Carlo signal squared over noise for all possible circular correction cuts.  It should be noted that the data has an excess of events with circular corrections of 1 which is not predicted by the Monte Carlo.  These events occurred primarily with higher rigidities and could be a result of unexpected scattering or a failure of the FastFit algorithm at high rigidities.  The best cut predicted from Monte Carlo is to remove events with circular correction greater than 0.26, which also removes this anomaly.
	
\begin{figure}[htp]
\centering
\includegraphics[width=14cm]{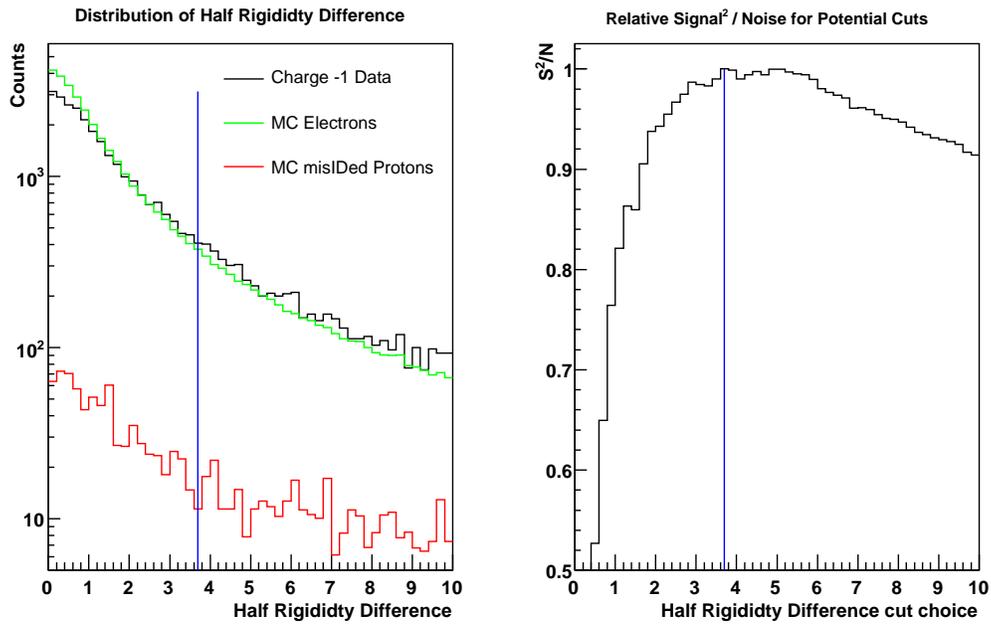}
\caption[Half Rigidity Cut]{Determining the cut on the difference in fit momentum for the first and last 3 tracker hits (half rigidities).  \emph{Left:} The distribution of the half rigidity difference for data and Monte Carlo events.  \emph{Right:} The expected signal squared to noise ratio versus the half rigidity difference cut choice.  To maximize the expected signal squared to noise, events with a half rigidity difference of 4.75GeV/c or greater were cut.}
\label{fig:da:hrigdiff_var}
\end{figure}

	A large difference between the rigidity of the first three tracker hits and the last three tracker hits indicates a poorly reconstructed track, a hard scattering event, or delta ray production.  The difference between half-rigidities is plotted in Fig.~\ref{fig:da:hrigdiff_var}.  Misidentified protons are more likely to have high half rigidity difference and removing events with a half rigidity difference of 4.75 GeV or greater maximizes the number of electrons squared over the number of misidentified protons in the Monte Carlo sample.
	
	After selection, the remaining events are segregated into Z=$+1$ and Z=$-1$, weighted by the inverse live time of the detector at the time of the event, and binned by their momentum.  The average live time for events that passed all cuts was 64\%.  The primary effect of selection cuts on the Z=$-1$ spectrum is to remove high energy events that are likely charge-misidentified protons.  In particular, without the circular correction and FastFit rigidity difference cuts, there is an excess of high energy events that cannot be accounted for by the Monte Carlo charge sign identification prediction.

\begin{figure}[htp]
\centering
\includegraphics[width=14cm]{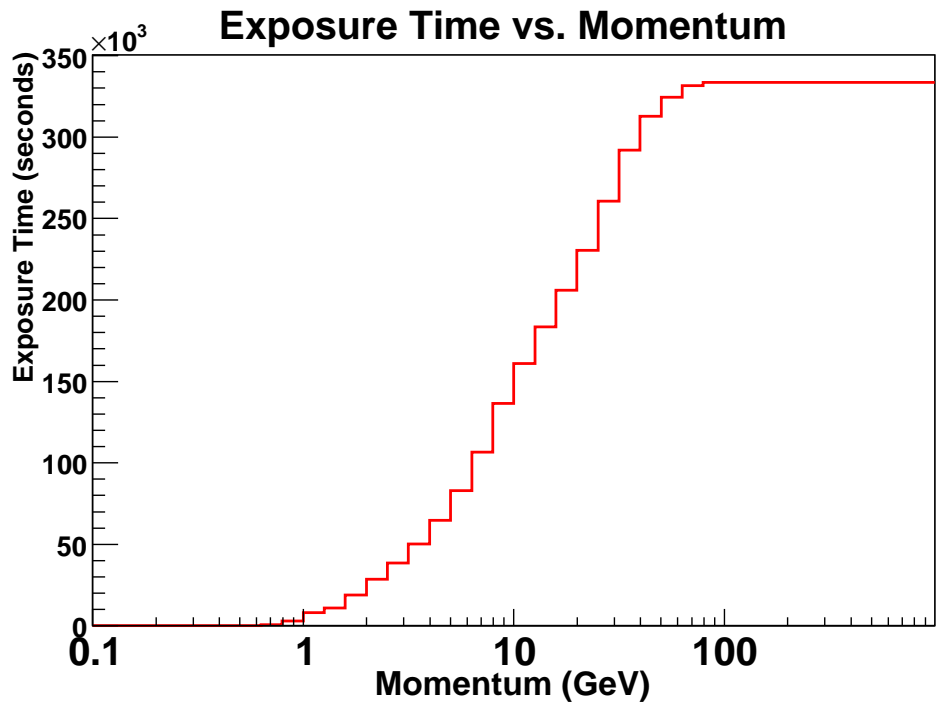}
\caption[Exposure Time]{The exposure time for each momentum bin due to the geomagnetic cutoff.}
\label{fig:da:exposure}
\end{figure}

\begin{figure}[htp]
\centering
\includegraphics[width=12cm]{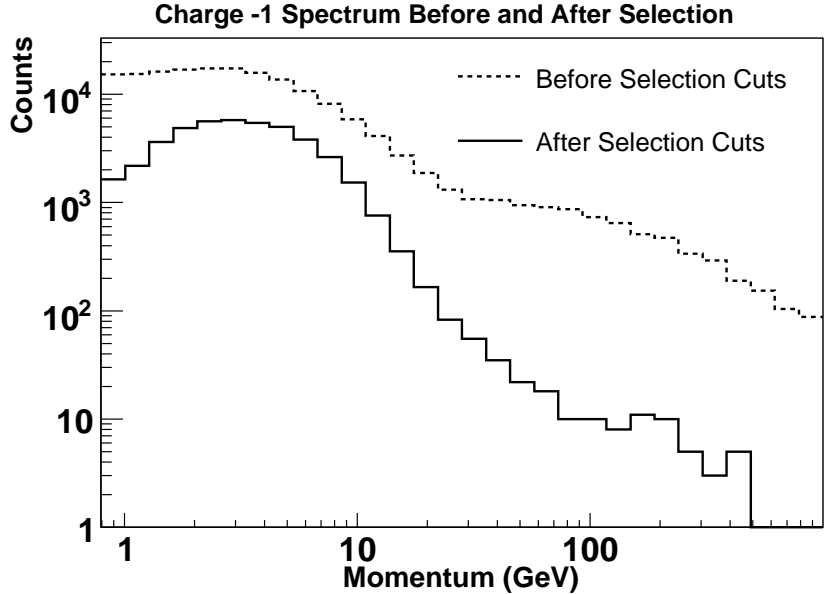}
\caption[Z=$-1$ Spectra Before and After Selection]{This shows the effect of the selection cuts on the preselected Z=$-1$ data.  The deviation from power law at high momenta is primarily from misidentified protons.}
\label{fig:da:beforeafterselection}
\end{figure}

	Finally, the number of counts in each momentum bin must be turned into a detector count rate by dividing by the time the detector was exposed to particles of that momentum.  The exposure time for each bin was different as a result of the geomagnetic cutoff: high momentum particles were always able to reach the detector while low momentum cosmic rays could only reach the detector when the shuttle was near the geomagnetic poles.  Exposure time is tracked during the cut process, Fig.~\ref{fig:da:exposure}.   The counts in each momentum bin were divided by the exposure time of that bin to yield the count rate in a detector unhampered by geomagnetic cutoff.  The Z=$-1$ spectrum before and after cuts is shown in Fig.~\ref{fig:da:beforeafterselection}.  

\section{Acceptance}

	The relationship between the cosmic ray flux and the count rate in the detector depends on the gathering power, resolution, and efficiency of the detector.  These are estimated with the Monte Carlo, producing an acceptance matrix.  When applied to a cosmic ray flux, the acceptance matrix gives the expected count rate in the detector.

\begin{figure}
\centering
\includegraphics[width=15cm]{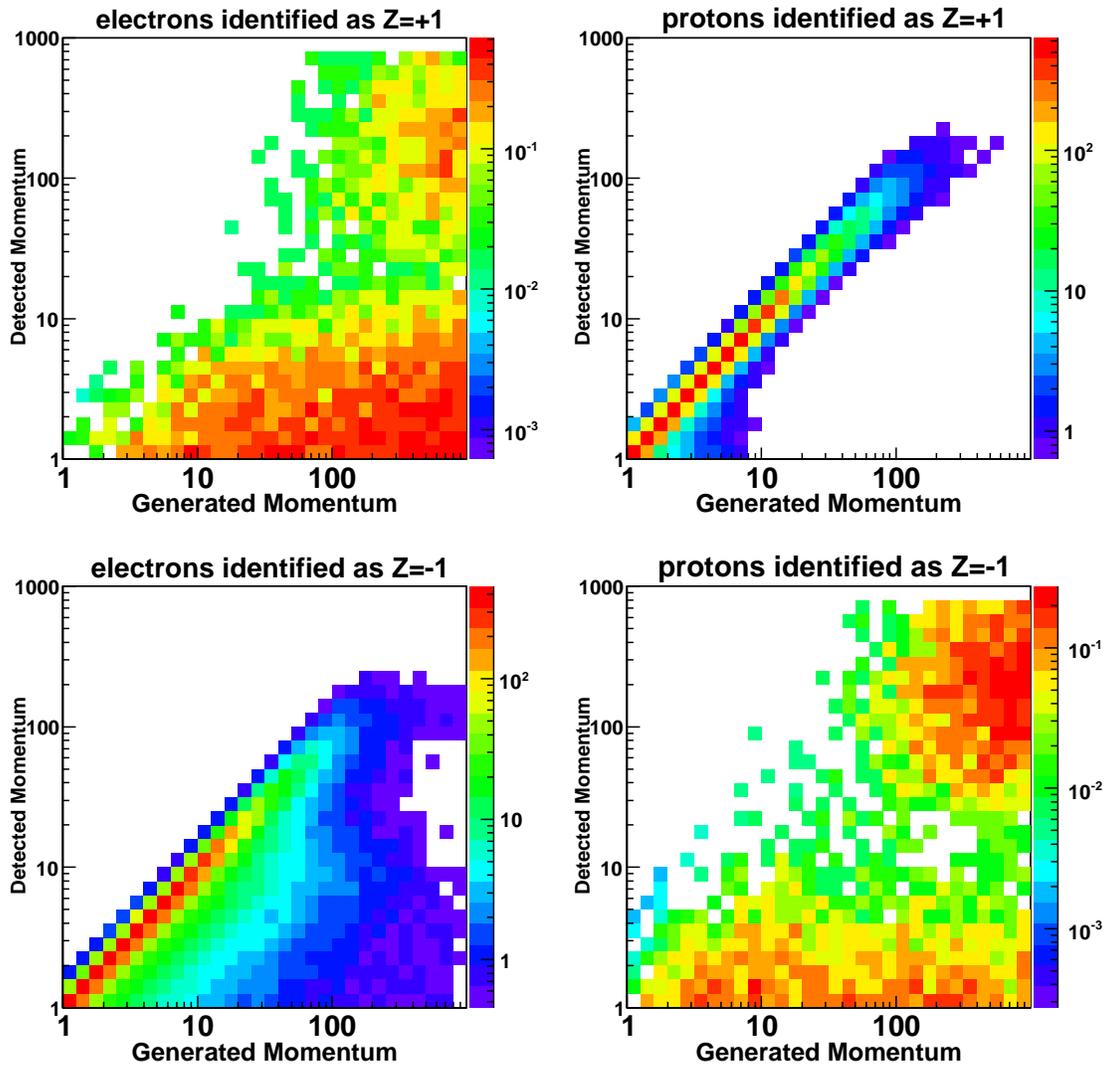}
\caption[Acceptance Matrices]{The acceptance matrices.  Units are Rate/Flux in $\mathrm{Counts\ s^{-1}\ (Particles\ s^{-1}\ sr^{-1}\ cm^{-2})^{-1}}$.  Notice that particles detected with the correct sign charge are also mostly detected with the correct momentum, though with decreasing efficiency at higher energies.  Also notice that high momentum protons are often misidentified as high momentum electrons.}
\label{fig:da:acceptance2d}
\end{figure}

	The acceptance of the detector was calculated with the Monte Carlo, in the manner described in \cite{ref:Telescope}.  Monte Carlo events were generated over a rectangle 250 cm by 90 cm, 100 cm above the detector in simulation with momenta directions distributed isotropically over the half-sphere towards the detector.  Events that would have triggered the detector were run through the same analysis chain as real data.  The matrix given by counting how many particles were created with one momentum and detected at another is both the gathering power and the resolution of the detector.  Monte Carlo protons which are detected as electrons are recorded in another matrix, so that the misidentified proton rate can be estimated from the proton background.  The layout of the matrices can be seen in Fig.~\ref{fig:da:acceptance2d}.
	
	Both statistical and systematic uncertainties were included in the acceptance matrix.  Each matrix element was assigned an uncertainty of $\sqrt{N_{counts}}$ if there were more than 10 counts from Monte Carlo in that element, and a Poisson uncertainty in accordance with counting statistics if there were less than 10 counts.

	By comparing the Monte Carlo detector simulation with prescaled events  \cite{ref:TriggerSystematics}, the simulation was found to be over-efficient by $13\pm3.5$ percent, mostly due to uncertainties in trigger efficiency and particle interactions \cite{ref:AMS01}.  Thus the value of each acceptance matrix element was decreased by 13 percent, and a 3.5 percent systematic uncertainty was added in quadrature to the statistical uncertainty.

	Finally, the off diagonal elements of the acceptance matrix were adjusted to account for exposure time.  Low energy events could not be misidentified as high energy events during the time that low energy particles could not reach the detector due to the geomagnetic cutoff.  Therefore, acceptance matrix elements that corresponded to lower momenta particles being mis-measured as having higher momenta were scaled by the ratio of the lower momentum exposure time over the higher momentum exposure time.  This accounted for the times when the detector was exposed to high momentum particles, but not low momentum particles.

%% file: numbers.tex
\newcommand{\electronspassedcuts}{$74,216$ }
\newcommand{\protonspassedcuts}{$6,563,097$ }
\newcommand{\mcprotoncount}{$1.8\times 10^{8}$ }
\newcommand{\mcelectroncount}{$8.8\times 10^{7}$ }

%% file: SusySignals.tex
\chapter{SUSY Signals}

	The signals of neutralino annihilation in cosmic ray flux are the charged products of the annihilation that have made their way through the galaxy to Earth.  A series of programs are used to calculate the annihilation rate and product spectra from the SUSY parameters and combine these into a cosmic ray flux at the edge of the Solar System.  The flux is then modified to account for the effect of the solar wind as particles travel to Earth.

\section{Generation}

	The expected neutralino annihilation rate at a certain point in the galaxy can be written $(\rho^2/m_{\chi}^2)\langle\sigma_{\chi\chi}v\rangle$, where $\rho$ is the dark matter density, $m_{\chi}$ is the neutralino mass, and $\langle\sigma_{\chi\chi}v\rangle$ is the neutralino annihilation cross section convolved with the velocity distribution of the neutralinos.  The program DarkSUSY \cite{ref:darksusy}, which uses the ISASUGRA~\cite{ref:isasugra} package, calculates $m_{\chi}$ and $\langle\sigma_{\chi\chi}v\rangle$ for various annihilation channels from a set of SUSY parameters.  For nonrelativistic neutralinos, the annihilation cross section is proportional to $1/v$, so the value $\sigma_{\chi\chi}v$ is independent of velocity.  This makes $\langle\sigma_{\chi\chi}v\rangle$  effectively independent of the velocity distribution of neutralinos \cite{ref:talkgondolo}.  The annihilation channels studied were $\chi\chi\rightarrow W^{+}W^{-}$, $\chi\chi\rightarrow ZZ$, $\chi\chi\rightarrow b\bar{b}$, $\chi\chi\rightarrow t\bar{t}$, and $\chi\chi\rightarrow \tau^{+}\tau^{-}$.  In all of the cases studied, the sum of the cross sections for these processes make up more than 80\% of the total annihilation cross section, and thus account for the majority of the products. These results from DarkSUSY are shown in Table \ref{table:ss:dstable}.

\begin{figure}[htp]
\centering
\includegraphics[width=16cm]{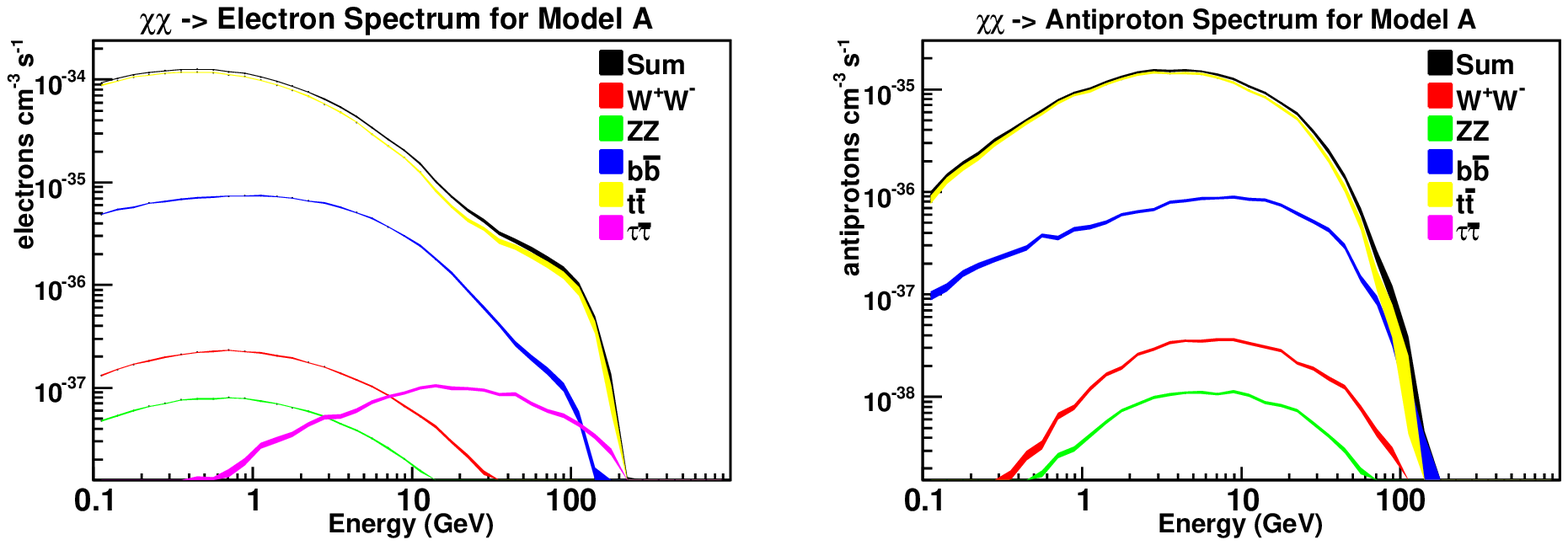}
\includegraphics[width=16cm]{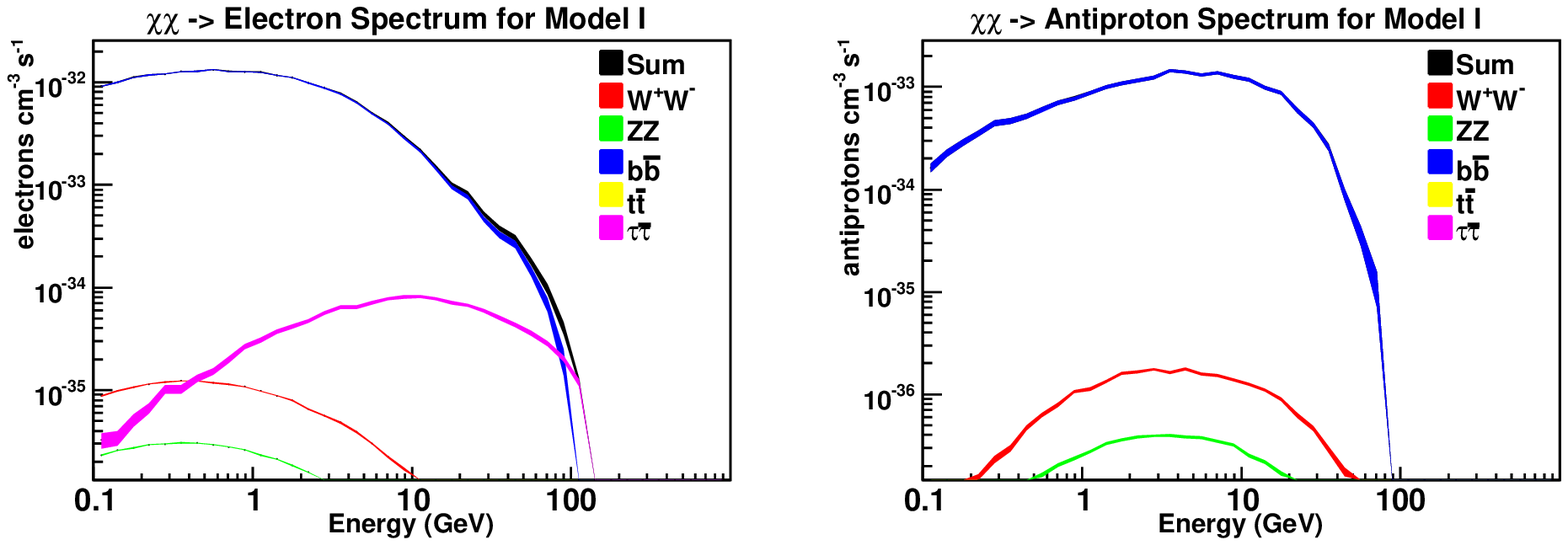}
\caption[Neutralino Annihilation Spectra]{Neutralino annihilation spectra, broken down by branching process.}
\label{fig:ss:dmsignalcomponents}
\end{figure}

	The program PYTHIA~\cite{ref:Pythia} was used to generate the electron and antiproton energy spectra from the decay of the annihilation products listed above.  PYTHIA reproduces the LEP results in the 100--200GeV range, so it can be expected to be reliable for neutralino annihilation events occurring with these energies.  PYTHIA simulates the decay of the annihilation products into stable particles, giving a Monte Carlo spectra of the results of neutralino annihilation.  These spectra were weighted by the cross section for that annihilation pathway and summed to give the total neutralino annihilation spectrum.  Some example electron and antiproton spectra from neutralino annihilation, broken down by component process, are shown in Fig.~\ref{fig:ss:dmsignalcomponents}.

\section{Propagation}

	As products from neutralino annihilation diffuse through the galaxy, their energy spectrum changes.   The program GALPROP \cite{ref:galprop} integrated over the annihilation rate throughout the galaxy, weighted by dark matter density squared, and accounted for the spectral shape change from energy loss.  The diffusive energy loss equations are dependent on a host of propagation parameters.  The set of parameters used in GALPROP for this analysis were from the publicly available model\footnote{Available from \textsc{http://galprop.stanford.edu/} as of April 2007.} ``galdef\_50p\_599278" obtained by a global fit to cosmic ray measurements, both charged and gamma rays \cite{ref:galpropparams}.  
	
\begin{figure}[htp]
\centering
\includegraphics[width=14cm]{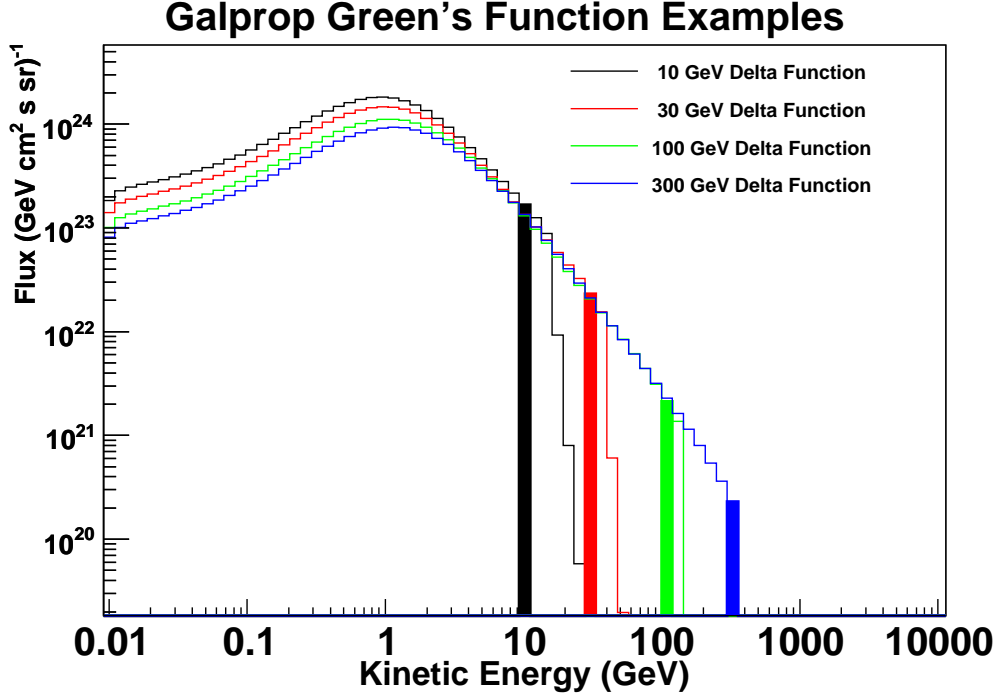}
\caption[Propagation Greens Functions]{Electron propagation Green's functions calculated by GALPROP; the cosmic ray spectra that would be seen at the solar system from monoenergetic sources distributed in an isothermal halo.  The highlighted bins are the initial energies.}
\label{fig:ss:galpropgreens}
\end{figure}

	 GALPROP was used to generate propagation Green's functions to improve calculation speed, and these Green's functions were used to determine the cosmic ray flux from the SUSY models.  For each momentum bin used in this analysis, GALPROP was run with a monoenergetic injection spectrum, weighted by a squared isothermal halo density distribution, and the results were sampled the Solar System's position in the galaxy producing a matrix of (input energy, output spectrum).  The spectrum at Earth from monoenergetic sources are illustrated in Fig.~\ref{fig:ss:galpropgreens}.  This propagation matrix was multiplied by the annihilation spectra of the SUSY models, calculated by PYTHIA, to yield the cosmic ray flux at the Solar System due to neutralino annihilation.  

	Energy loss limits the distance that neutralino annihilation products can travel before becoming so low in energy as to be undetectable.  This analysis is sensitive to electrons from neutralino annihilations to a range of roughly two kiloparsecs from Earth, as discussed in Appendix C.
	
\section{Solar Modulation}

	Finally, the cosmic ray fluxes at the solar system must be adjusted to account for the solar modulation of low energy particles before they reach Earth.  As described in Chapter 5, the force-field approximation was used to modulate the dark matter annihilation signals.  An uncertainty on $\Phi$ was introduced because of the limitations of the force field approximation and the different measurements of $\Phi$ that were reported at the time of the experiment.

\begin{figure}[htp]
\centering
\includegraphics[width=14cm]{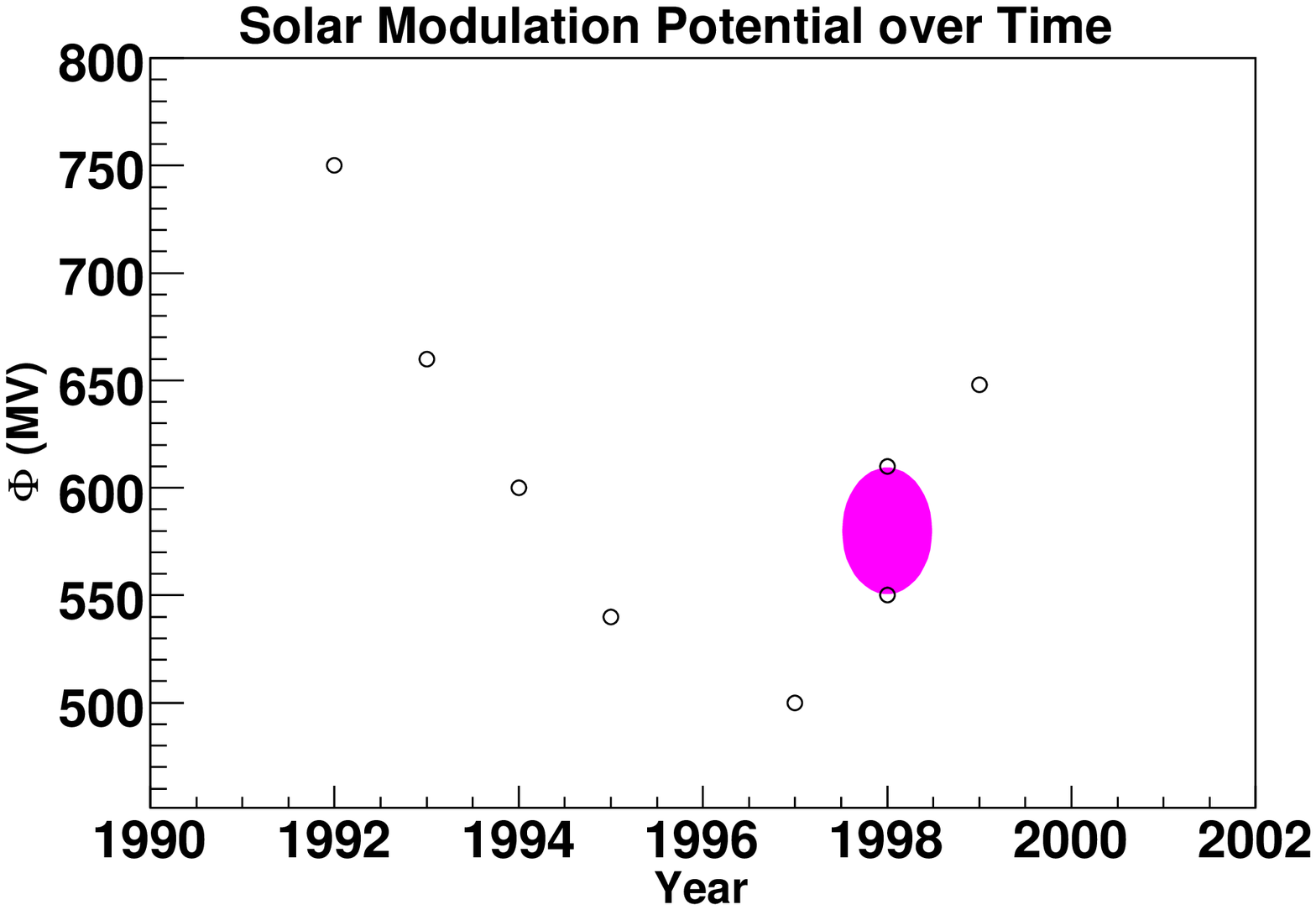}
\caption[Solar Modulation over Time]{Solar Modulation from the BESS experiment \cite{ref:BESS1}, \cite{ref:BESS2} and other experiments \cite{ref:galpropparams}.  The $\Phi$ used in this analysis is highlighted.}
\label{fig:ss:solarmod}
\end{figure}

\begin{figure}[htp]
\centering
\includegraphics[width=16cm]{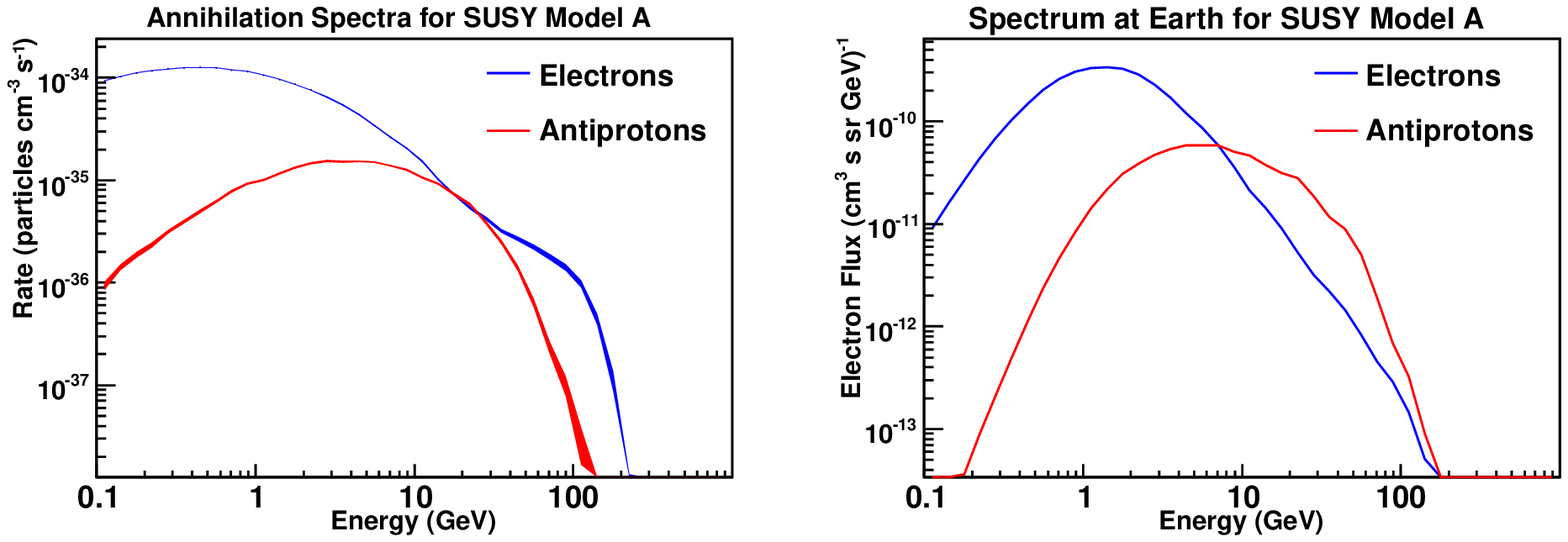}
\includegraphics[width=16cm]{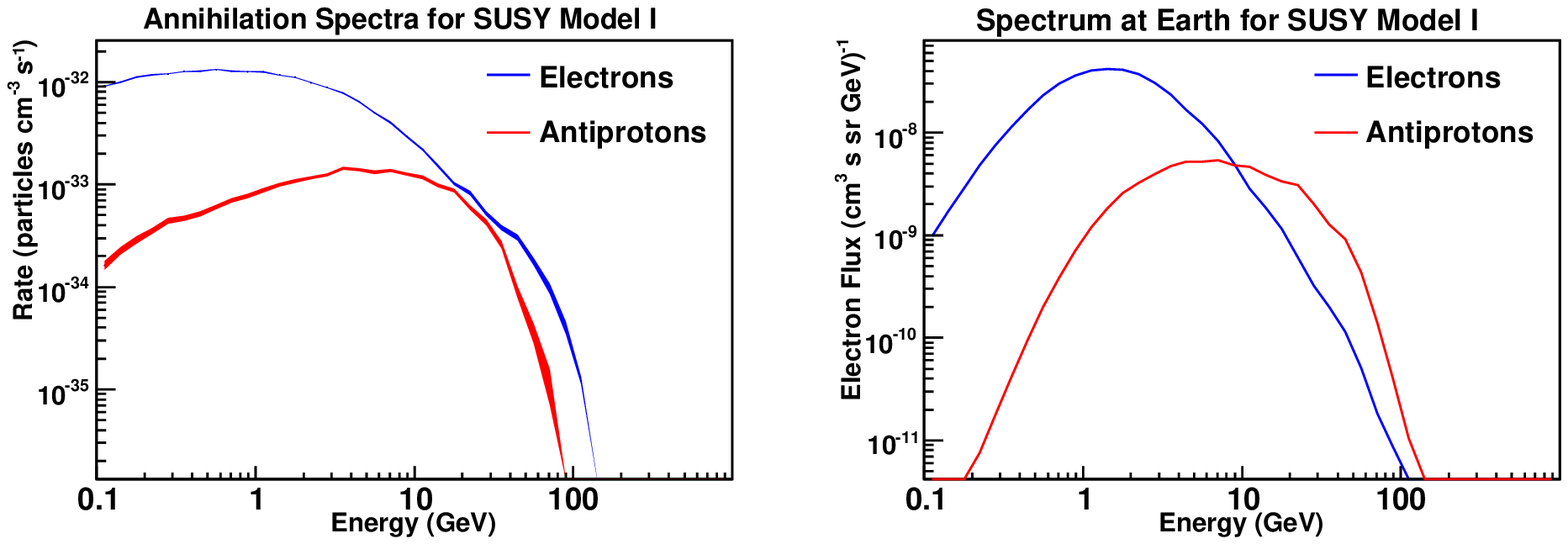}
\caption[Annihilation Spectra above Earth]{A comparison of electron and antiproton spectra at the point of annihilation and above Earth, after propagation and solar modulation.  It can be seen that electrons are more strongly affected by the diffusive energy loss than antiprotons.}
\label{fig:ss:signalpropagation}
\end{figure}

	Measurements of $\Phi$ from a number of experiments around the time AMS-01 flew are shown in Fig.~\ref{fig:ss:solarmod}.  In 1998, different fits of the BESS experiment proton data yielded different $\Phi$ values \cite{ref:BESS1}, \cite{ref:BESS2}, \cite{ref:galpropparams}.  A value of $580$ MV, the mean of the two measurements for 1998, with a Gaussian uncertainty of $30$ MV was used in this analysis.  This uncertainty in $\Phi$ translates into a theoretical uncertainty in both the background and signal spectra at low energies.  A few annihilation spectra and resulting cosmic ray fluxes at Earth are shown in Fig.~\ref{fig:ss:signalpropagation}.

%% file: Backgrounds.tex
\input{alphabet_numbers.tex}

\chapter{Backgrounds}

	The majority of events recorded by AMS-01 were, for this study, background cosmic rays, and these needed to be distinguished from neutralino annihilation signals.  The proton background, generated by GALPROP, was fit to the Z=$+1$ data and used to calculate the misidentified proton spectrum.  These misidentified protons were subtracted from the Z=$-1$ data to yield the data from only electrons and antiprotons.  This was fit with the electron background to understand the background in the case of no dark matter signal.

\section{Generation of Backgrounds}

	The program GALPROP predicts the cosmic ray backgrounds given the propagation parameters and the particle injection spectrum.  In GALPROP, the electron and proton injection spectra are taken to be a power law in energy: $NE^{-\gamma}$, with normalization $N$ and spectral index $\gamma$.  While the injected particles are thought to come from supernova, neither $N$ nor $\gamma$ are theoretically well predicted so they must be determined from data.

\begin{figure}[htp]
\centering
\includegraphics[width=15cm]{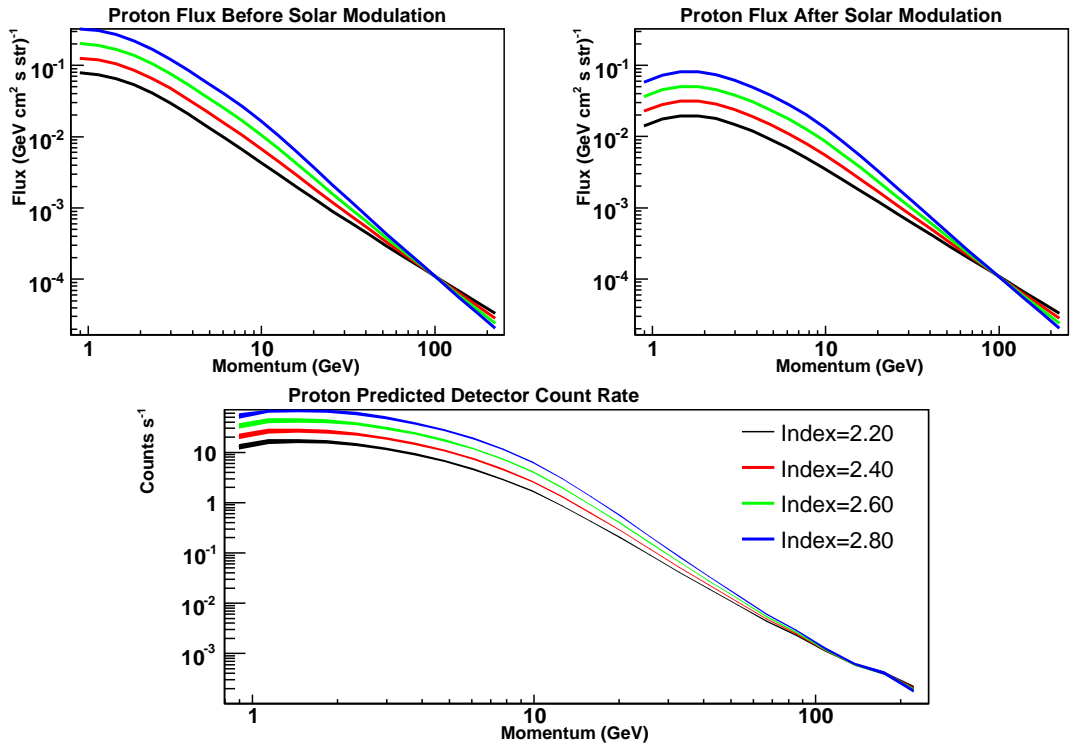}
\caption{The effect of the injection index on the proton background.}
\label{fig:bg:proton_bgsm}
\end{figure}

\begin{figure}[htp]
\centering
\includegraphics[width=15cm]{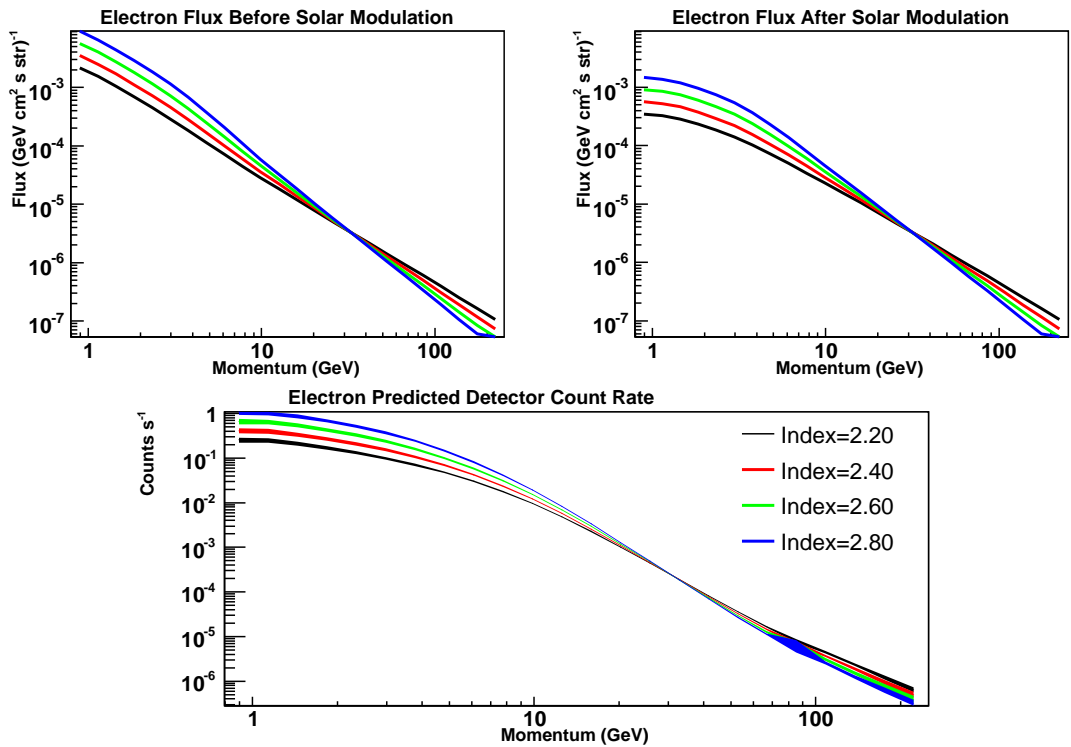}
\caption{The effect of the injection index on the electron background.}
\label{fig:bg:electron_bgsm}
\end{figure}

	The most recent values for the background injection spectral indices found in GALPROP model ``galdef\_50p\_599278" for the energy range used in this analysis are $2.36$ for protons and $2.50$ for electrons.  These are normalized to a flux of $4.9\times 10^{-9} (\mathrm{cm^2\ sr\ s\ MeV})^{-1}$ at 100 GeV for protons and $4.0\times 10^{-10} (\mathrm{cm^2\ sr\ s\ MeV})^{-1}$ at 34.5 GeV for electrons.  These four parameters were determined in \cite{ref:galpropparams} using the results of several cosmic ray experiments.  However, not only was agreement between experiments poor, but additionally the parameters are dependent on solar modulation \cite{ref:moskocommun}.  Therefore, both the indices and normalizations of the backgrounds were refit to the AMS-01 data for this analysis.  The effects of the injection index on the backgrounds are shown for protons in Fig.~\ref{fig:bg:proton_bgsm} and for electrons in Fig.~\ref{fig:bg:electron_bgsm}.
	
	A library of backgrounds with injection indices ranging from $2$ to $3$ with a spacing of $0.005$ was generated to speed the fitting of the backgrounds to AMS-01 data.  To generate a background with a specific injection index, a linear interpolation between the two closest library backgrounds was used instead of a computationally costly re-calculation of GALPROP's diffusive equations.  The library index spacing was well below the uncertainty in the fit parameters, so this method added negligible uncertainty to the background.  No library was needed to vary the normalization; the background was simply rescaled.
	
\section{Background Fits}

	The backgrounds generated by GALPROP were modulated to account for solar effects and multiplied by the acceptance matrix to obtain the expected background spectra in the detector.  The proton background fit to the AMS-01 Z=$+1$ spectrum yielded a normalization of \bgprotonfitnorm, as a fraction of the initial GALPROP normalization, and an injection index of \bgprotonfitgamma (Fig.~\ref{fig:bg:protonfit}) with a $\chi^2/N$ of 7.98 for $N=13$.  The fit method is described in detail in Appendix A.

\begin{figure}[htp]
\centering
\includegraphics[width=14cm]{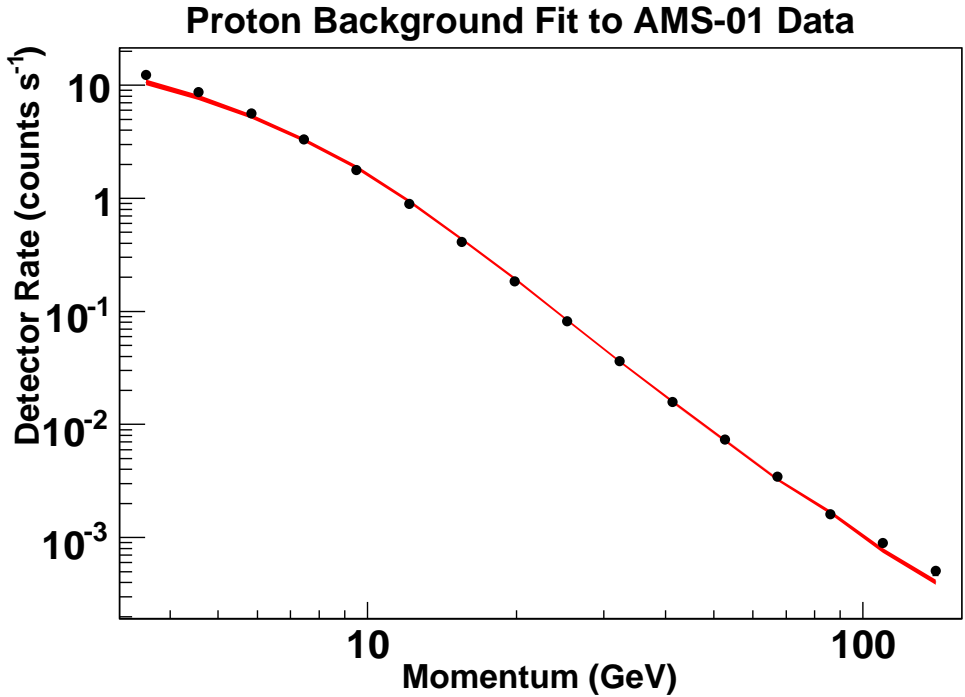}
\caption[Proton Background]{The fit to the proton background.  The statical errors on the data are too small to be seen.}
\label{fig:bg:protonfit}
\end{figure}

\begin{figure}[htp]
\centering
\includegraphics[width=14cm]{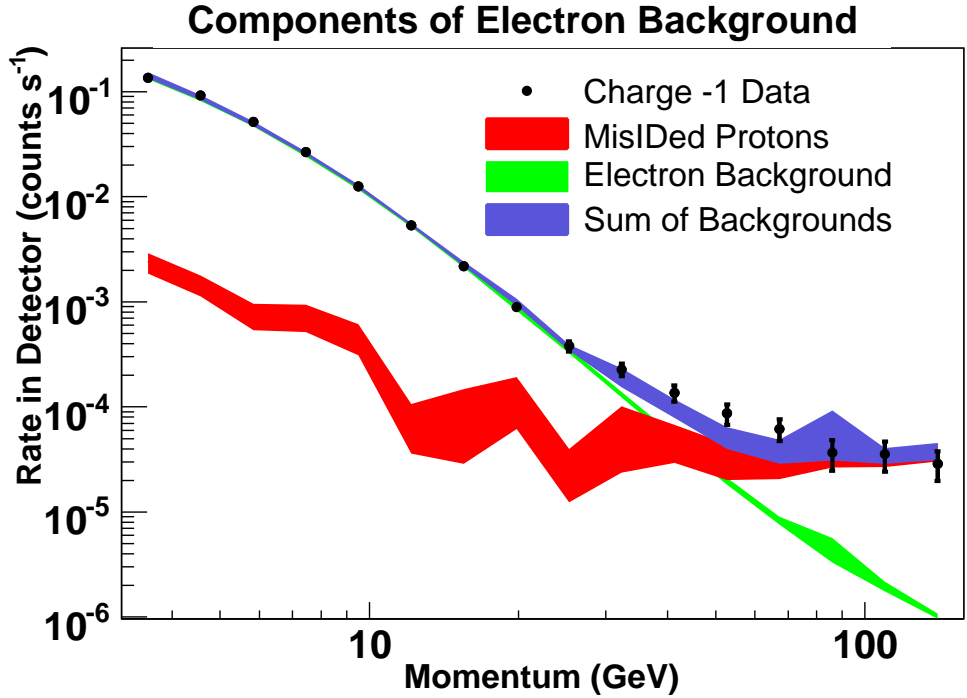}
\caption{Components of electron fit.}
\label{fig:bg:electroncomponents}
\end{figure}

\begin{figure}[htp]
\centering
\includegraphics[width=14cm]{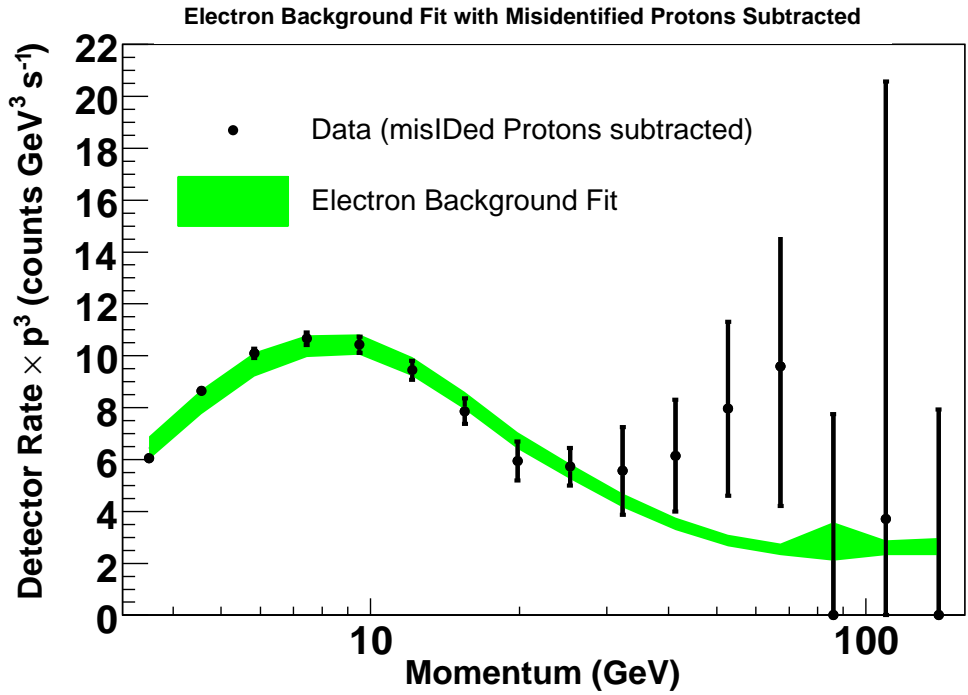}
\caption{Electron fit after misidentified protons have been subtracted.}
\label{fig:bg:electron_fit_midsubtracted}
\end{figure}

	The resultant proton background was then used with the acceptance matrix to predict the rate of charge misidentified protons in the AMS-01 Z=$-1$ spectrum.  Fig.~\ref{fig:bg:electroncomponents} shows the combination of the fit of the electron background and the misidentified protons, yielding a normalization and spectral index for the electron background of \bgelectronfitnorm and \bgelectronfitgamma with a $\chi^2/N$ of 2.04 with $N=13$.  These are the best fit values for the electron background assuming no dark matter signal.  For clarity, the same fit is shown in Fig.~\ref{fig:bg:electron_fit_midsubtracted} with the misidentified protons subtracted from the data and the rate of each momentum bin multiplied by the momentum cubed to remove the power law slope of the background.
	
\begin{figure}[htp]
\centering
\includegraphics[width=16cm]{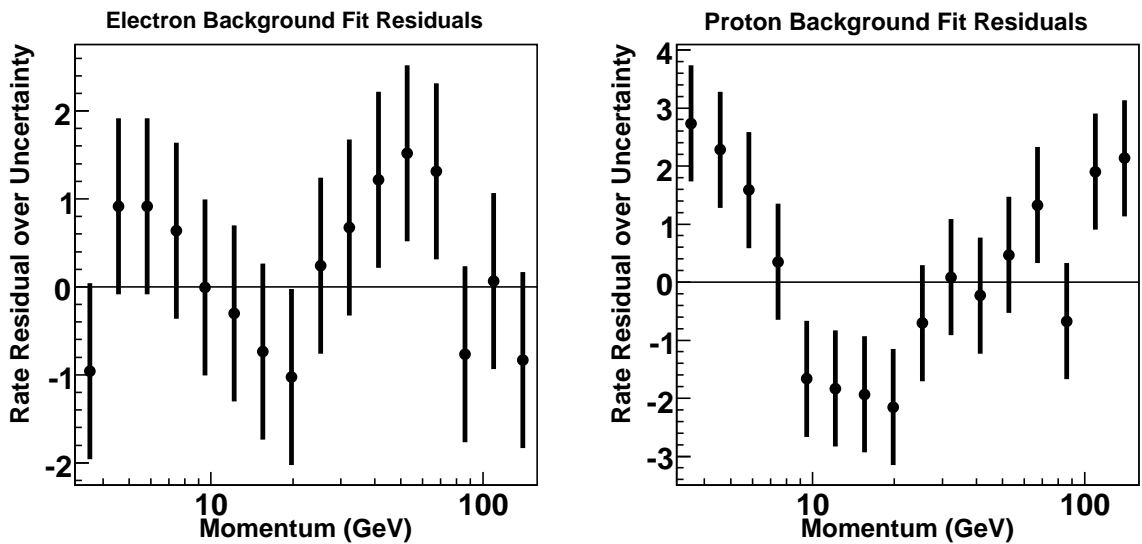}
\caption[Background Fit Residuals]{Residuals to proton and electron background fit.  Each residual is normalized to the uncertainty of that bin.}
\label{fig:bg:residuals}
\end{figure}

	The residuals from the proton and electron background fits are shown in Fig.~\ref{fig:bg:residuals}.  The proton spectrum shows some structure that is not captured by the fit.  This could be the result of imprecision in the GALPROP background model, oversimplification of the solar modulation model, or imperfect simulation of detector systematics in the Monte Carlo.  The proton systematic only entered the dark matter fit at second order, through the misidentified proton subtraction.  The electron spectrum also suggests extra structure at the $1\sigma$ level, and while this could be a result of neutralino annihilation signal, the similar behavior of the proton residuals makes this doubtful.  The same systematic that affects the protons likely affects the electrons, but is less significant because of lower statistics.

%% file: alphabet_numbers.tex
\newcommand{\bgprotonfitnorm}{$0.67\pm 0.02$}
\newcommand{\bgprotonfitgamma}{$2.43\pm 0.01$ }
\newcommand{\bgelectronfitnorm}{$0.70\pm 0.04$ }
\newcommand{\bgelectronfitgamma}{$2.70\pm 0.04$ }

%% file: Limits.tex
\chapter{Limits}

	Neutralino annihilation signals for Models A--M (Table \ref{table:lim:bestfits}) and the electron background simultaneously were fit to the AMS-01 Z=$-1$ data.  All models were consistent with no dark matter signal, so limits were found on the maximum boost factor for each SUSY model that would still be consistent with the data.
	
\section{Fitting}

	The signal from each SUSY model plus the electron background was fit to the AMS-01 Z=$-1$ data using the methods detailed in Appendix A, allowing the background normalization, background index, and signal boost factor to vary. The results of this fit are shown in Table~\ref{table:lim:bestfits}.  All of the fits were consistent with no dark matter signal (a boost factor of zero).  In the cases where the fit boost was less than zero, the best fit was taken to be the fit with no dark matter, as a negative signal is unphysical.

	A number of sources contributed significantly to uncertainties on the fits and subsequent limits on dark matter signals:  Both the signal and background were subject to uncertainties from the solar modulation approximation (Chapter 7), as well as the systematic errors of how well the Monte Carlo simulated the detector (Chapter 6).  The misidentified proton subtracted data, aside from the statistical uncertainties (Chapter 6), suffered at high energies from the uncertainties in the proton background fit that yielded the misidentified proton rate (Chapter 8).  The momentum range over which certain sources of uncertainty dominate is discussed in Chapter 11.

\begin{table}[htp]
\centering
\input{alphabet_bestfittable.tex}
\caption{Best fits of signal to Z=$-1$ data}
\label{table:lim:bestfits}
\end{table}

\section{Limits on Boost Factor}

	A three dimensional confidence region was then found for the background normalization, injection index, and signal boost, as described in Appendix A.  An example of these confidence regions are shown in Fig.~\ref{fig:lim:conf} and all regions can be found in Appendix C.  This yielded a 90\% confidence limit on the maximum boost factor that would be compatible with each SUSY model studied, listed in Table~\ref{table:lim:limits}.  This can also be interpreted as the upper limit on the density squared of dark matter in our region of the galaxy.  The spectral shape expected for these upper limits can be seen in Fig.~\ref{fig:lim:limfit} and in Appendix C.

\begin{figure}[htp]
\centering
\includegraphics[width=12cm]{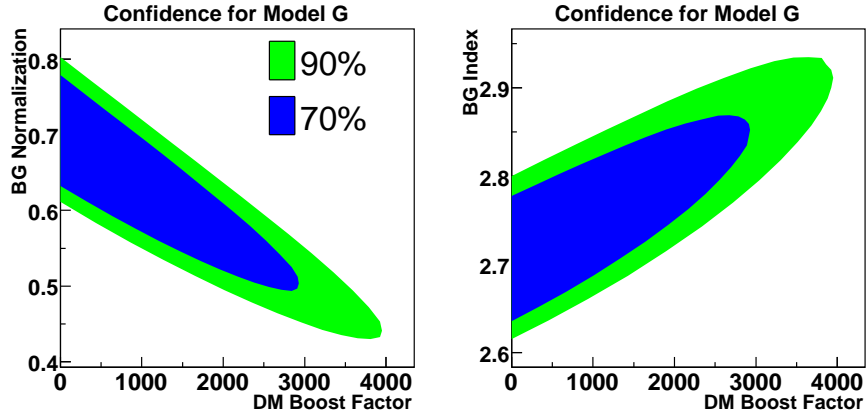}
\caption[Confidence region for Model G]{3-d confidence region for Model G, as projected onto 2-d axes.}
\label{fig:lim:conf}
\end{figure}

\begin{table}[htp]
\centering
\input{alphabet_boundtable.tex}
\caption{Limits on boost factor for SUSY Models}
\label{table:lim:limits}
\end{table}

\begin{figure}[htp]
\centering
\includegraphics[width=12cm]{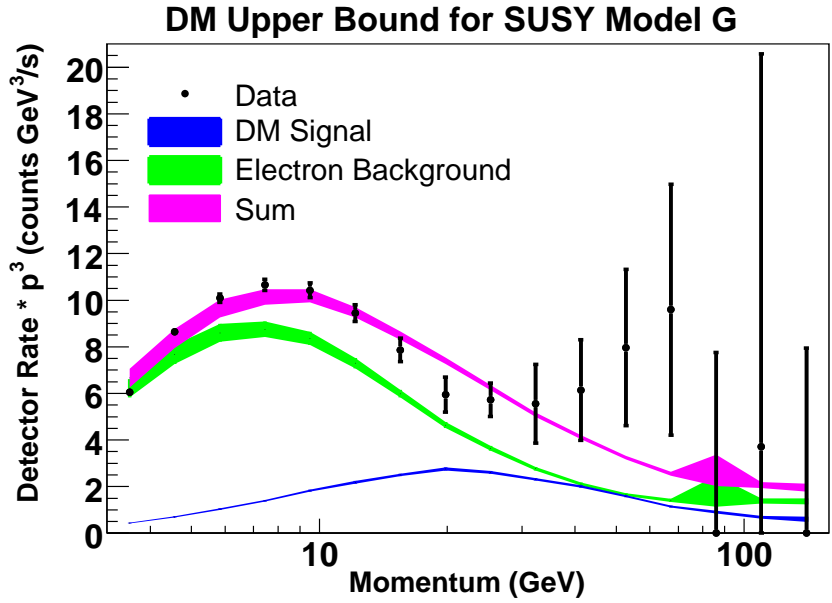}
\caption{Spectrum for Model G at 90\% maximum boost limit.}
\label{fig:lim:limfit}
\end{figure}

%% file: alphabet_bestfittable.tex
\begin{tabular}{|c|r|r|r|r|}
\hline
\multicolumn{5}{|c|}{Central Fit Values} \\
\hline
Model & Boost & Background & Background & $\chi^2/N$ \\
 &  & Normalization & Index ($\gamma$) & (N=13) \\
\hline
A & $-2.4 \times 10^{4}$ & 0.804 & 2.646 & 1.95 \\
B & $-6.6 \times 10^{3}$ & 1.466 & 2.427 & 0.86 \\
C & $-4.4 \times 10^{3}$ & 0.762 & 2.670 & 2.00 \\
D & $-1.7 \times 10^{3}$ & 0.709 & 2.701 & 2.04 \\
E & $980$ & 1.333 & 2.432 & 1.43 \\
F & $300$ & 0.686 & 2.716 & 2.05 \\
G & $-1.8 \times 10^{3}$ & 0.836 & 2.630 & 1.91 \\
H & $6.0 \times 10^{4}$ & 0.635 & 2.752 & 2.01 \\
I & $340$ & 0.860 & 2.619 & 1.87 \\
J & $3.4 \times 10^{3}$ & 0.646 & 2.743 & 2.03 \\
K & $310$ & 0.619 & 2.763 & 1.98 \\
L & $14$ & 0.715 & 2.697 & 2.04 \\
M & $40$ & 0.618 & 2.765 & 1.97 \\
\hline
\end{tabular}

%% file: alphabet_boundtable.tex
\begin{tabular}{|c|r|r|} 
\hline
 & Boost 90\% & $\langle\rho^2\rangle$ 90\% \\
& confidence level  & confidence level \\
Model & Upper Limit &  Upper Limit $\mathrm{(GeV\ cm^{-3})^2}$ \\
\hline
A & $7.2 \times 10^{4}$ & $6.5 \times 10^{3}$ \\
B & $2.7 \times 10^{3}$ & $240$ \\
C & $2.4 \times 10^{4}$ & $2.1 \times 10^{3}$ \\
D & $1.0 \times 10^{5}$ & $9.2 \times 10^{3}$ \\
E & $510$ & $46$ \\
F & $5.2 \times 10^{3}$ & $460$ \\
G & $3.9 \times 10^{3}$ & $360$ \\
H & $2.8 \times 10^{5}$ & $2.5 \times 10^{4}$ \\
I & $630$ & $57$ \\
J & $1.9 \times 10^{4}$ & $1.7 \times 10^{3}$ \\
K & $1.2 \times 10^{3}$ & $110$ \\
L & $370$ & $33$ \\
M & $150$ & $13$ \\
\hline
\end{tabular}

%% file: Results.tex
\chapter{Results and Interpretation}

	The bound set on the dark matter boost factor for a given SUSY model is a limit on how clumpy dark matter can be if that model accurately describes particle physics.  This means that measurements of supersymmetry can be used to make statements about models of dark matter distribution, and vice versa.  The results of this analysis were interpreted and compared to other experiments.

\section{Results}

\begin{figure}
\centering
\includegraphics[width=14cm]{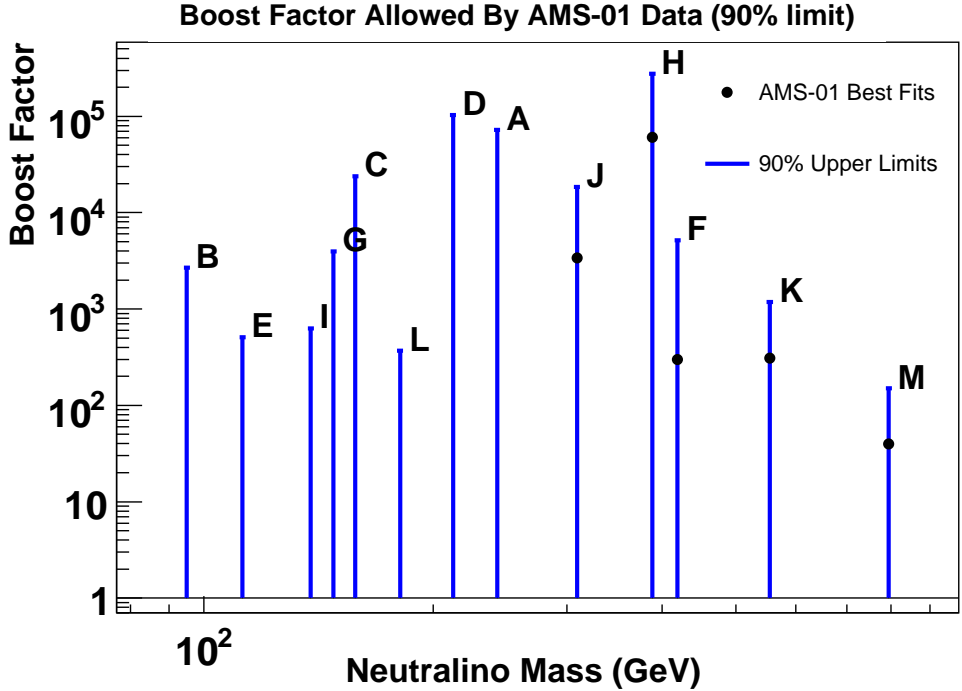}
\caption[Boost Factor Upper Limits]{Upper limits on boost factor obtained from this analysis at each benchmark model in Supersymmetric space.  Only best fits with positive boost factors are shown.}
\label{fig:res:limitvsmass}
\end{figure}

	The 90\% confidence upper limit on the boost factor for each of the mSUGRA models proposed in \cite{ref:sspointsupdated} is shown in Fig.~\ref{fig:res:limitvsmass}, along with the neutralino mass for each model.  Most boost factor limits lie between $10^2$ and $10^5$.  The momentum bins most influential to these limits are in the range 10--30 GeV/c, where a balance is struck between the larger backgrounds at low momenta and the low statistics at high momenta.  While at first it might seem that the signal strength should decrease with increasing neutralino mass, making limits worse at higher masses, no such trend is seen.  This is likely because the decreasing number density of neutralinos is offset by a greater annihilation product multiplicity and energy.

\section{Interpretation}

	These limits can be used to see if a supersymmetric model is compatible with a dark matter distribution.  From an astrophysics point of view, the limits could be used to turn the knowledge of what supersymmetry model governs our universe into an upper bound on how clumpy nearby dark matter can be.  For example, if SUSY Model E were found to be an accurate description of physics, then the distribution of dark matter in our galaxy must have a boost factor of less than 500.

	From a particle physics point of view, a measurement of the local dark matter distribution would exclude certain supersymmetric models from being responsible for all dark matter.  If dark matter was found to be particularly clumpy with a boost factor of 1000 or more then benchmark models E, I, L, and M could not be responsible for all of the dark matter in our galaxy.  This would mean that either those models were not correct, or that there was an additional component to dark matter besides neutralinos.  As no measurement or reliable prediction of boost factor exists, further astrophysical work is required to use these limits.

\section{Comparison with Other Annihilation Experiments}

	The HEAT experiment was a balloon based spectrometer that was flown in 1994, 1995, and 1998.  Its design was similar to that of AMS, with a magnet, tracker, and TOF to determine particle charge and momentum, and additionally included a transition radiation detector to allow for separation between electrons and positrons.  The cosmic ray electron and positron fluxes were measured from 5--50 GeV \cite{ref:HEAT1}.

\begin{figure}[htp]
\centering
\includegraphics[width=16cm]{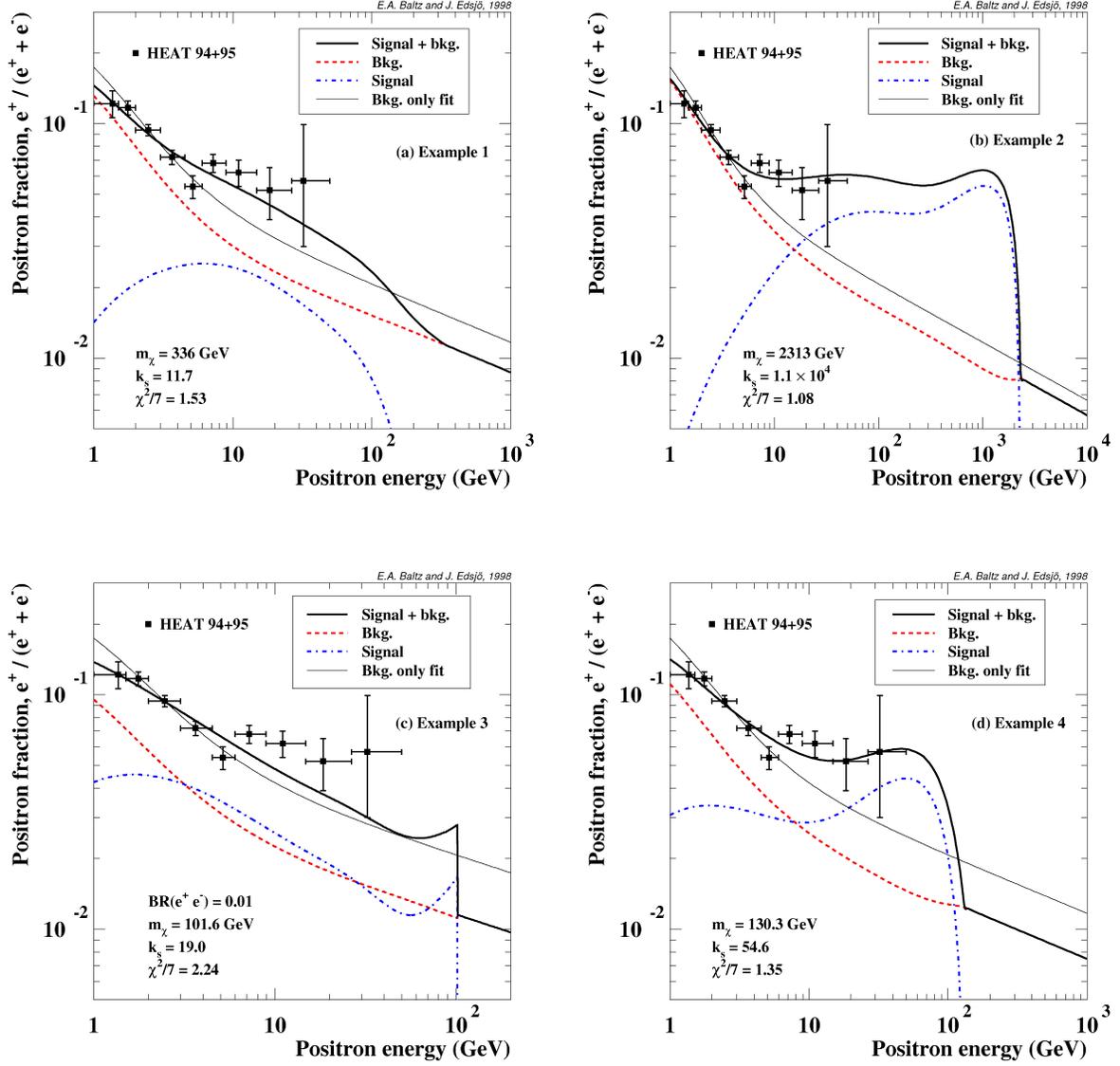}
\caption[HEAT MSSM Fits]{Best fit MSSM models for the HEAT data, from \cite{ref:HEAT2}}
\label{fig:res:heatfit}
\end{figure}

	In \cite{ref:HEAT2}, the positron to electron ratio measured in the HEAT experiment was analyzed for signs of neutralino annihilation.  The ratio was in excess of expectations from background, especially at higher energies.  A number of MSSM models and boost factors were presented as possible explanations of the excess, the best four of which are shown in Fig.~\ref{fig:res:heatfit}.  These models, labeled HEAT1--HEAT4 here, are parametrized along with the best fit boost factor from HEAT in Table~\ref{table:res:heatparams}.

\begin{table}[htp]
\centering
\begin{tabular}{|c|r|r|c|c|c|r|r|c|}
\hline
Model & $\mu$ & $M_2$ & $\tan\beta$ & $m_A$ & $m_0$ & $A_b/m_0$ & $A_t/m_0$ & Heat Boost Fit \\ 
\hline
HEAT1 & -852.3 & -670.1 & 13.1 & 664.0 & 1940.6 & -1.56 & -1.60 & 11.7\\
HEAT2 & 2319.9 & 4969.9 & 40.6 & 575.1 & 2806.6 & 2.00 & 0.27 & $1.1\times 10^4$\\
HEAT3 & -1644.3 & -202.2 & 54.1 & 181.7 & 2830.9 & 2.57 & 2.50 & 19.0\\
HEAT4 & -221.8 & -324.5 & 1.01 & 792.2 & 2998.7 & 1.71 & 0.67 & 54.6\\
\hline
\end{tabular}
\caption{MSSM Parameters for HEAT models}
\label{table:res:heatparams}
\end{table}

\begin{figure}[htp]
\centering
\includegraphics[width=14cm]{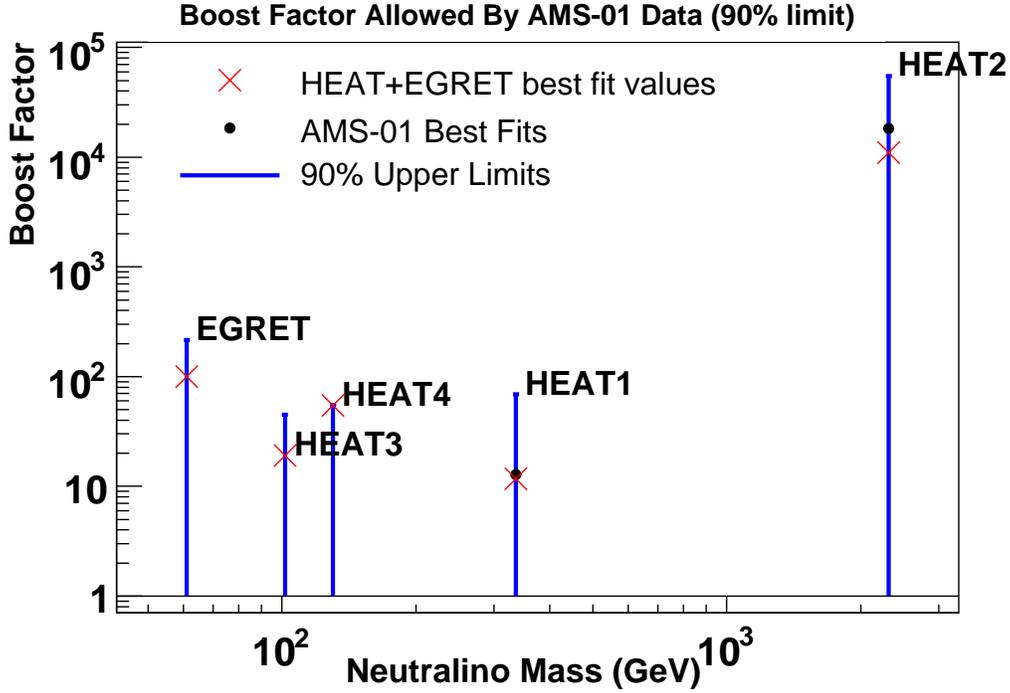}
\caption[Boost Factor Comparisons with HEAT and EGRET]{The AMS-01 boost factor upper limits on models suggested in \cite{ref:HEAT2} and \cite{ref:EGRET2}, as compared to the best fit values from HEAT and EGRET.}
\label{fig:res:otherexplimits}
\end{figure}

	The HEAT best fit boosts and the upper limits from this AMS-01 analysis are shown in Fig.~\ref{fig:res:otherexplimits} and Table~\ref{table:res:heattable1}.  Models HEAT1--HEAT3 are consistent with the limits from this analysis, but the HEAT best fit boost factor for model HEAT4 lies on the edge of the 90\% confidence upper limit from AMS-01 data, disfavoring this model.  While in general the positron to electron ratio of the HEAT experiment was more sensitive than the AMS-01 data, the HEAT excess began at $10$ GeV, where AMS still had good sensitivity.  Model HEAT4 had an excess at this energy, which conflicts with the AMS measurement.  It can be noted that for the model HEAT1, the best fit boost factor from HEAT and the present analysis are quite similar (Table~\ref{table:res:heattable2}).

\begin{table}[htp]
\centering
\input{otherexps_boundtable.tex}
\caption{AMS-01 boost factor upper bounds on HEAT and EGRET suggested models.}
\label{table:res:heattable1}
\end{table}

\begin{table}[htp]
\centering
\input{otherexps_bestfittable.tex}
\caption{AMS-01 best fit boost factors from HEAT and EGRET suggested models.}
\label{table:res:heattable2}
\end{table}

	The EGRET experiment was a gamma ray telescope that ran from 1991 to 2000 and observed the distribution of cosmic gamma rays in the range 0.03--10 GeV.  It found an excess of gamma rays in all sky directions, possibly the result of WIMP annihilation.  The excess was greater towards the center of the galaxy than would be expected from a isothermal halo distribution, possibly suggesting a more complicated distribution of dark matter \cite{ref:EGRET1}.

\begin{figure}
\centering
\includegraphics[width=12cm]{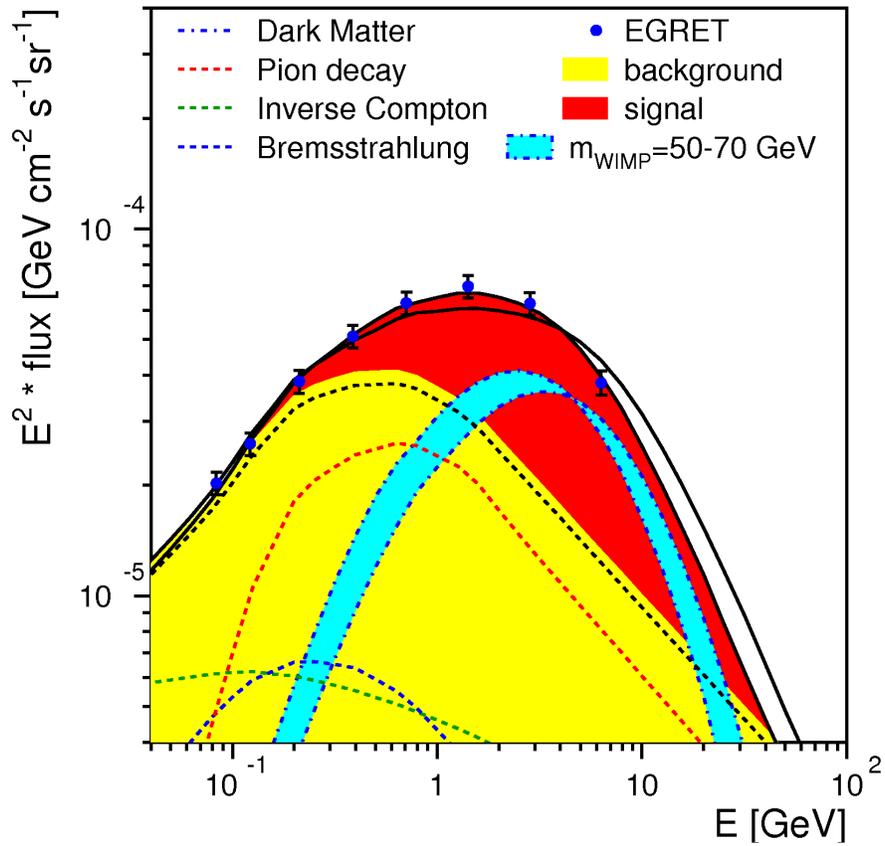}
\caption[EGRET gamma-ray excess]{A favored mSUGRA model fit the EGRET gamma ray data \cite{ref:EGRET2}}.
\label{fig:res:egretfit}
\end{figure}

	An analysis of what mSUGRA parameters would lead to dark matter annihilation signals consistent with EGRET was done in \cite{ref:EGRET2}.  A favored mSUGRA model with a $(m_0, m_{\frac{1}{2}}, \tan\beta, sgn(\mu))$ of (1400 GeV, 175 GeV, 51, $+1$) is shown in Fig.~\ref{fig:res:egretfit}.   The EGRET data suggested a boost factor on this model ranging from 100 to $10^3$ \cite{ref:EGRET1}.  The boost factor limit from this analysis for this model is roughly 200 (Fig.~\ref{fig:res:otherexplimits}).  This limit is compatible with the EGRET result, however, EGRET was sensitive to dark matter in the center of the galaxy where the boost factor is expected to be lower due to the destruction of dark matter clumps by tidal forces, while AMS-01 was only sensitive to dark matter in the nearby volume of the galaxy.  Additionally, AMS-01 is not sensitive to the ringlike halo substructures proposed in \cite{ref:EGRET1}, so a better understanding of halo structure is needed to properly compare boost factors from the two experiments.
	
\section{Comparison with Direct Detection Experiments}

\begin{figure}
\centering
\includegraphics[width=12cm]{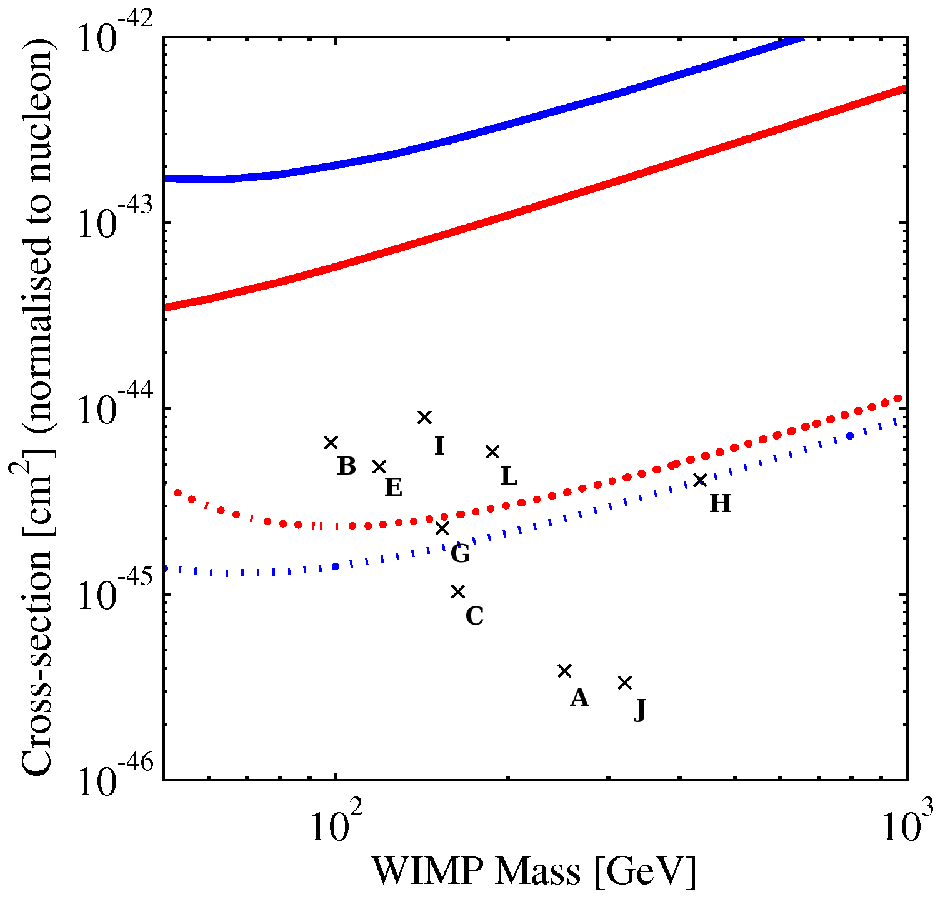}
\includegraphics[width=12cm]{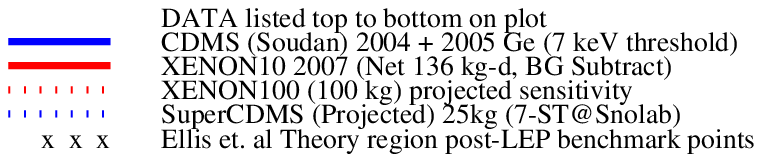}
\caption[Direct Detection Limits]{Current and projected spin-independent neutralino-nucleon cross section limits from direct experiments as compared to SUSY points studied in this analysis.  Note models D,F,K, and M all lie below $10^{-46} \mathrm{cm}^2$.  From \cite{ref:directplotgen}}.
\label{fig:res:directplot}
\end{figure}

	Direct dark matter detection experiments rely on nuclear recoil from collisions with dark matter particles.  Therefore, the rate seen in these detectors is proportional to the dark matter-nucleus cross section and the density of dark matter at Earth.  The current and projected limits on the mass and nuclear recoil cross section of dark matter from two recent experiments, CDMS \cite{ref:cdmsnospin} and XENON \cite{ref:xenon10}, are shown in Fig.~\ref{fig:res:directplot}.  Also shown are the masses and cross sections predicted by the SUSY benchmark models used in this analysis.  While these models cannot currently be probed, it appears that direct experiments in the near future will have the sensitivity required to test some of these models.  However, in the case of clumpy dark matter, whether or not a direct signal is seen depends strongly on whether the Earth lies in an overdensity or underdensity of dark matter.  If dark matter is in small dense clumps, it is likely that the Earth is in an underdensity, making a smaller than expected signal in direct detection experiments.

\section{Comparison with Collider Experiments}
	
\begin{figure}
\centering
\includegraphics[width=12cm]{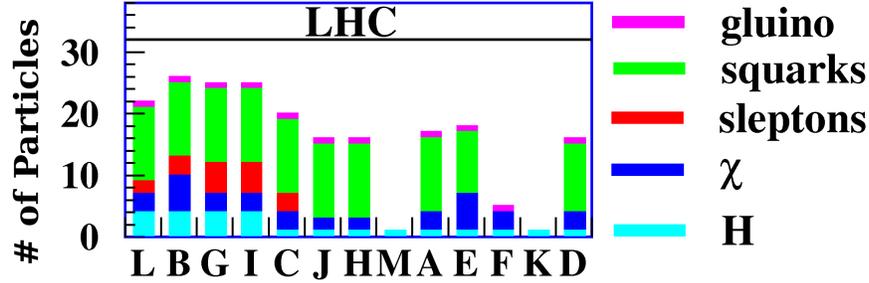}
\caption[LHC SUSY Visibility]{Number of observable particles at the LHC for the sample mSUGRA points assuming unlimited luminosity, from \cite{ref:sspointsupdated}}.
\label{fig:res:lhccount}
\end{figure}

	The supersymmetric benchmark models used in this analysis may be visible in future collider experiments, as is discussed extensively in \cite{ref:sspointsupdated}.  Of particular interest are the upcoming measurements at the LHC, which may detect new particles from these SUSY models as seen in Fig.~\ref{fig:res:lhccount}.  For most models, a number of supersymmetric particles would be visible at the LHC allowing detailed parameter measurement, but for models M and K, only a Higgs particle would be seen.  These models would be indistinguishable at the LHC, meaning that the supersymmetric parameters would not be measurable.  This analysis was able to limit the boost factor of models M and K to under around 1000, so future indirect dark matter searches that are three orders of magnitude better would be able to distinguish these models while the LHC could not.  The next chapter discusses ways that indirect dark matter searches could be improved.	

%% file: otherexps_boundtable.tex
\begin{tabular}{|c|r|r|} 
\hline
 & Boost 90\% & $\langle\rho^2\rangle$ 90\% \\
& confidence level  & confidence level \\
Model & Upper Limit &  Upper Limit $\mathrm{(GeV\ cm^{-3})^2}$ \\
\hline
HEAT1 & $68$ & $6.1$ \\
HEAT2 & $5.5 \times 10^{4}$ & $4.9 \times 10^{3}$ \\
HEAT3 & $44$ & $4.0$ \\
HEAT4 & $54$ & $4.9$ \\
EGRET & $210$ & $19$ \\
\hline
\end{tabular}

%% file: otherexps_bestfittable.tex
\begin{tabular}{|c|r|r|r|r|}
\hline
\multicolumn{5}{|c|}{Central Fit Values} \\
\hline
Model & Boost & Background & Background & $\chi^2/N$ \\
 &  & Normalization & Index ($\gamma$) & (N=13) \\
\hline
HEAT1 & $13$ & 0.644 & 2.745 & 2.03 \\
HEAT2 & $1.8 \times 10^{4}$ & 0.604 & 2.777 & 1.87 \\
HEAT3 & $110$ & 1.506 & 2.412 & 0.92 \\
HEAT4 & $24$ & 0.832 & 2.629 & 1.92 \\
EGRET & $180$ & 0.959 & 2.606 & 1.53 \\
\hline
\end{tabular}

%% file: Future.tex
\chapter{Recommendations for Future Work}

	The primary limitations of this analysis are discussed here, along with how limits could be improved by additional experimental and theoretical work.

\section{Limitations of this Method}

\begin{figure}[htp]
\centering
\includegraphics[width=14cm]{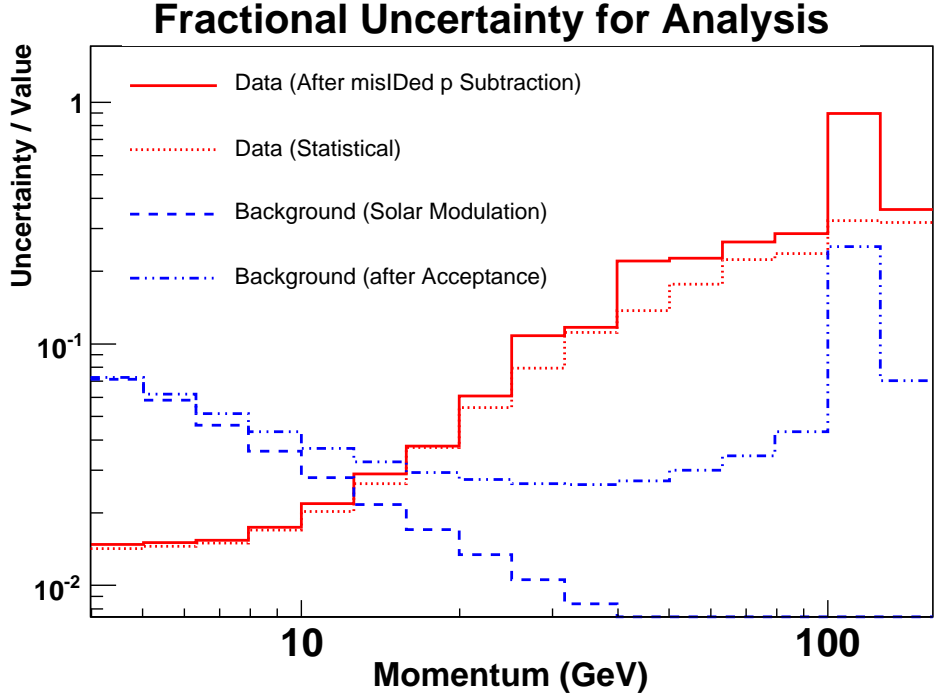}
\caption{Relative uncertainties from theoretical background and data}
\label{fig:fut:uncplot}
\end{figure}

	In order to improve the limits obtained in this analysis, it is necessary to examine the origin of the uncertainties in the background, data, and signal.  These uncertainties can overwhelm dark matter annihilation signals and make boost factor upper limits worse.  Fig.~\ref{fig:fut:uncplot} shows the uncertainty fraction (uncertainty divided by value) for the backgrounds and data during several steps of this analysis.  This plot shows that this analysis was limited by low data statistics and resolution at high energies, and astrophysical uncertainties at lower energies.

	In this plot, the dotted line shows the statistical uncertainty from the limited number of counts.  The much greater uncertainty at higher energies (when compared to lower energies) comes from the power law nature of the background, and the only way to improve these limits is to have an experiment that either runs longer or has a larger detector area.  The uncertainty will decrease as the square root of the run time or area.  The solid line above this is the uncertainty after the misidentified protons have been subtracted.  The way to improve or eliminate this uncertainty is to design a detector that can distinguish protons from electrons at high energies.
The dashed line shows the uncertainty caused by the uncertainty in the $\Phi$ parameter in the force field approximation to solar modulation.  This will hopefully improve as models for the effect of solar fluctuations are refined.  Near constant monitoring of cosmic ray fluxes would undoubtedly be useful to accurately determine how solar modulation changes over time.  The dash-dotted line indicates uncertainty in the background after the acceptance matrix is applied.  Uncertainty increases at high momenta here mostly as a result of momentum resolution and misidentified proton issues, and these could be reduced with more Monte Carlo simulation. It is clear, however, that the statistical uncertainties dominate at higher energies.  The sources of uncertainties in the annihilation signal are the same as in the background.

	One major source of uncertainty not shown results from the galactic propagation model.  This is difficult to quantify and not included in the limits in this paper, but recent work \cite{ref:satalk} suggests that uncertainty in the propagation parameters would only make the limits set here larger by a few percent.

\section{Improvements by the AMS-02 Experiment}
	
	Data from the AMS-02 experiment, for which AMS-01 was a test run, would result in a considerable improvement in this analysis.  It would improve uncertainties on three fronts:

	1) The active lifetime of AMS-02 would be at least 3 years, being mounted to the International Space Station.  This would give a statistical improvement of more than an order of magnitude over the AMS-01 experiment.

	2) The AMS-02 experiment's superconducting magnet would have a bending power over 5 times that of the AMS-01 experiment's permanent magnet and the silicon tracker would be eight layers compared to AMS-01's six, improving resolution at high energies \cite{ref:AMS02magnet}.

	3) AMS-02 would include a transition radiation detector which would allow separation of protons from electrons at energies well above 200 GeV \cite{ref:AMS02TRD}.  This would decrease the amount of misidentified protons that must be subtracted from the electron data.  

	AMS-02 would additionally be able to look at more particle spectra.  With a much longer running time and good particle mass separation from the transition radiation detector, the positron and antiproton spectra could be measured separately from the proton and electron spectra.  These antiparticle spectra have the same signal strength as the electron spectrum but with much less background, making them more attractive spectra to search for dark matter annihilation remnants.  AMS-02 would also feature an electromagnetic calorimeter that would be sensitive to photons, and could search for photons from neutralino annihilation in a manner similar to the EGRET experiment.

\section{Improvements by Other Future Experiments}

	Other experiments may also be able to improve dark matter limits.  While space is the ideal location to observe cosmic rays, high altitude balloon experiments can also take good measurements of cosmic ray spectra. In particular, it has been shown that antiproton spectra can be measured with the balloon experiments BESS \cite{ref:BESS3}.  Antiprotons have a very small cosmogenic background \cite{ref:AMS01}, and as seen in Fig.~\ref{fig:ss:signalpropagation}, while they are produced less copiously than electrons or positrons in dark matter annihilation, they are also less affected by galactic propagation.  Thus they have a similar signal strength to electrons or positrons.  While BESS had run times too brief and maximum measurable momenta that were too low to be optimal for detecting dark matter annihilation signals, larger balloon experiments that could be run for longer periods of time may have a good chance of detecting dark matter signals.

	Beyond antiprotons, antideuterons are expected to have an even smaller background.  For certain SUSY models, the expected antideuteron flux from annihilation is twenty times that of the expected background from secondary production.  The general antiparticle spectrometer (GAPs) experiment, still in the prototype stage, is designed to detect the absorption of low energy antideuterons from cosmic rays in a gas cell.  Thus it could be very sensitive to neutralino annihilation, depending strongly on the particular SUSY model \cite{ref:GAPS}.

	Future gamma ray telescopes have the ability to probe dark matter substructure more finely.  Because of the directional nature of gamma rays, the density distribution of a dark matter clump can be measured from it's luminosity.  The Gamma-ray Large Area Telescope (GLAST) may be sensitive to clumps with boost factors as low as five under the right conditions \cite{ref:GLAST}.

	In conclusion, the future is bright for dark matter experiments.  This analysis has shown that the boost factor of many SUSY models must be less than $10^4$.  Indirect experiments planned for the next few years, between longer running times, better understanding of backgrounds, and antiparticle identification will likely be $10^4$ times more sensitive, allowing the discovery or exclusion of those models.  If dark matter is produced by one of the SUSY models that was permitted to have a boost factor of greater $10^4$, however, then we may have to wait for the following generation of dark matter experiments, or look for another way of detecting dark matter.

%% file: Fitting.tex
\chapter{Fitting}

	Fits and confidence intervals in this analysis are done using the log likelihood method assuming Gaussian errors.  The figure of merit for a parameter set with this method is the logarithm of the likelihood of the data given the parameters, which is:
\begin{equation}
f=\sum_{i\in bins}(\frac{(x_i-d_i)^2}{\sigma_i^2}-2\ln{\sigma_i})
\end{equation}
for bins $i$ with data values $d_i$ and predicted values $x_i$, with combined data and prediction errors $\sigma_i$.  This figure is equal to the $\chi^2$ of the fit plus an additional term that favors  good fits with small uncertainties over poor fits with large uncertainties.

	Finding the best fit parameters involves finding the parameters that minimize the log likelihood $f$.  The program MINUIT \cite{ref:Minuit} uses a gradient descent method that efficiently finds the parameters at the function minimum. Uncertainties in the fit parameters are taken to be $\sigma_\alpha=(d^2f/d\alpha^2)^{-1}$ for parameter $\alpha$, which assumes a Gaussian form for the errors.
	
	The difference between the log likelihood for a point and the minimum log likelihood, $\Delta f$, is related to how much less likely a parameter set is than the most likely parameter set.  This allows contours of confidence intervals to be mapped out.  For Gaussian uncertainties with 3 parameters, a $\Delta f$ of 3.67 indicates the contour of a $70\%$ confidence ellipsoid, and a $\Delta f$ of 6.25 indicates the contour of a $90\%$ confidence ellipsoid \cite{ref:Minuit}.

%% file: Geomagnetic.tex
\chapter{The Geomagnetic Cutoff}

Here the formula for the geomagnetic cutoff is derived more briefly than in \cite{ref:stormer}, but with the same result.  The goal is to find at what momentum particles are trapped in the Earth's field, and at what momentum are they free.

\begin{figure}[h]
\centering
\includegraphics[width=11.5cm]{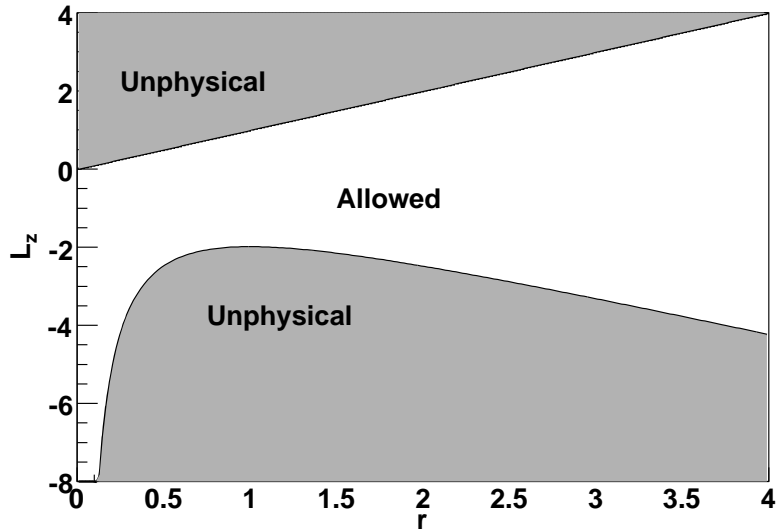}
\caption[Allowed $L_z$]{Allowed region of $L_z$ and $r$.}
\label{fig:apb:lzlimit}
\end{figure}

	Beginning with the Lagrangian for a charged particle in an electromagnetic field:
\begin{equation}
\mathcal{L}=-mc^2\sqrt{1-\beta^2}-q\Phi+q\beta\cdot\vec{A}
\end{equation}
	The Earth's magnetic field can be approximated as a dipole.
	For a magnetic dipole with moment $\vec{M}$, the potentials are:
\begin{equation}
\Phi=0\ \ \ \ \vec{A}=\frac{\mu_0}{4\pi}\frac{\vec{M}\times\vec{r}}{|r|^3}
\end{equation}
	So the Lagrangian becomes:
\begin{equation}
\mathcal{L}=-mc^2\sqrt{1-\beta^2}+\frac{q\mu_0}{4\pi|r|^3}\vec{\beta}\cdot(\vec{M}\times\vec{r})
\end{equation}
	Expressing this in polar coordinates with the the magnetic dipole pointing towards $\phi=0$ with magnitude $M$, this can be written:
\begin{equation}
\mathcal{L}=-mc^2\sqrt{1-\beta^2}+\dot{\phi}\frac{q\mu_0M}{4\pi cr}\sin^2\theta
\end{equation}
	The angular momentum $L_z$ of this system is conserved.
\begin{equation}
\frac{d\mathcal{L}}{d\dot{\phi}}=-\dot{\phi}\gamma mc^2r^2\sin^2 \theta+\frac{q\mu_0 M}{4\pi cr}\sin^2 \theta = L_z
\end{equation}
	Notice that $\dot{\phi}\gamma mc^2 r\sin\theta = p_{\phi}$, the momentum in the $\phi$ direction.   Define a new angle, the East-West angle $\phi_{EW}$, that is the angle that $p$ makes with the zenith in the $\phi$ direction, so that $p_{\phi}=p \cos{\phi_{EW}}$.  The angular momentum can thus be rewritten:
   
\begin{equation}
L_z=-pr\sin \theta \cos \phi_{EW} + \frac{q\mu_0M}{4\pi cr}\sin^2\theta
\end{equation}
	Replace $q$ by $Q|Z|$ where $Q$ is $+1$ or $-1$ according to the sign of the particle's charge, and Z is the magnitude of the charge.  Then rescale $r$ and $L_z$ by $(\frac{|Z|\mu_0M}{4\pi cp})^{1/2}$ to make them unitless.  This factor will return later. 
\begin{equation}
L_z=-r\sin \theta \cos \phi_{EW} + \frac{Q}{r}\sin^2\theta
\end{equation}
	$\cos \phi_{EW}$ can range from $-1$ for particles coming directly from the East to $+1$ for particles coming from the West, and $\cos \theta$ can range from $-1$ to $+1$.     
	Thus, the maximum and minimum $L_z$ for a particle at radius $r$ are given by (Fig.~\ref{fig:apb:lzlimit}):
\begin{equation}
-r-\frac{1}{r}<L_z<r
\end{equation}

	Particle trajectories in this figure are lines of constant $L_z$.  It can be seen that particles with $L_z$ less than $-2$ lie in one of two areas, those with $r<1$ correspond to particles trapped within the Earth's field and those with $r>1$ are particles that come from space that are turned away by the field.  Particles with $L_z>-2$ are not trapped.  Therefore, the minimum radius particles from space can penetrate into the Earth's field can be found by solving:
\begin{equation}
-2<-r\sin \theta \cos \phi_{EW} + \frac{Q}{r}\sin^2\theta
\end{equation}
if this is put in terms of $1/r$, it becomes
\begin{equation}
\frac{Q}{r^2}\sin^2\theta +2r-r\sin \theta \cos \phi_{EW} >0
\end{equation}
choosing the appropriate region for r,
\begin{equation}
\frac{1}{r}<\frac{1+\sqrt{1-Q \sin^3\theta\cos\phi_{EW}}}{Q \sin^2\theta}
\end{equation}
unscaling r, this becomes
\begin{equation}
\frac{1}{r}(\frac{|Z|\mu_0M}{4\pi cp})^{1/2}<\frac{1+\sqrt{1-Q \sin^3\theta\cos\phi_{EW}}}{Q \sin^2\theta}
\end{equation}
and so the bound on momenta of free particles at a given radius is
\begin{equation}
p>\frac{|Z|\mu_0M}{4\pi cr^2}\frac{\sin^4\theta}{(1+\sqrt{1-Q \sin^3\theta\cos\phi_{EW}})^2}
\end{equation}

The quantity $(|Z|\mu_0M)/(4\pi cr^2)$ for a particle near the Earth's surface is about $|Z| \times 59.6 \mathrm{GeV/c}$, and it is customary to speak of (geomagnetic) latitude $\lambda$ instead of $\theta$, where $\lambda=\theta+\pi$, so the geomagnetic cutoff is given by:

\begin{equation}
\frac{p}{|Z|}>(\mathrm{59.6[GeV/c]} )\frac{\cos^4\lambda}{(1+\sqrt{1-Q \cos^3\lambda\cos\phi_{EW}})^2}
\end{equation}

%% file: Propagation.tex
\chapter{Propagation}
	This appendix outlines the approximations used to estimate the volume of the galaxy probed by indirect dark matter experiments using electrons and antiprotons.

	The primary mechanisms of energy loss for electron cosmic rays are synchrotron radiation, inverse Compton scattering, ionization, and bremsstrahlung.  The energy loss from each process is (from \cite{ref:logairv2}):
\begin{eqnarray*}
\mathrm{Synchrotron\ and}\ \ \ \ \ && \nonumber \\
\mathrm{Inverse\ Compton\ Scattering}\ \frac{dE}{dt} & = & -9.4\times10^{-8} E^2\ \mathrm{GeV^{-2}eV\ s^{-1}} \\
\mathrm{Ionization}\ \frac{dE}{dt} & = & -2.5\times10^{-7}\ \mathrm{eV\ s^{-1}} \\
\mathrm{Bremsstrahlung}\ \frac{dE}{dt} & = &-3.7\times10^{-7}E\ \mathrm{GeV^{-1}eV\ s^{-1}} \\
\end{eqnarray*}
	For high energy electrons, the dominant form of energy loss is synchrotron and inverse Compton scattering.  From this process alone, an electron with an energy typical of an annihilation product, 100 GeV, would be reduced to an energy below the range of this analysis, 1 GeV, after about ten million years of travel ($3\times 10^{14}$ s).

	As they travel through the galaxy, cosmic rays will be scattered by sudden magnetic field changes due to stars or other galactic objects.  In our neighborhood of the galaxy, there is roughly one star every cubic parsec, so it can be supposed that cosmic rays will have a mean free path of roughly one parsec ($\lambda=3\times 10^{18}$ cm).  The velocity of cosmic rays is nearly $c$, but they will travel in helices along the predominant magnetic field lines.  Assuming equipartition of velocity between the circular and liner part of the helix, the cosmic rays will move along the field lines with a velocity of $c/\sqrt{3}$.  This yields a diffusion constant of:
\begin{equation}
D=v\lambda=\frac{c}{\sqrt{3}}\approx5\times10^{28}\mathrm{cm^2s}
\end{equation}
The diffusion constant used in GALPROP, which is calculated more carefully, is $10^{29} \mathrm{cm^2s}$.

	The distance traveled after a certain time $t$ in a diffusive process is $\sqrt{Dt}$.  Therefore, the range over which annihilation electrons can be detected can be estimated by $\sqrt{10^{29} \mathrm{cm^2 s^{-1}}\ 3\times10^{14} \mathrm{s}}\approx 2\ \mathrm{kiloparsecs}$.  Antiprotons are not nearly as affected by synchrotron radiation and Compton scattering, and may travel much greater distances before losing their energy.  However, the galactic disk is only a few kiloparsecs thick and when a particle leaves the galactic disk it is no longer subject to scattering by magnetic fields around stars and may leave the galaxy entirely.  Therefore, despite having less energy loss than electrons, antiprotons from sources further than a few kiloparsecs will be greatly attenuated.

	The solar system is about 8.5 kiloparsecs from the center of the galaxy, so an indirect search using electrons and antiprotons with a range of a few kiloparsecs probes only a relatively local region of the galaxy.  This region is large enough, however, to guarantee that there is dark matter within range, as the stars used in determining the local dark matter density are within this region.

%% file: ExtraPlots.tex
\chapter{Additional Plots}

\section{Confidence Intervals}
\begin{figure}[h]
\centering
\includegraphics[width=11.5cm]{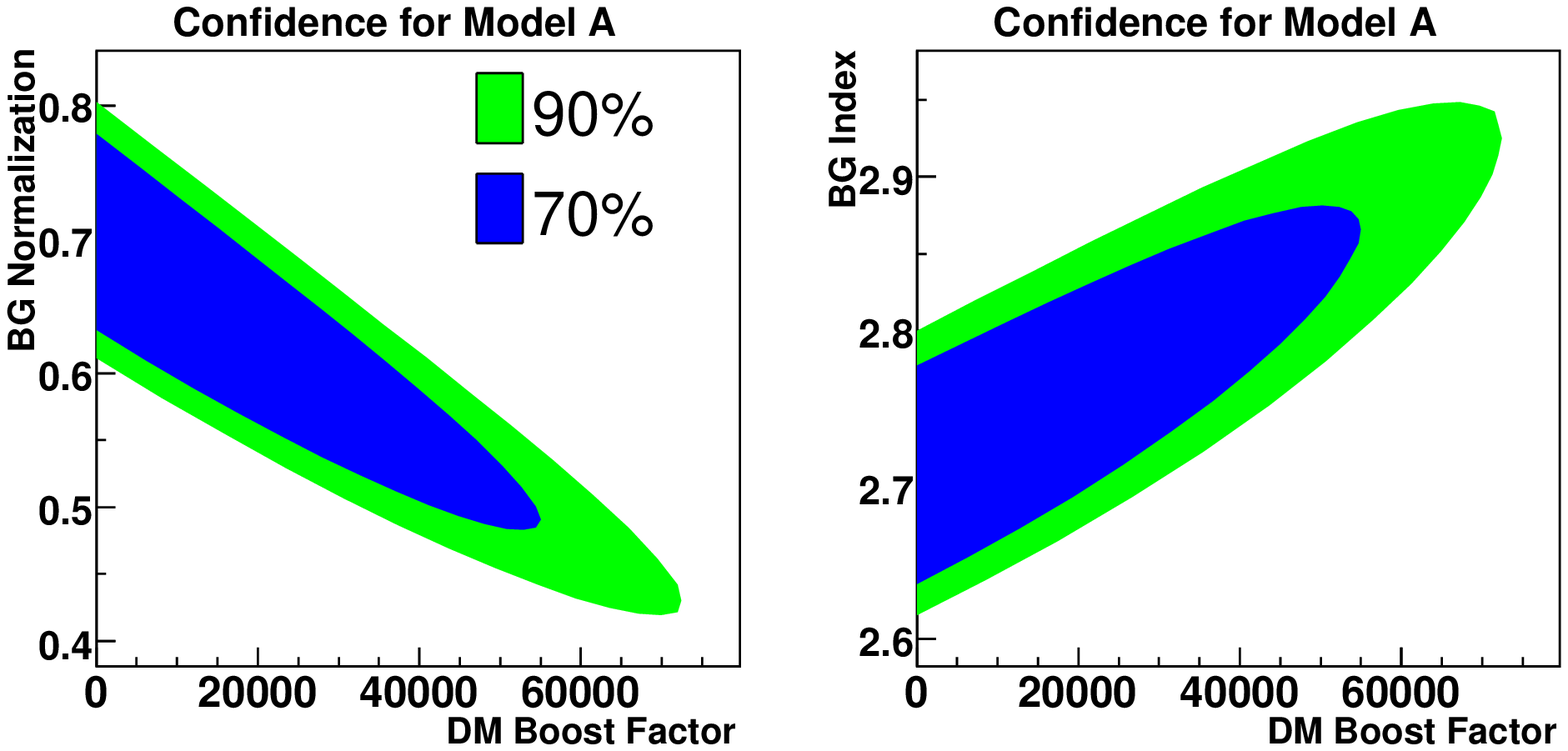}
\includegraphics[width=11.5cm]{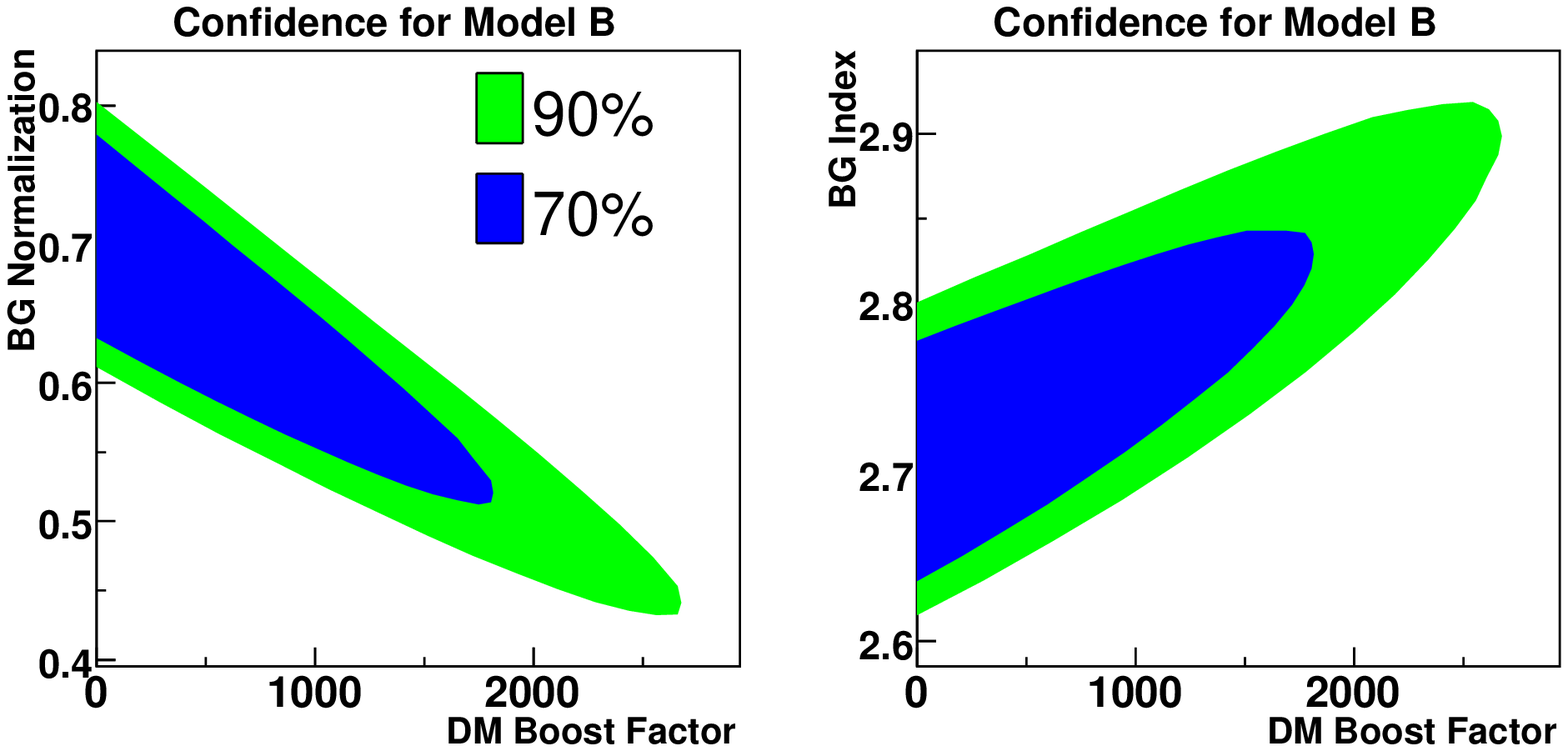}
\caption{Confidence regions for Model A and B.}
\end{figure}

\begin{figure}[h]
\centering
\includegraphics[width=11.5cm]{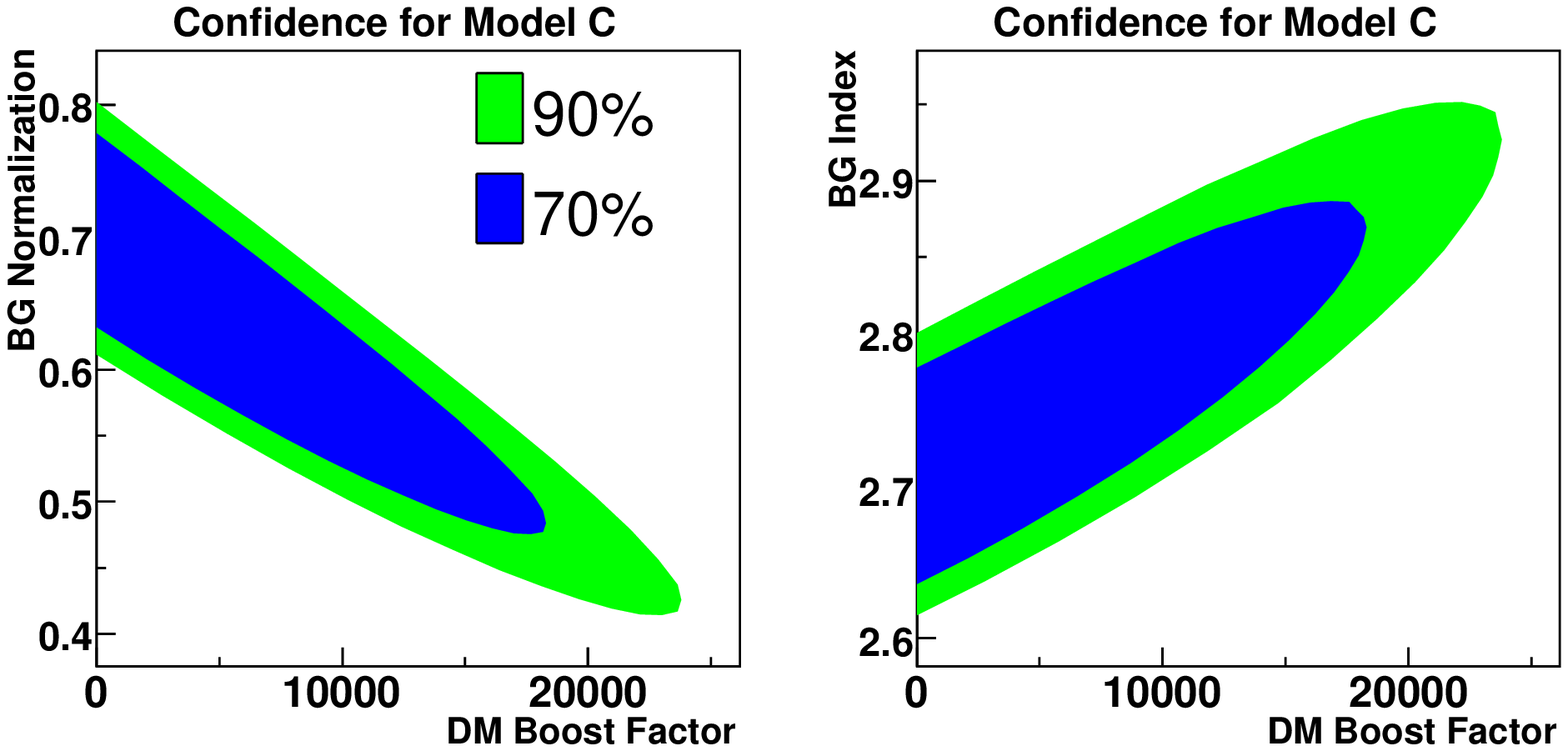}
\includegraphics[width=11.5cm]{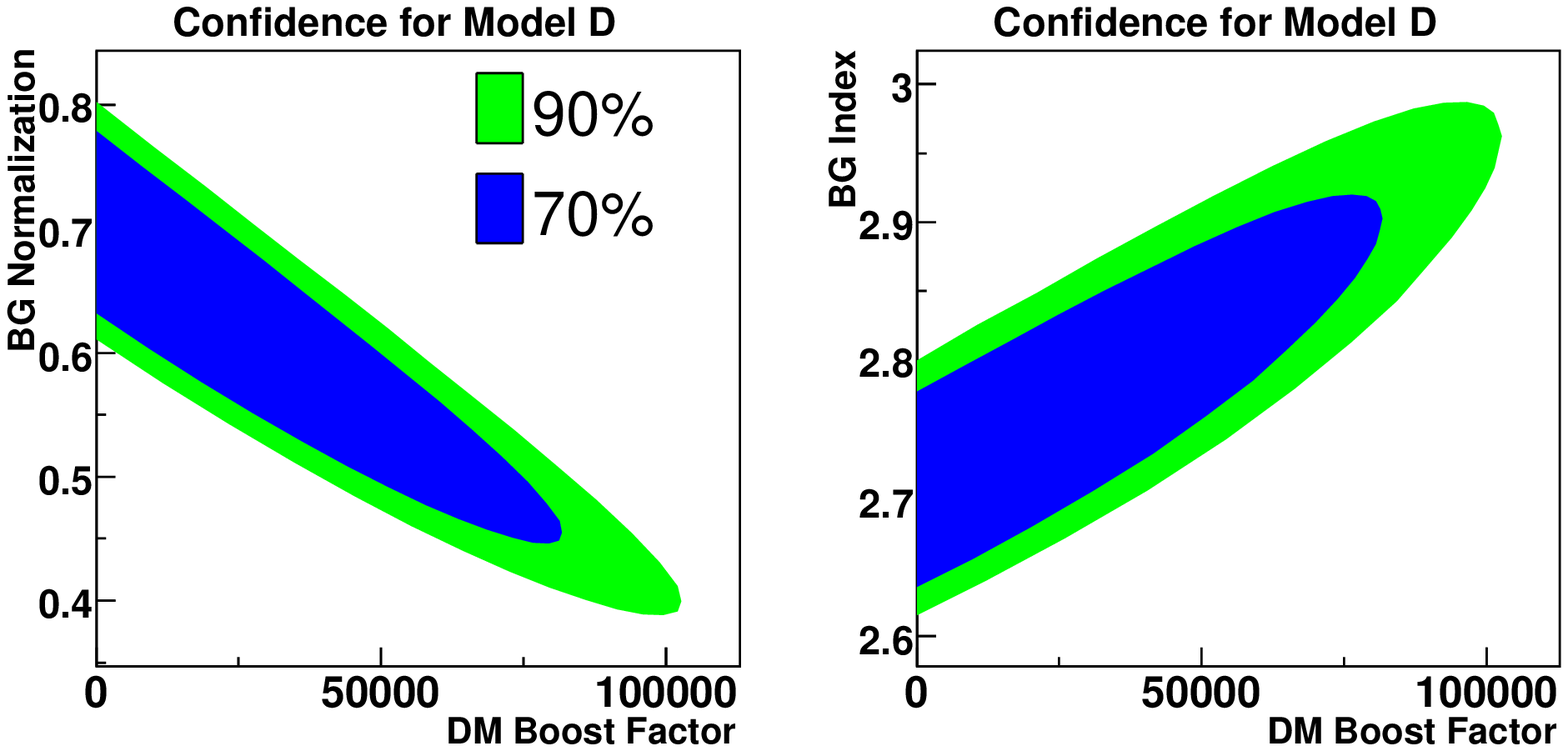}
\includegraphics[width=11.5cm]{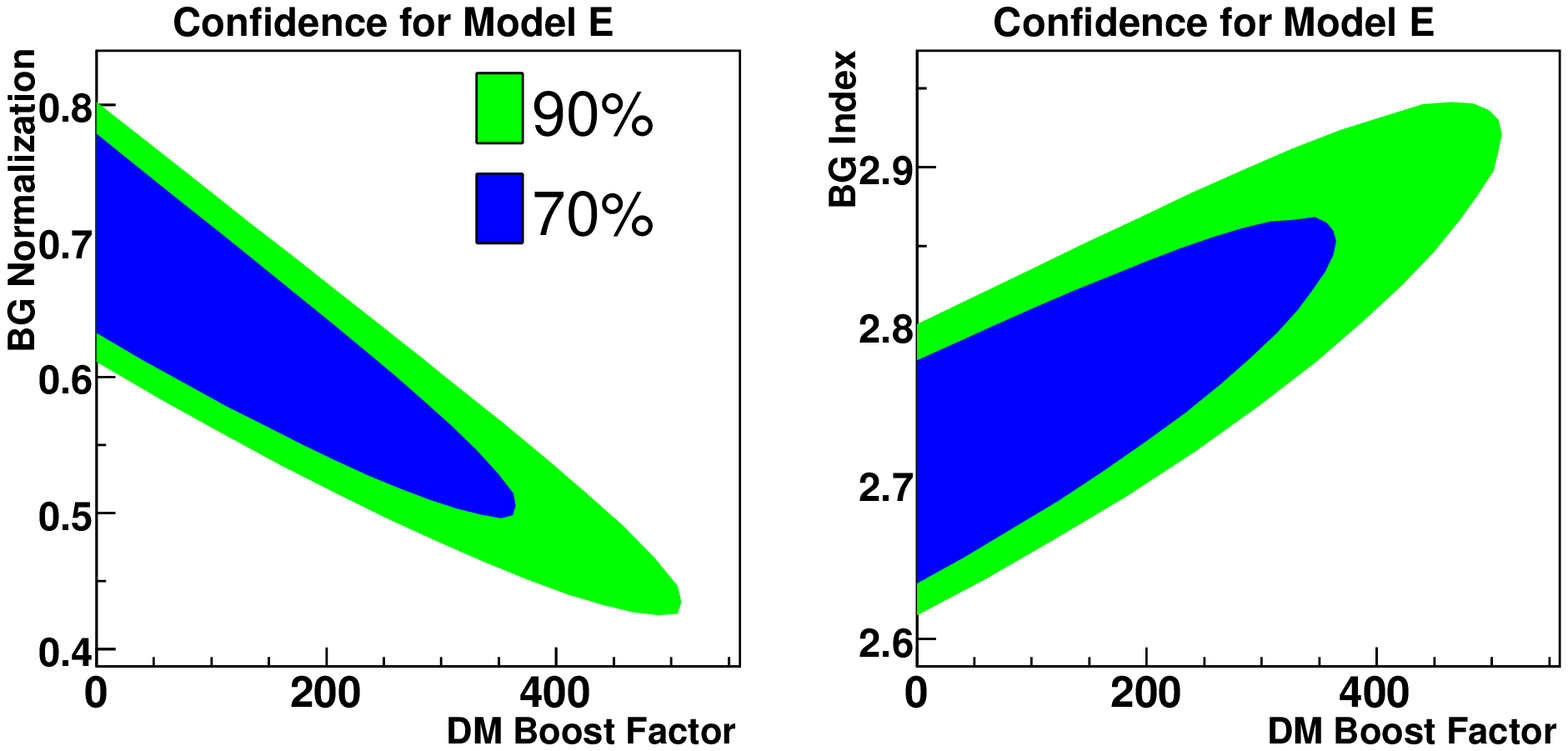}
\includegraphics[width=11.5cm]{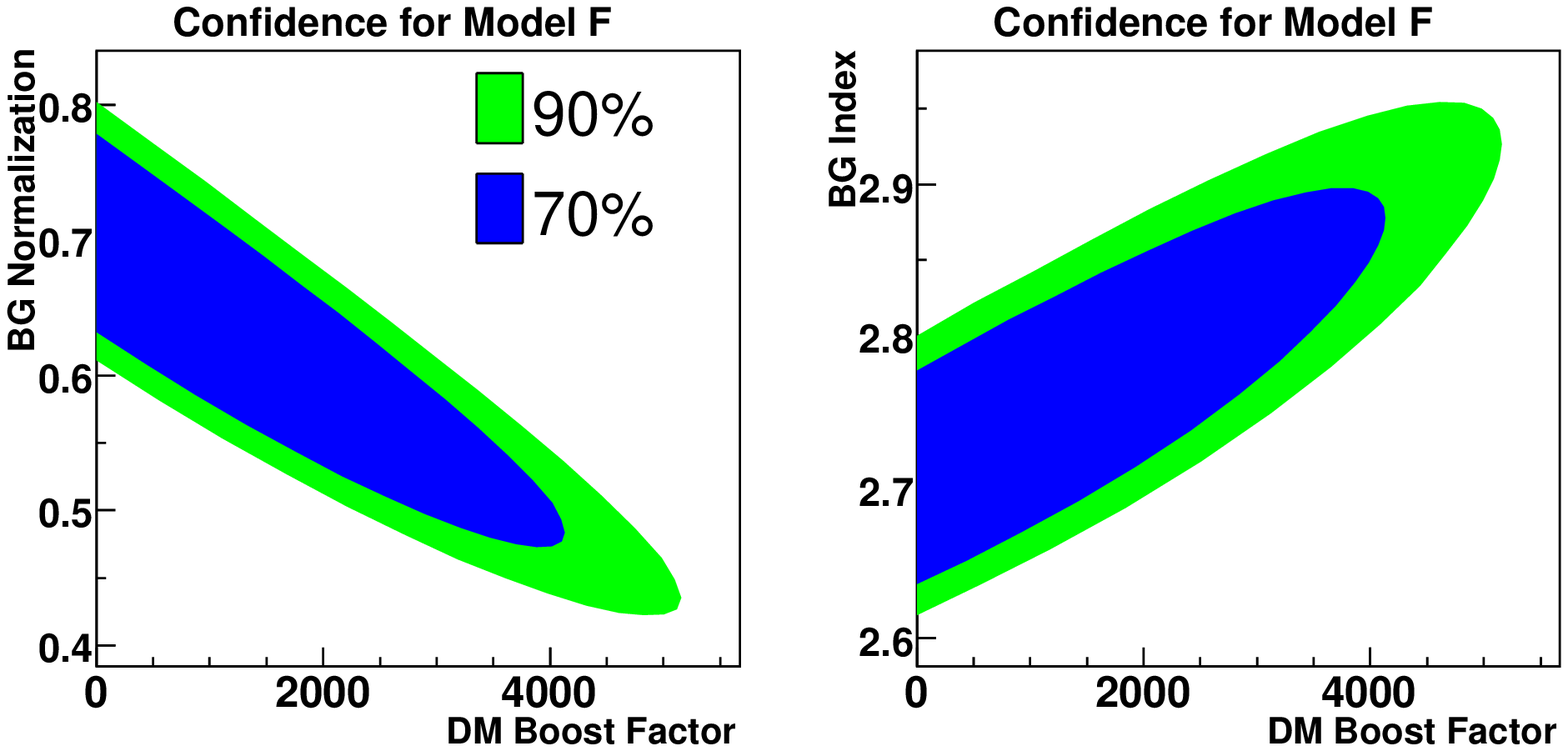}
\caption{Confidence regions for Models C to F.}
\end{figure}

\begin{figure}[h]
\centering
\includegraphics[width=11.5cm]{alphabet_2dconf_6.eps}
\includegraphics[width=11.5cm]{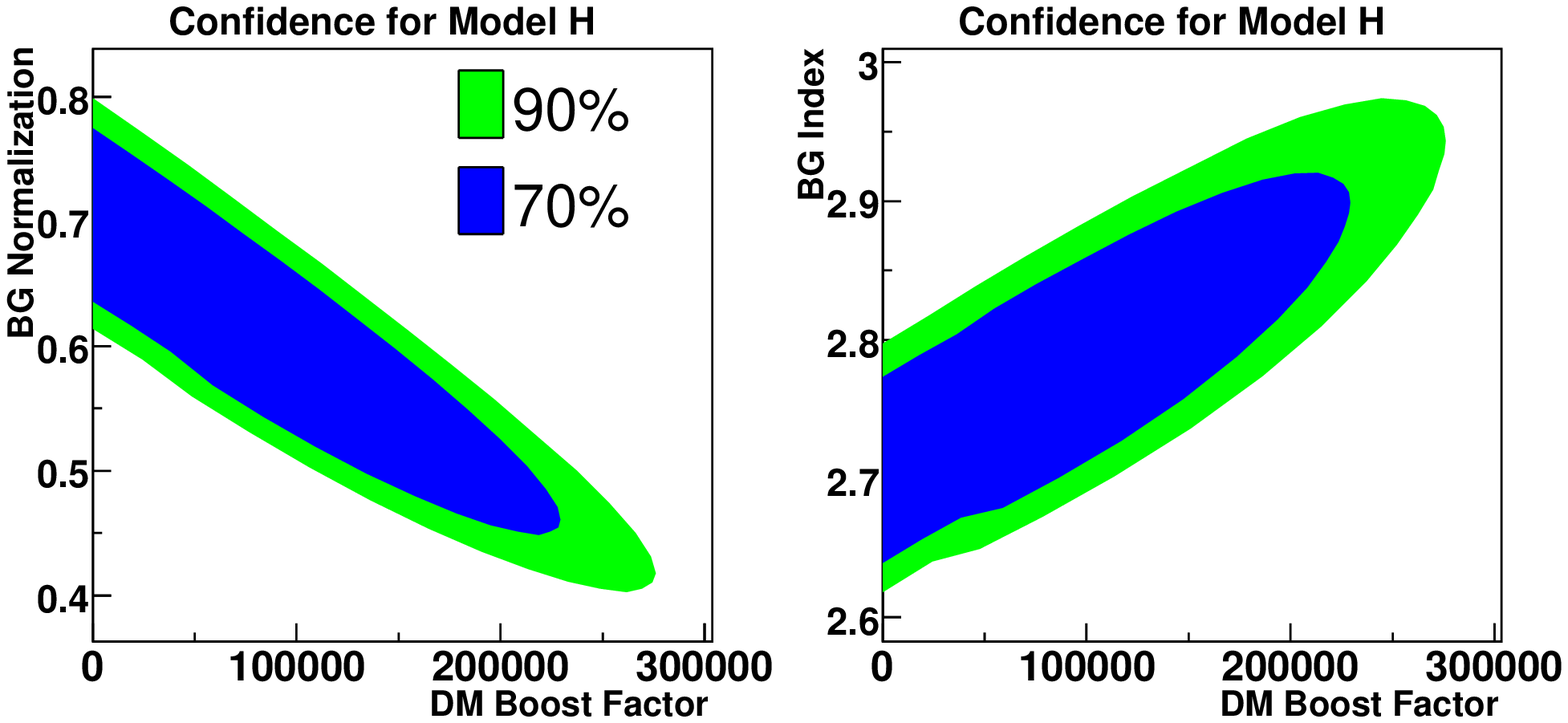}
\includegraphics[width=11.5cm]{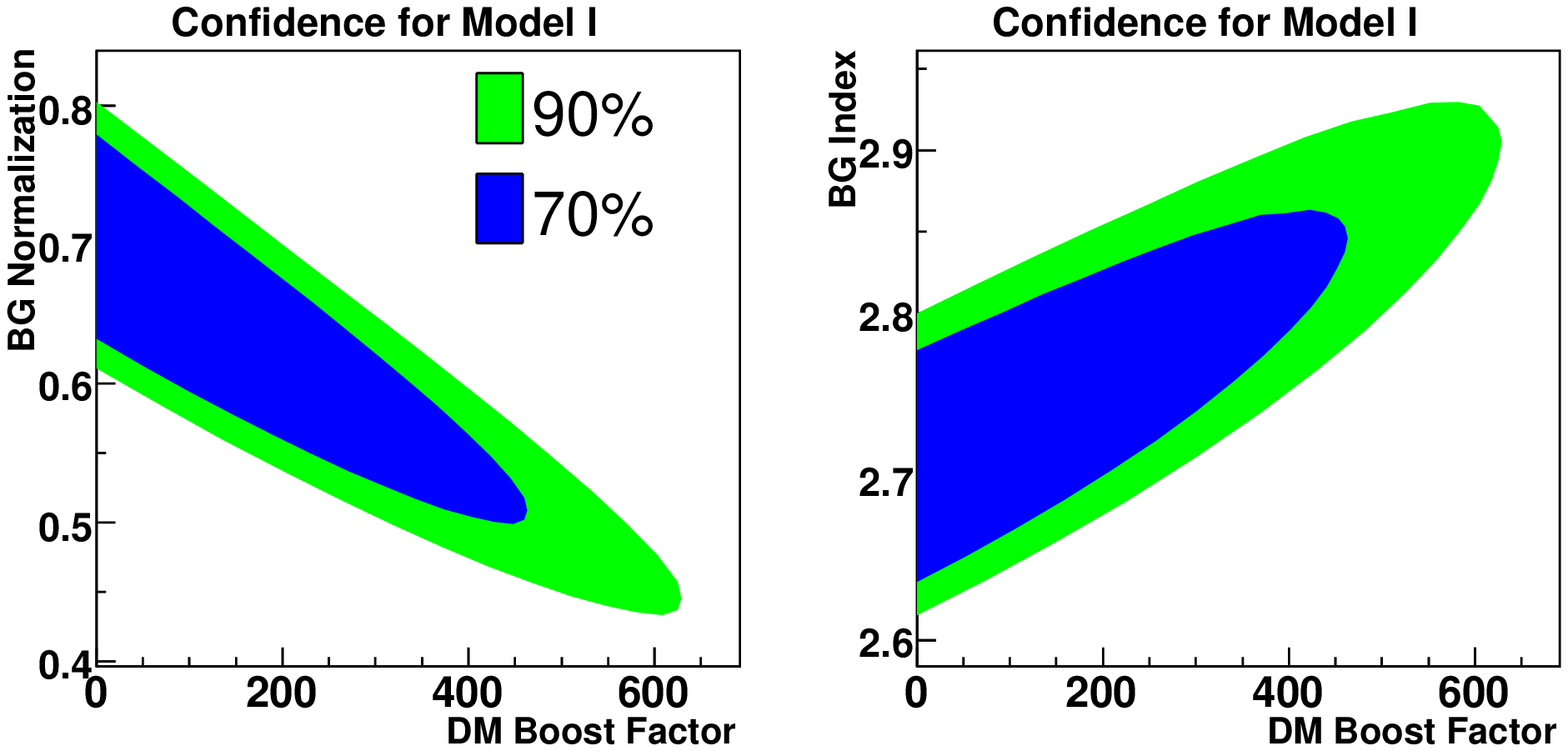}
\includegraphics[width=11.5cm]{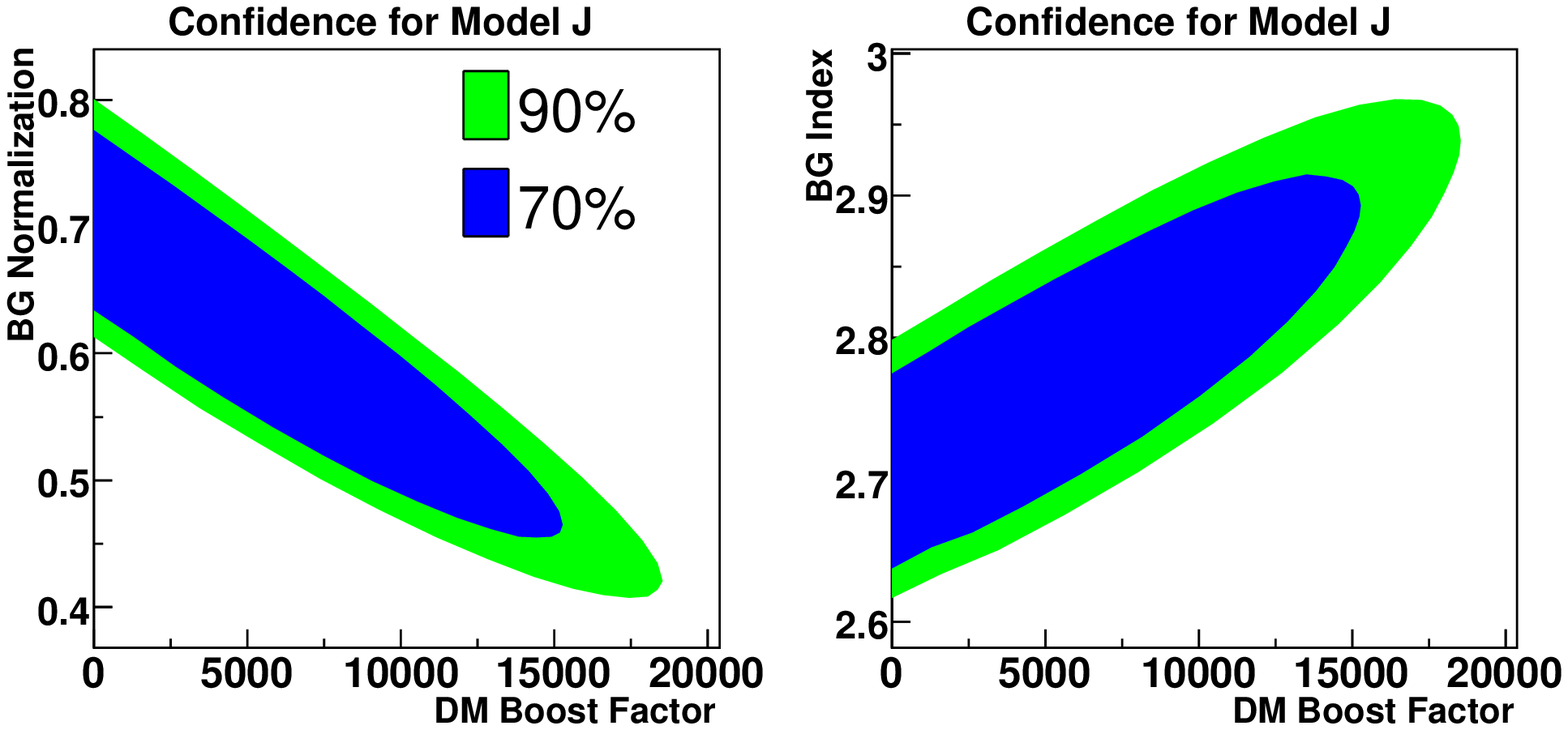}
\caption{Confidence regions for Models G to J.}
\end{figure}

\begin{figure}[h]
\centering
\includegraphics[width=11.5cm]{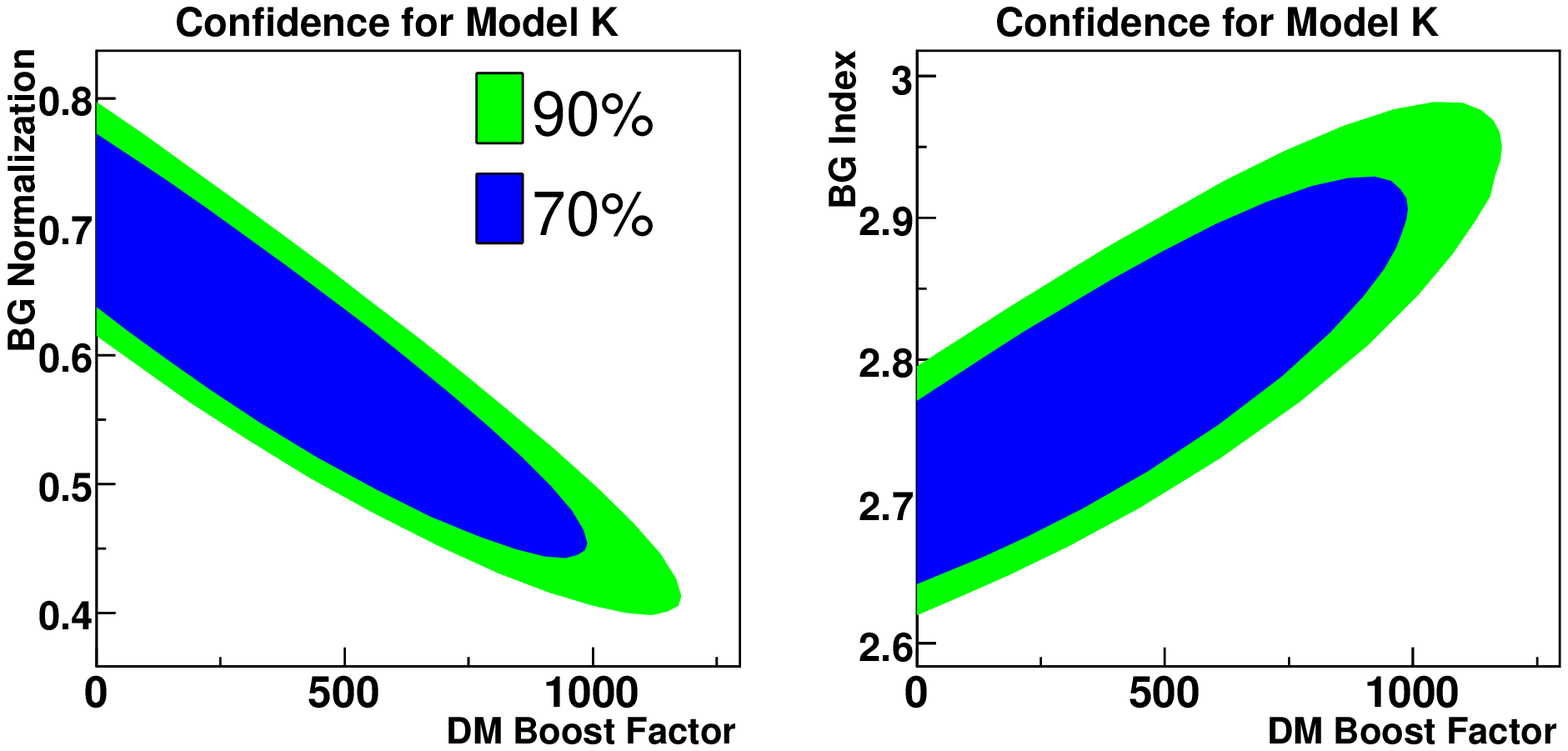}
\includegraphics[width=11.5cm]{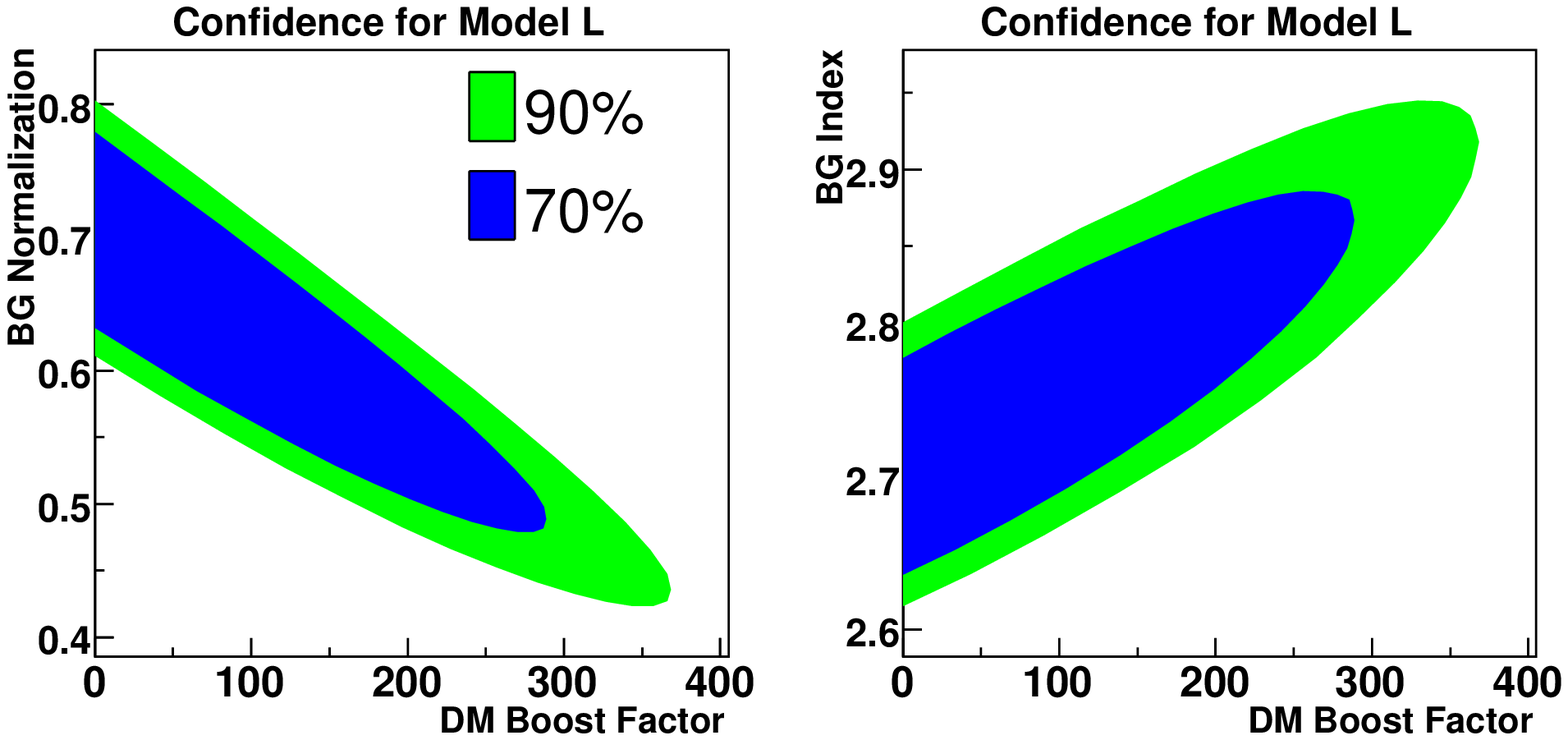}
\includegraphics[width=11.5cm]{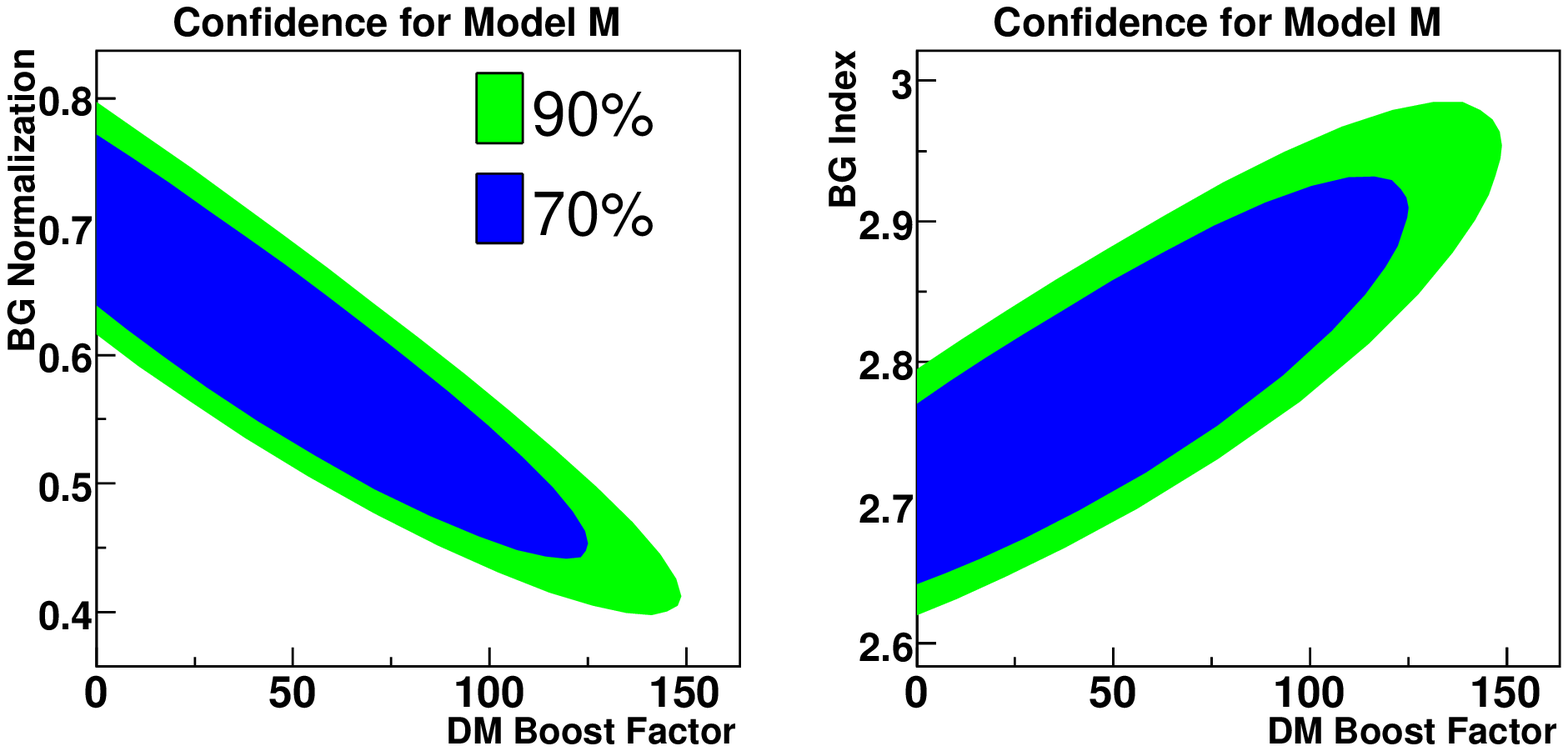}
\caption{Confidence regions for Models K to M.}
\end{figure}
\clearpage

\section{Maximum DM Boost Spectra}

\begin{figure}[htp]
\centerline{\hbox{\includegraphics[width=9cm]{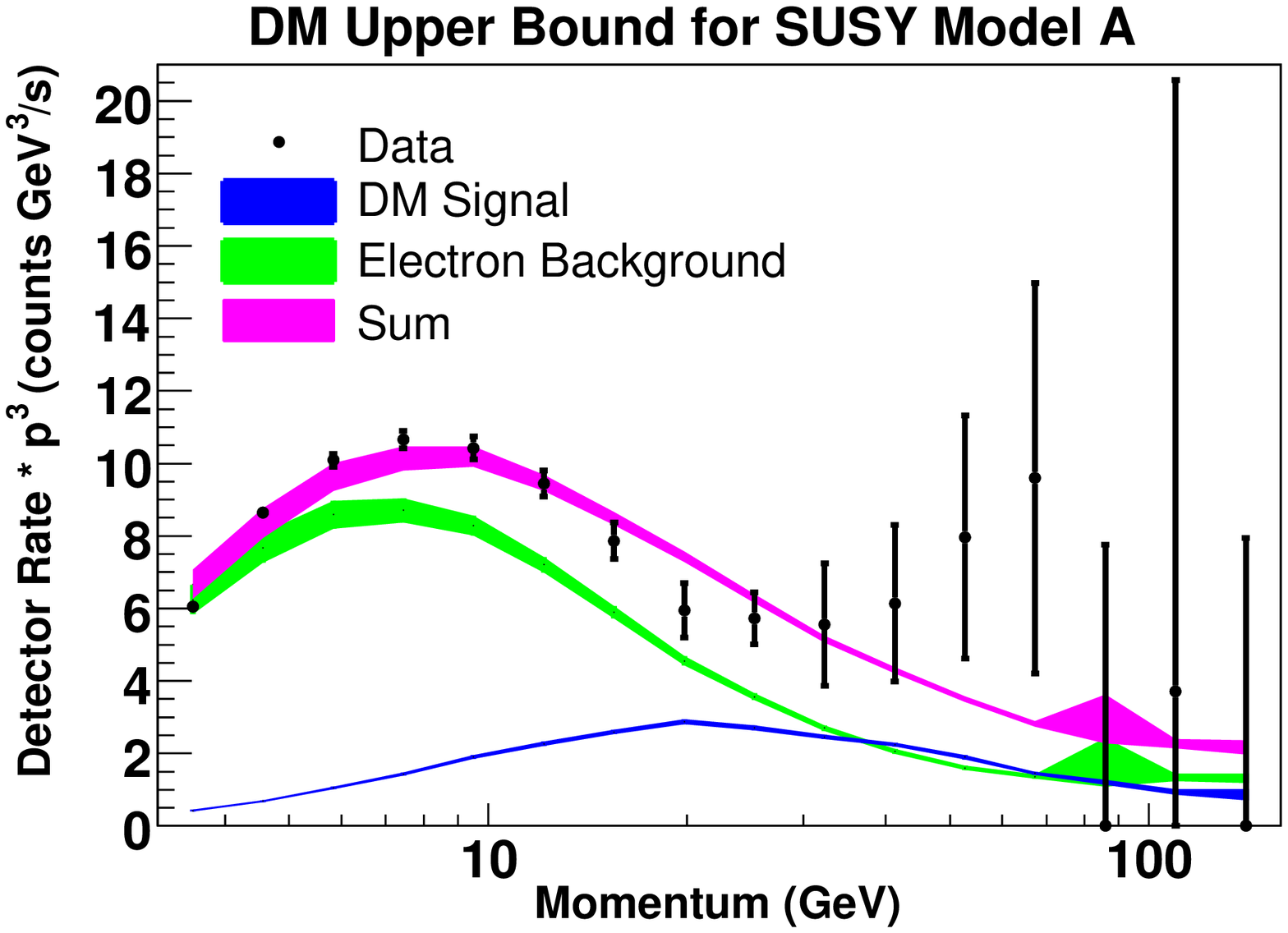} 
	\includegraphics[width=9cm]{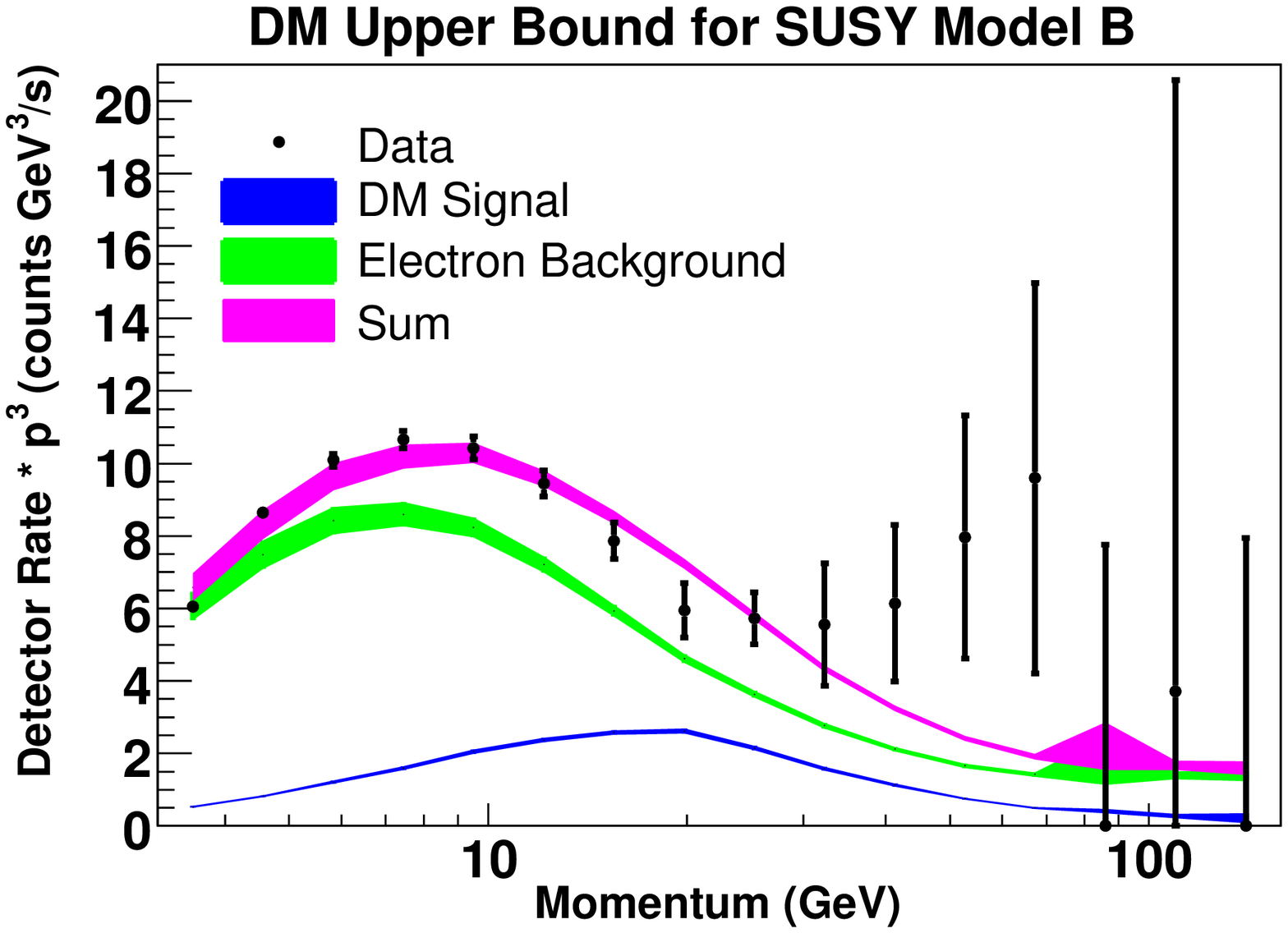}}}
\centerline{\hbox{\includegraphics[width=9cm]{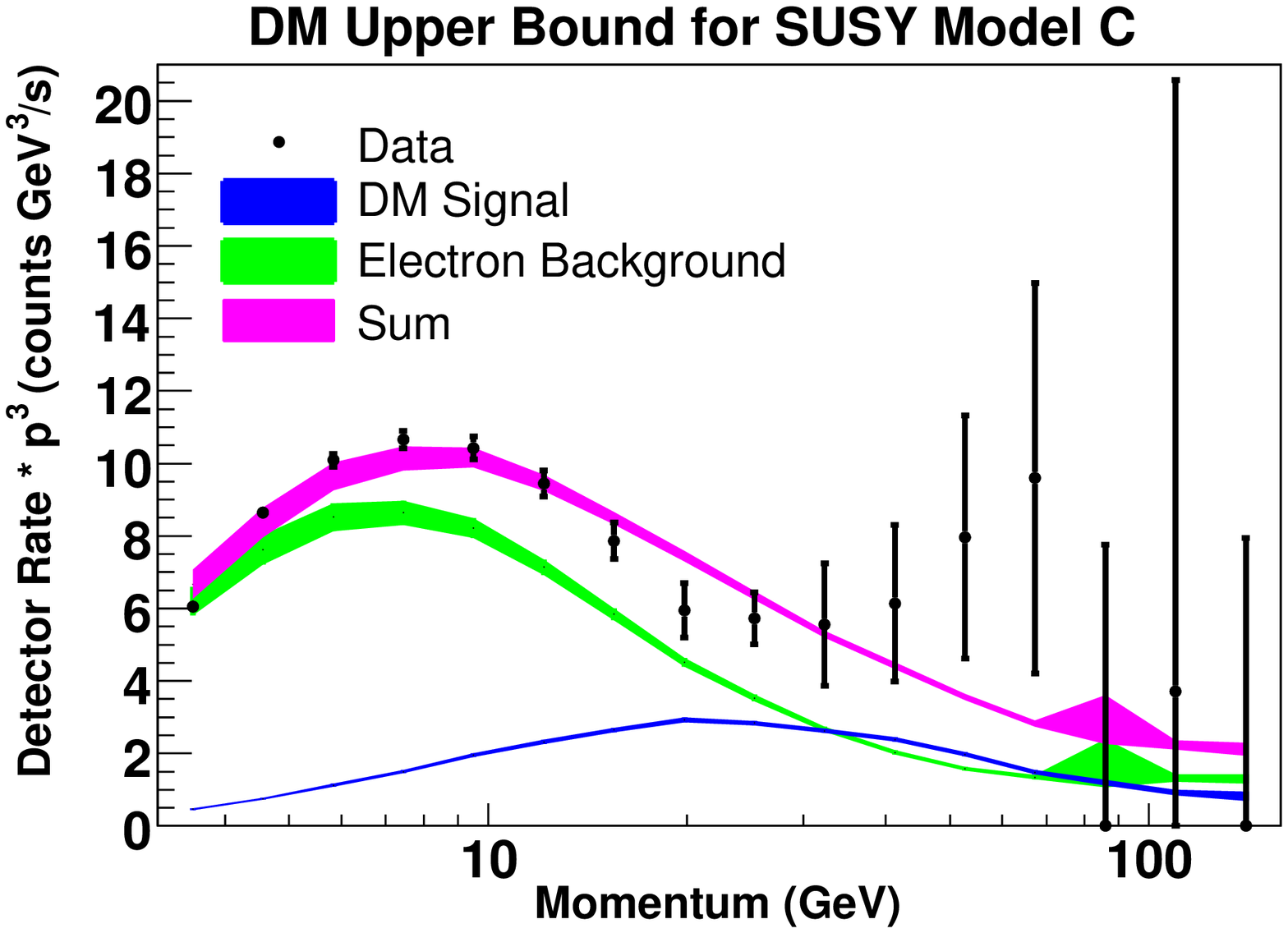} 
	\includegraphics[width=9cm]{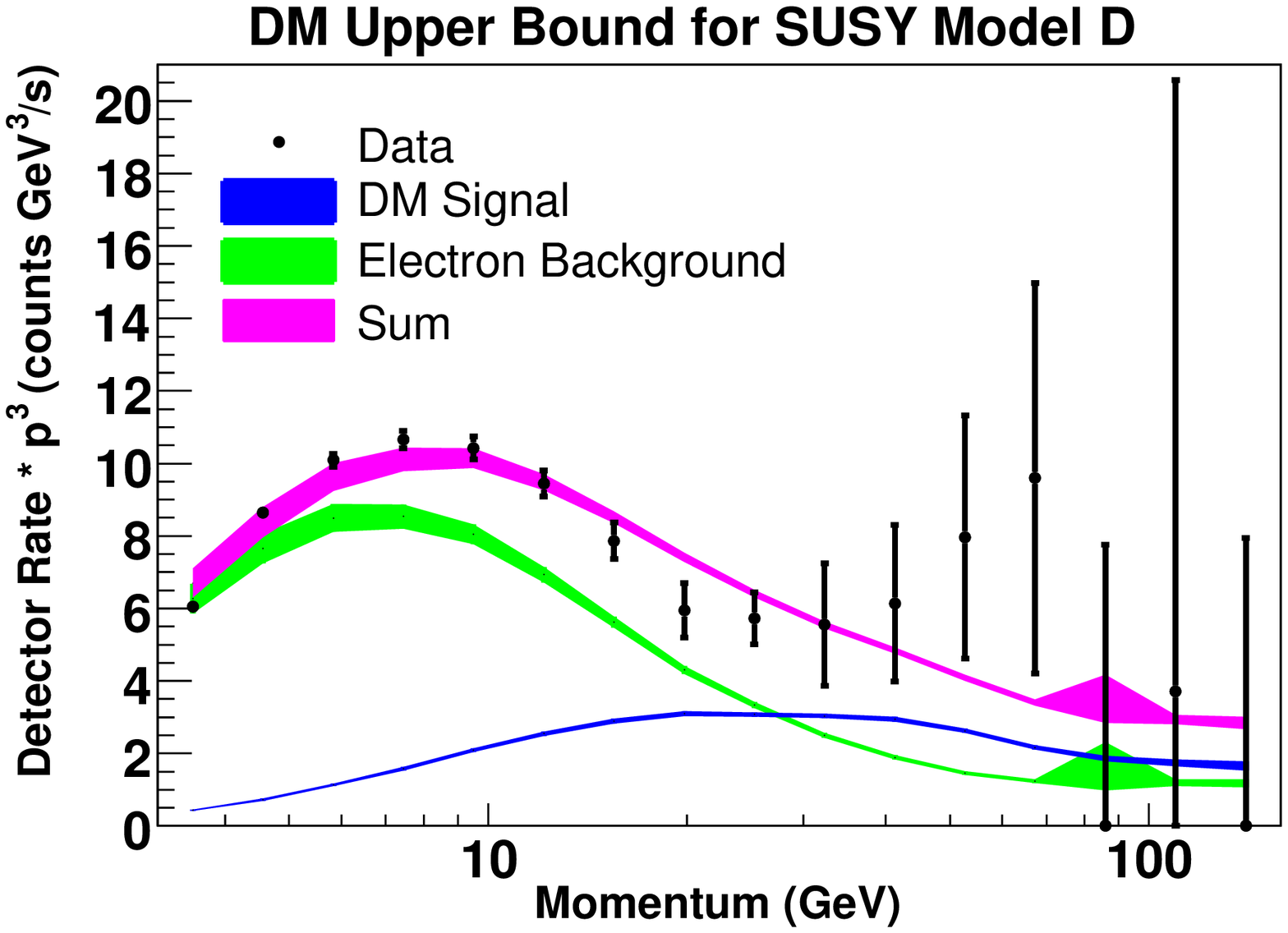}}}
\caption{Spectra for Models A and D at 90\% maximum boost limit.}
\end{figure}

\begin{figure}[htp]
\centerline{\hbox{\includegraphics[width=9cm]{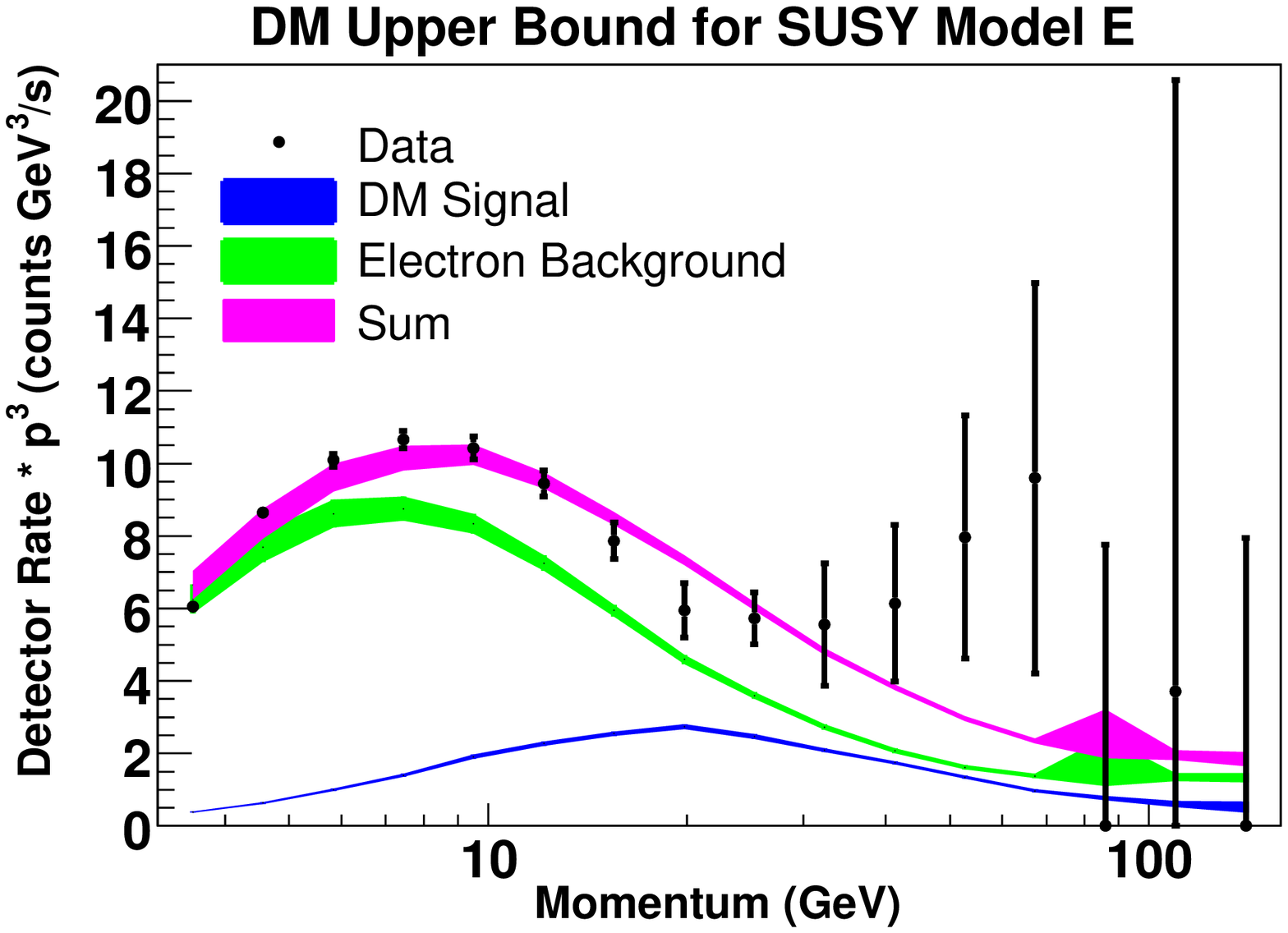} 
	\includegraphics[width=9cm]{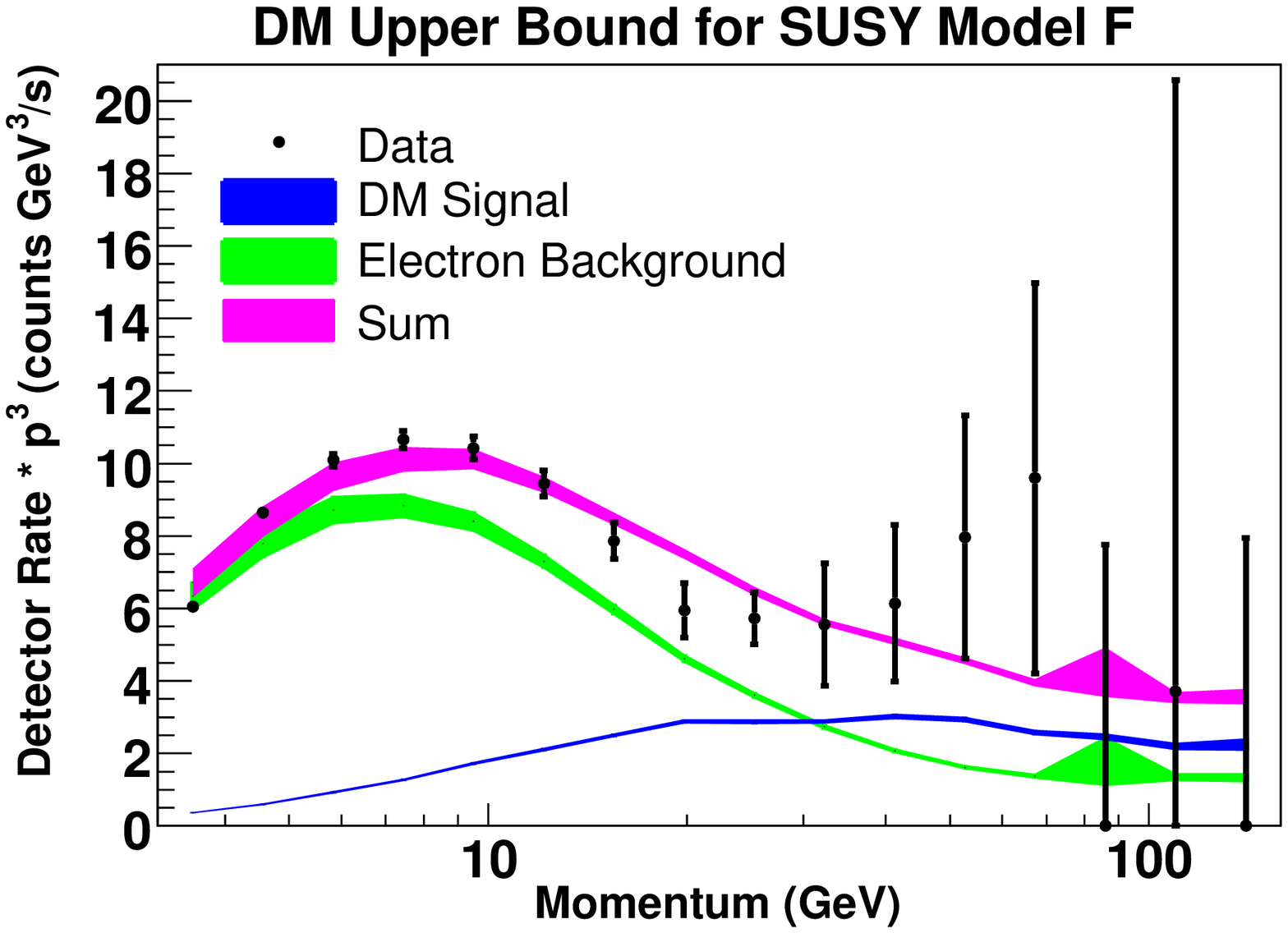}}}
\centerline{\hbox{\includegraphics[width=9cm]{alphabet_dm_predicted_spectrum_6_alternate.eps} 
	\includegraphics[width=9cm]{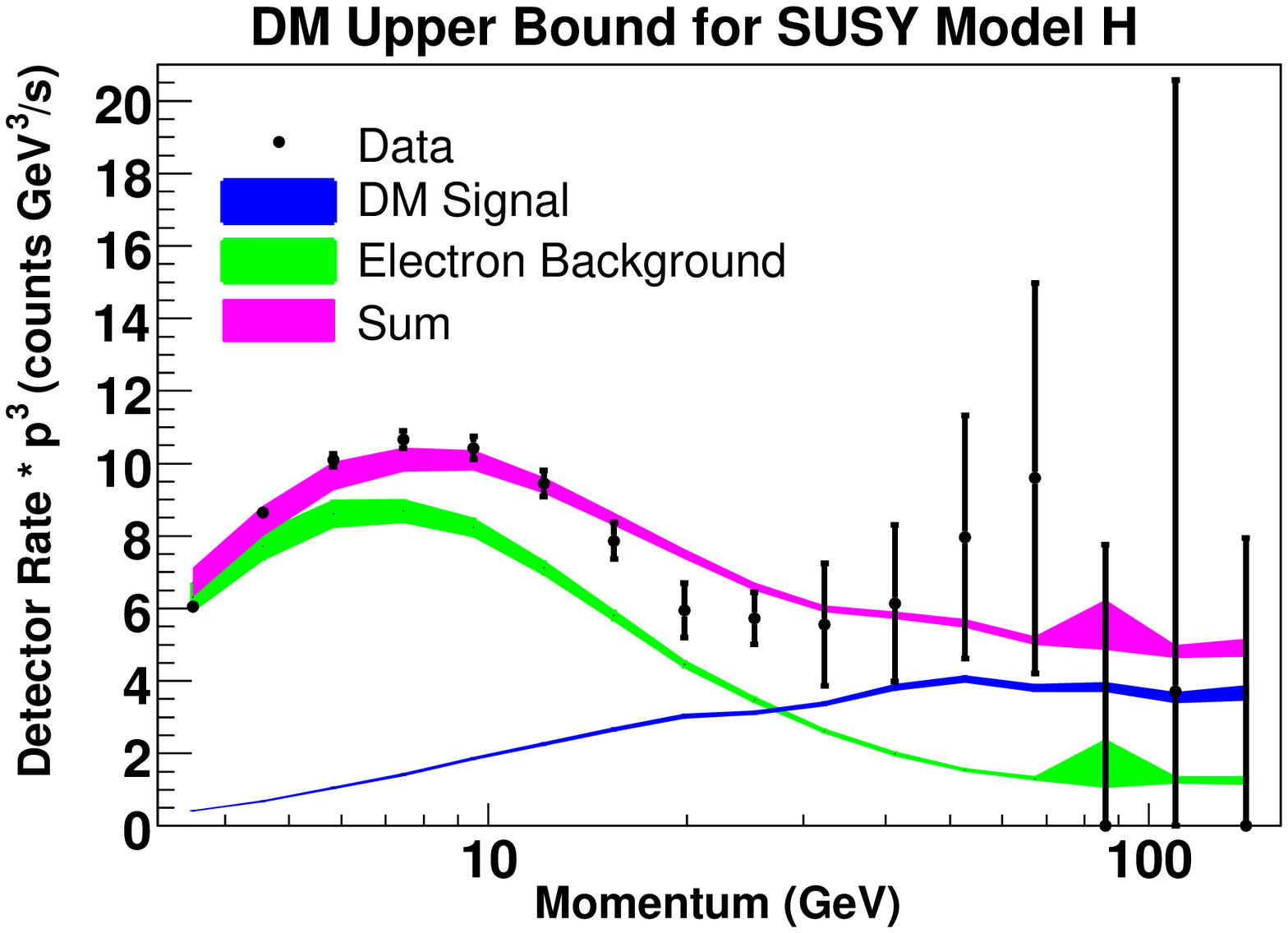}}}
\centerline{\hbox{\includegraphics[width=9cm]{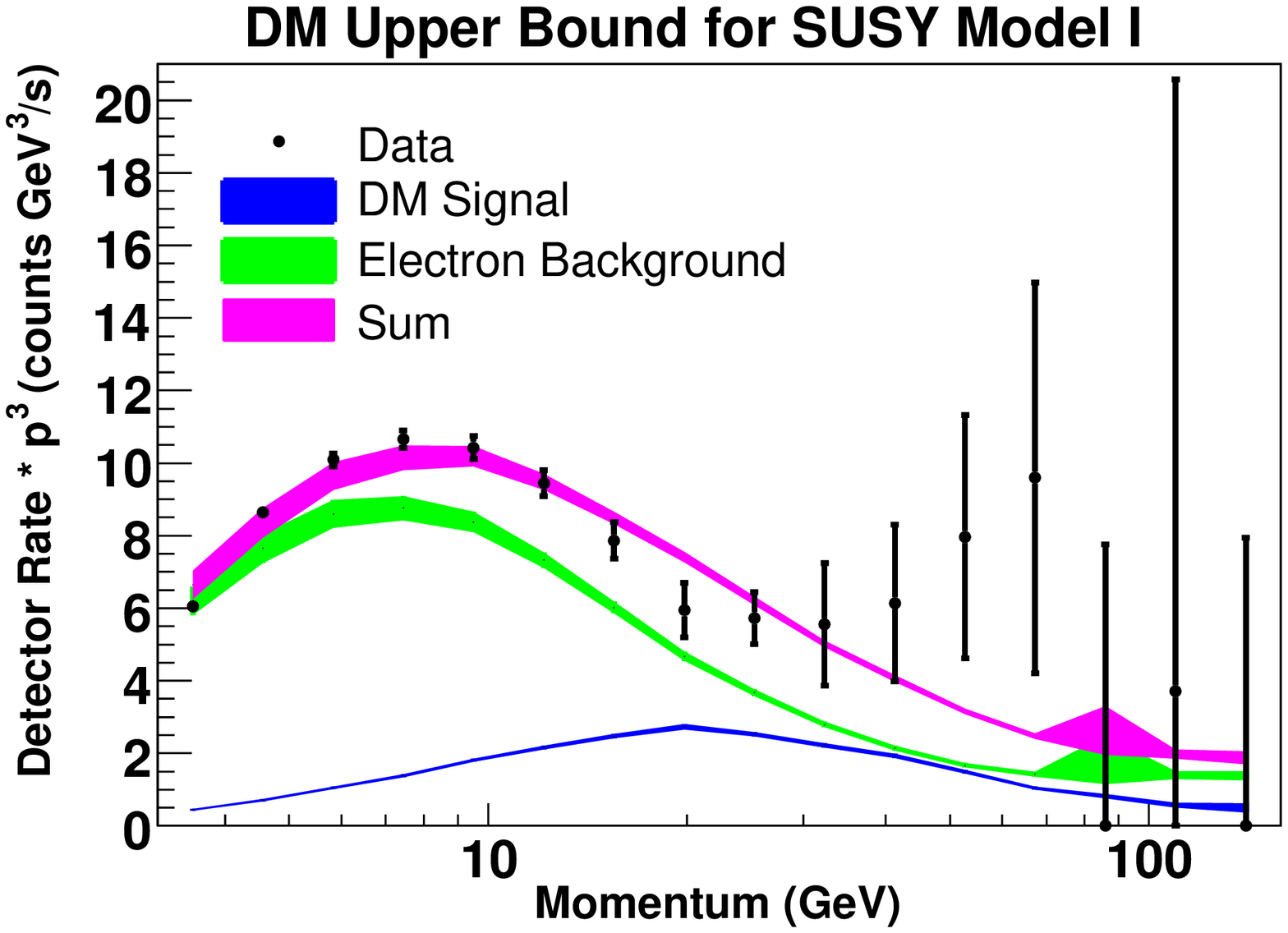} 
	\includegraphics[width=9cm]{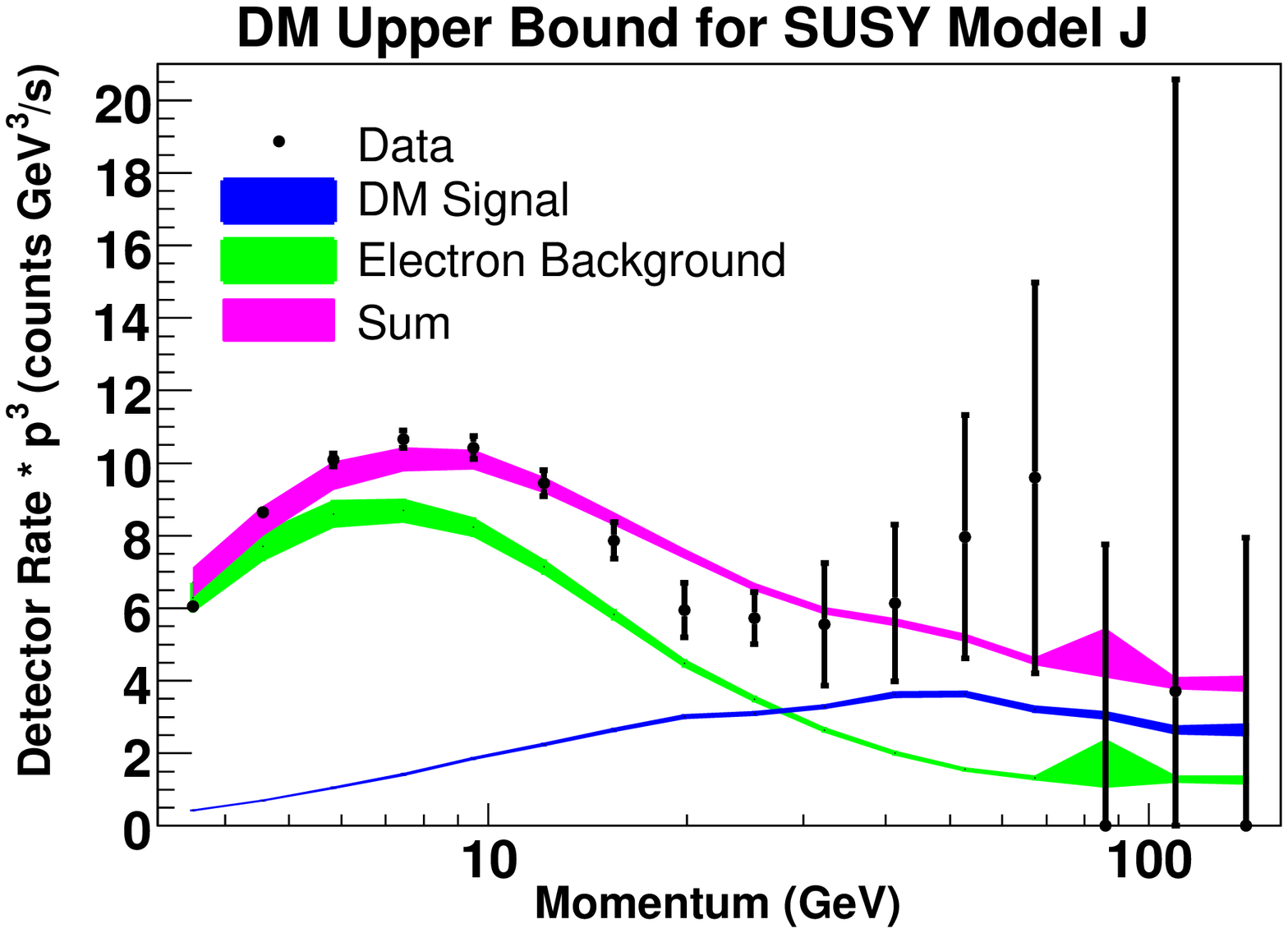}}}
\caption{Spectra for Models E to J at 90\% maximum boost limit.}
\end{figure}

\begin{figure}[htp]
\centerline{\hbox{\includegraphics[width=9cm]{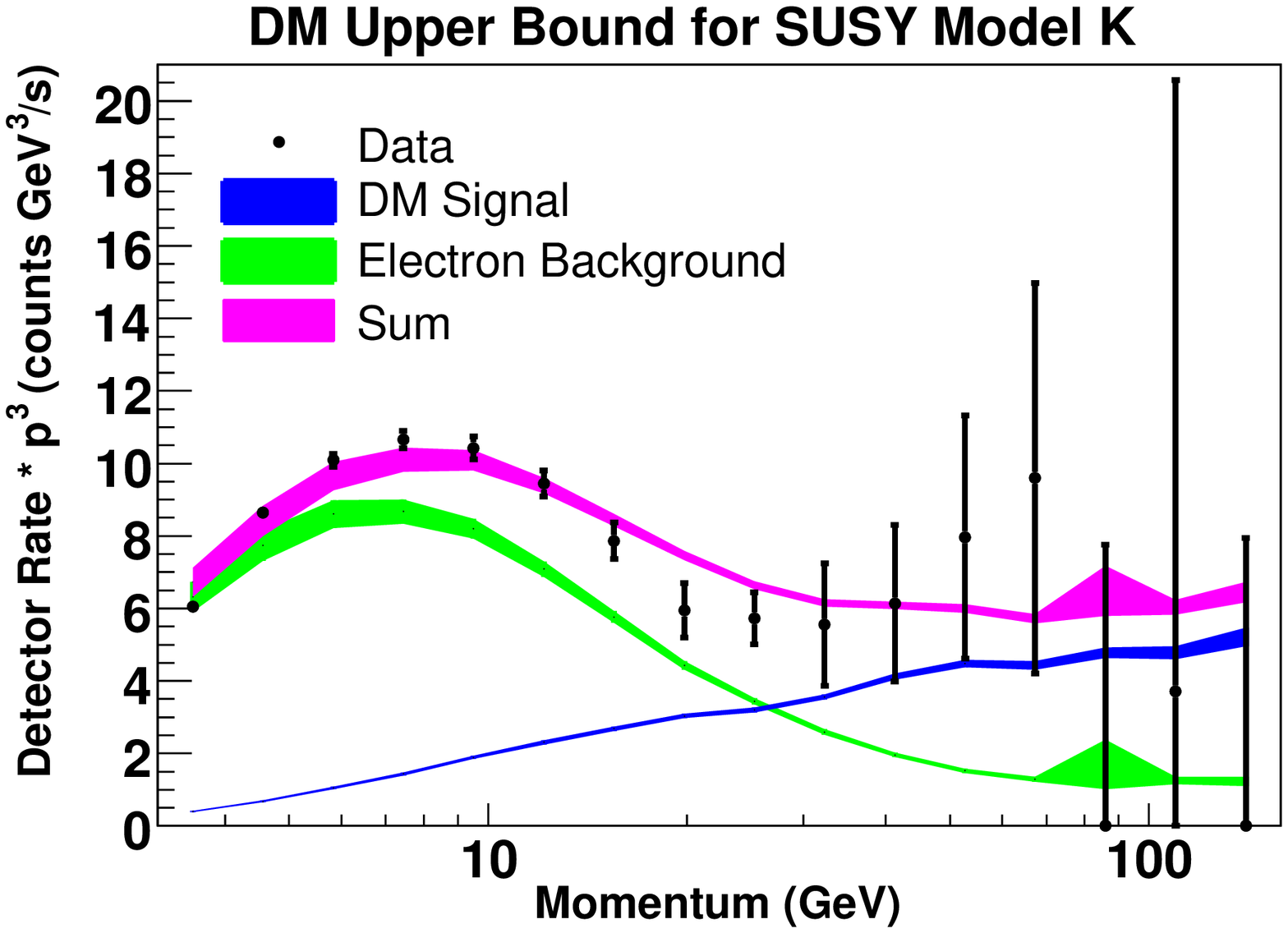} 
	\includegraphics[width=9cm]{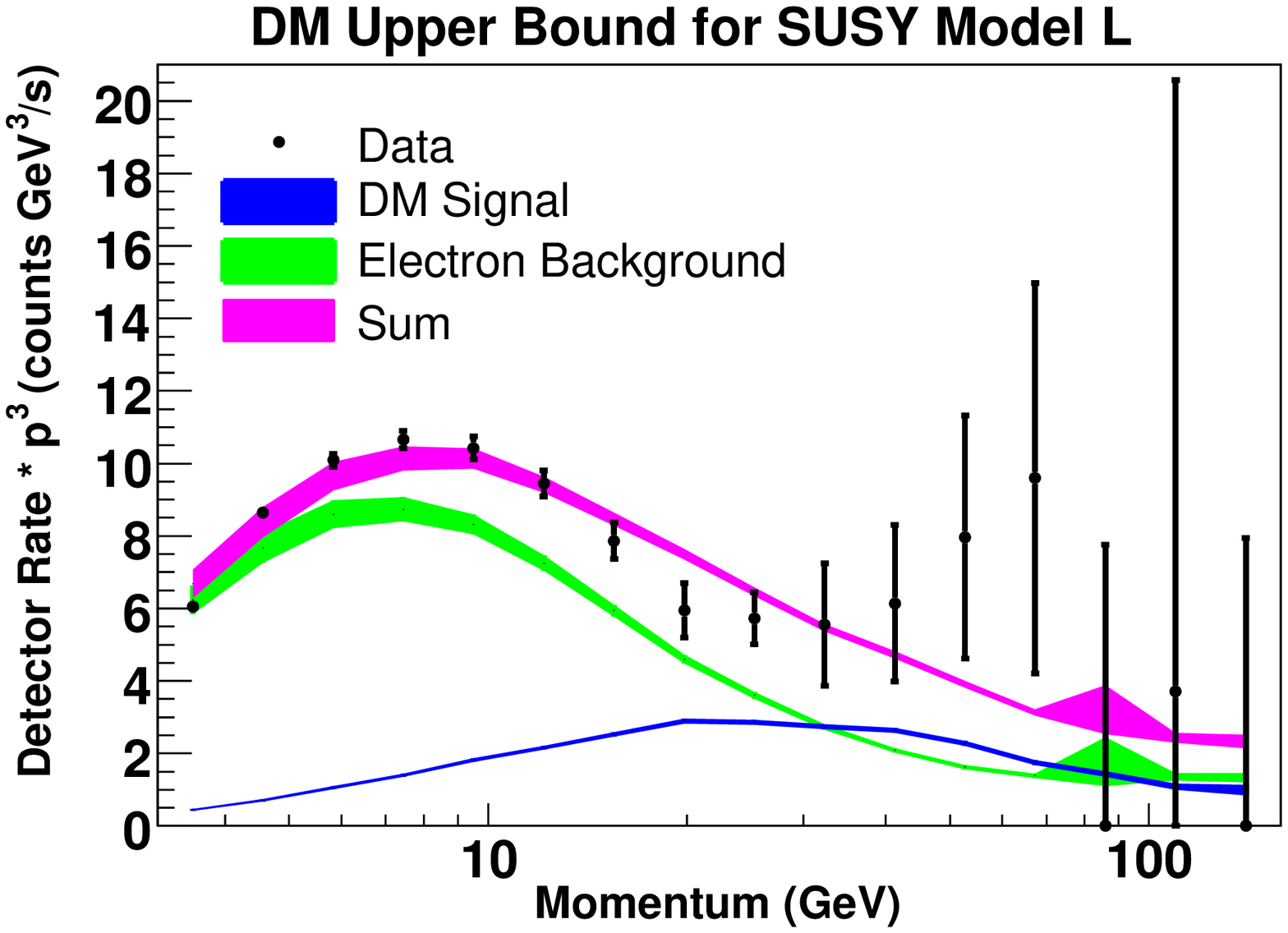}}}
\centerline{\includegraphics[width=9cm]{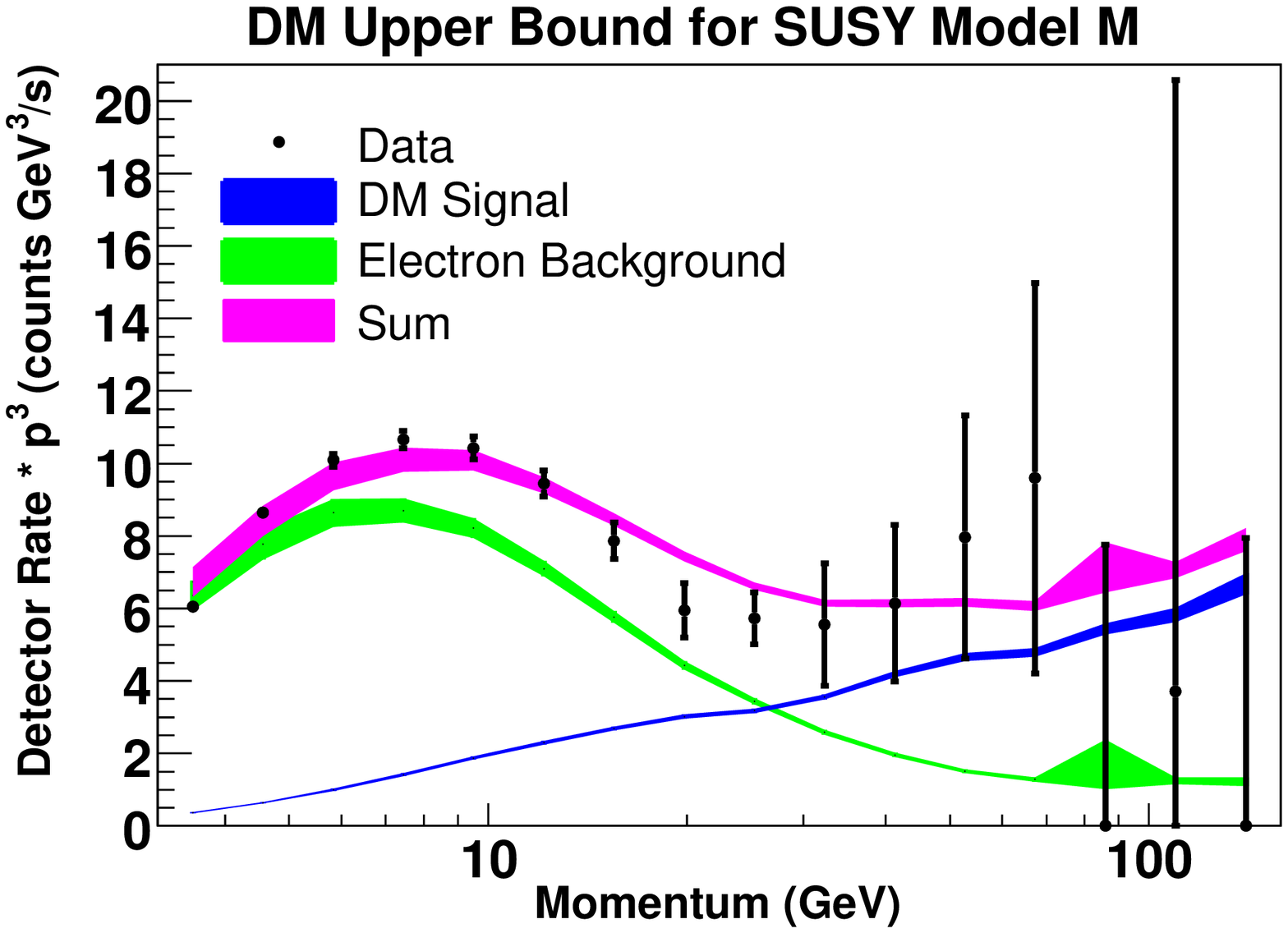} }
\caption{Spectra for Models K to M at 90\% maximum boost limit.}
\end{figure}